\documentclass[traditabstract,longauth]{aa} % for a referee version

\usepackage{graphicx}
\usepackage{txfonts}
\usepackage{natbib}
\usepackage{threeparttable}
  \def\PRODIMO{{\sc ProDiMo}}

  \def\Rin{R_{\rm in}}

  \def\amin{{a_{\rm min}}}
  \def\amax{{a_{\rm max}}}

\begin{document}

   \title{Gas modelling in the disc of HD 163296}
   \titlerunning{Gas modelling in the disc of HD 163296}
   \author{I.~Tilling\inst{1}
     \and P.~Woitke\inst{2,3,4}
     \and G.~Meeus\inst{5}
     \and A.~Mora\inst{6}
     \and B.~Montesinos\inst{7}
     \and P.~Riviere-Marichalar\inst{7}
     \and C.~Eiroa\inst{5} 
     \and W.-F.~Thi\inst{8}
     \and A.~Isella\inst{9}          
     \and A.~Roberge\inst{10}
     \and C.~Martin-Zaidi\inst{8}
     \and I.~Kamp\inst{11}
     \and C.~Pinte\inst{8}
     \and G.~Sandell\inst{12}
     \and W.~D.~Vacca\inst{12}
     \and F.~M{\'e}nard\inst{8}
     \and I.~Mendigut{\'i}a\inst{7}
     \and G.~Duch{\^e}ne\inst{8,13}
     \and W.~R.~F.~Dent\inst{14}
     \and G.~Aresu\inst{11}
     \and R.~Meijerink\inst{15}
     \and M.~Spaans\inst{11}
     %\and C.~A.~Grady\inst{14}
}
   \institute{%1 %Woitke Rice Tilling Thi Phillips
             Institute for Astronomy, University of Edinburgh,
             Royal Observatory, Blackford Hill, Edinburgh EH9 3HJ, UK;
         \and %2
             University of Vienna, Dept. of Astronomy, T{\" u}rkenschanzstr. 17,A-1180 Vienna, Austria
         \and %3 %Woitke Wright
             UK Astronomy Technology Centre, Royal Observatory, Edinburgh,
             Blackford Hill, Edinburgh EH9 3HJ, UK;
         \and %4 %Woitke Poelman
             SUPA, School of Physics \& Astronomy, University of St.~Andrews,
             North Haugh, St.~Andrews KY16 9SS, UK;
         \and %5 %Meeus Eiroa
             Dep. de F\'isica Te\'orica, Fac. de Ciencias, UAM Campus 
             Cantoblanco, 28049 Madrid, Spain
	 \and %6 %Mora 
             ESA-ESAC Gaia SOC, P.O. Box 78. E-28691 Villanueva de 
             la Ca\~{n}ada, Madrid, Spain 
	 \and %7 %Montesinos Riviere-Marichalar Huelamo Mendigutia Barrado
             Departamento de Astrof{\'i}sica, Centro de Astrobiolog{\'i}a 
             (CAB, INTA-CSIC), ESAC Campus, P.O. Box 78, E-28691 Villanueva de la Ca\~nada, Madrid, Spain
         \and %8 %Menard Augereau Duchene Lebreton Pinte Thi 
                %Martin-Zaidi
  	     UJF-Grenoble 1 / CNRS-INSU, Institut de Plan{\'e}tologie et d’Astrophysique (IPAG) UMR 5274, Grenoble, F-38041, France
         \and %9
             Department of Astronomy, California Institute of Technology, MC 249-17, Pasadena, CA 91125, USA
	 \and %10 %Roberge Danchi
             NASA Goddard Space Flight Center, Exoplanets \& Stellar 
             Astrophysics, Code 667, Greenbelt, MD 20771, USA
         \and %11 %Kamp Podio Aresu
             Kapteyn Astronomical Institute, Postbus 800,
             9700 AV Groningen, The Netherlands;
	 \and %12 %Howard Sandell
             SOFIA-USRA, NASA Ames Research Center, Mailstop 211-3 Moffett 
             Field CA 94035 USA
         \and %13
             Astronomy Department, University of California, Berkeley, 
             CA 94720-3411, USA;
	 \and %14 %Dent de~Gregorio
             ESO-ALMA, Avda Apoquindo 3846, Piso 19, Edificio Alsacia, 
             Las Condes, Santiago, Chile
         \and %15 %Meijerink
             Leiden Observatory, Leiden University, P.O. Box 9513, NL-2300 RA, Leiden, The Netherlands
 }

   \date{Received; accepted}

   \abstract{We present detailed model fits to observations of the disc around the Herbig Ae star HD 163296. This well-studied object has an age of $\sim$\,4\,Myr, with evidence of a circumstellar disc extending out to $\sim$\,540\,AU. We use the radiation thermo-chemical disc code \PRODIMO\ to model the gas and dust in the circumstellar disc of HD 163296, and attempt to determine the disc properties by fitting to observational line and continuum data. These include new Herschel/PACS observations obtained as part of the open-time key program GASPS (Gas in Protoplanetary Systems), consisting of a detection of the [{\sc Oi}]\,63\,$\mu$m line and upper limits for several other far infrared lines. We complement this with continuum data and ground-based observations of the $^{12}$CO 3-2, 2-1 and $^{13}$CO J=1-0 line transitions, as well as the H$_2$ S(1) transition. We explore the effects of stellar ultraviolet variability and dust settling on the line emission, and on the derived disc properties. Our fitting efforts lead to derived gas/dust ratios in the range 9-100, depending on the assumptions made. We note that the line fluxes are sensitive in general to the degree of dust settling in the disc, with an increase in line flux for settled models. This is most pronounced in lines which are formed in the warm gas in the inner disc, but the low excitation molecular lines are also affected. This has serious implications for attempts to derive the disc gas mass from line observations. We derive fractional PAH abundances between 0.007 and 0.04 relative to ISM levels. Using a stellar and UV excess input spectrum based on a detailed analysis of observations, we find that the all observations are consistent with the previously assumed disc geometry.}


   \maketitle

%=====================================================================

\section{Introduction} %

Protoplanetary discs set the initial conditions for planet formation, and the processing of chemical species and dust grains as discs evolve may ultimately provide the seed for life. Indeed, conditions within circumstellar discs are conducive to a rich chemistry, with a large variety of conditions ranging from icy mantles on grain surfaces in the cold disc midplane to ionised gas bathed in stellar ultraviolet radiation at the disc surface. In order to attempt to probe the gas in discs it is necessary to observe  gas emission lines from a variety of chemical species (molecules, atomic species, ions), and couple these observations with detailed thermochemical disc modelling.

\begin{table*}
\centering
\caption{Observed line data with {\sc
         Herschel/PACS}. Detections are listed as
         $F_L\!\pm\!\sigma)$ whereas non-detections quote $<3\!\sigma$. 
         There is an additional absolute flux calibration error of 30\%. In the case of the [{\sc Cii}]\,158\,$\mu$m line, strong contamination from the surrounding background prevented us from estimating a meaningful upper limit.}
\begin{tabular}{lcccccc}
\hline
Species       &  $\lambda$\,[$\mu$m] & $\nu$\,[GHz] & Line Flux [$10^{-18}$\,W\,m$^{-2}$] & Continuum\,(RMS) [Jy] \\
\hline\hline
OI            &  63.18             & 4745.05  & 193.1$\,\pm\,5.8$ & $16.46\,\pm\,0.07$\\
OI            &  145.52            & 2060.15  & $<$ 8.5     & $22.12\,\pm\,0.03$\\
CII           &  157.74            & 1900.55  & --           & $23.90\,\pm\,0.05$\\ 
p-H$_2$O      &  89.99             & 3331.40  & $<$ 9.4      & $19.14\,\pm\,0.04$\\
o-H$_2$O      &  179.52            & 1669.97  & $<$ 14.5     & $22.28\,\pm\,0.08$\\
o-H$_2$O      &  180.42            & 1661.64  & $<$ 16.2     & $21.58\,\pm\,0.09$\\
o-H$_2$O      &  78.74             & 3810.01  & $<$ 15.0     & $21.00\,\pm\,0.10$\\
OH            &  79.11             & 3792.19  & $<$ 17.0     & $20.70\,\pm\,0.10$\\
OH            &  79.18             & 3788.84  & $<$ 17.0     & $20.60\,\pm\,0.10$\\
CO J=36-35    &  72.85             & 4115.20  & $<$ 11.6     & $17.30\,\pm\,0.03$\\
CO J=33-32    &  79.36             & 3777.63  & $<$ 22.8     & $18.12\,\pm\,0.07$\\
CO J=29-28    &  90.16             & 3325.12  & $<$ 11.1      & $19.10\,\pm\,0.05$\\
CO J=18-17    &  144.78            & 2070.68  & $<$ 13.1     & $22.05\,\pm\,0.04$\\
\hline
\end{tabular}
\label{tab:obsflux}
\end{table*}

One important disc property for planet formation models is the amount of gas present as it evolves. The dust characteristics are generally better-understood than the gas, and the ISM gas/dust mass ratio of 100 is often adopted to give an estimate of the gas mass. However, gas masses derived from  millimetre and sub-mm CO emission observations are typically lower than from dust observations with this assumption, sometimes by a factor $\sim$\,100 \citep{Thi2001}. This is variously attributed to gas depletion due to photoevaporation or planet formation, and CO freeze-out on grains \citep{Zuckerman1995, Kamp2000}. These low level rotational CO transitions are also strongly dependent on the disc size, and often optically thick under certain disc conditions, limiting their ability to probe the deeper disc layers \citep{Panic2009,Hughes2008}. In order to better constrain the disc gas mass it is necessary to observe additional gas species, such as those observable in the far-infrared. The FIR range is a crucial observing window which can help to resolve some of these ambiguities. The fine structure lines of atomic oxygen and ionised carbon probe the gas in the warm disc surface layers, and as products of CO photodissociation provide a good test for this proposed explanation of CO under-abundance. In addition, the high level rotational transitions of CO arise from warm gas in the inner disc, probing different disc regions to the low level mm CO lines.

In addition to the discrepancy between gas mass estimates from millimetre continuum and CO observations, there is uncertainty regarding the derived disc outer radii from CO and continuum observations. The outer radius suggested by model fits to dust emission is often found to be smaller than that of the gas disc. In the case of the Herbig Ae star AB Aurigae, \citet{Pietu2005} suggested that a change in dust grain properties, leading to reduced opacity in the outer disc, could explain this discrepancy. For HD 163296, \citet{Isella2007} attributed the apparent difference in disc radii to a sharp drop in surface density, opacity or dust-to-gas ratio beyond 200\,AU, while \citet{Hughes2008} proposed that the apparently conflicting dust and gas observations could be reconciled using a different treatment of the disc surface density at the outer edge, motivated by similarity solutions for the time evolution of accreting discs. This study attributed smaller derived disc radii from dust observations to detector sensitivity thresholds, arguing in favour of larger dust discs, albeit with an exponentially-tapered density at the outer edge. A steepening of the density profile in the outer disc has also been observed in the case of DM Tau, LkCa 15 and MWC 480 by \citet{Pietu2007}.

The Herschel open-time key program GASPS (Gas in Protoplanetary Systems) aims to characterise the gas in discs at each stage of their evolution, by observing the far-IR lines of a wide range of discs, from young gas-rich accretion discs to gas-poor debris discs \citep{Mathews2010}. This paper represents the second in-depth study of a Herbig disc as part of this program, following after a study focusing on the young Herbig Ae star HD 169142 \citep{Meeus2010}.

HD 163296 is an isolated Herbig Ae star with spectral type A1Ve, mass $\sim$\,2.5\,M$_{\odot}$ and stellar luminosity $\sim$\,38\,L$_{\odot}$ (this work), situated at a distance of $118.6^{+12.7}_{-10.0}$pc \citep{vanLeeuwen2007}. Scattered light and millimetre continuum observations indicate the presence of an inclined circumstellar disc extending out to a radius of 540AU \citep{Mannings1997, Grady2000, Wisniewski2008}. In addition, there is evidence of an asymmetric outflow perpendicular to the disc, with a chain of six Herbig-Haro knots tracing its mass-loss history \citep{Devine2000}. The derived disc dust mass is in the range $(5-17) \times 10^{-4}$M$_{\odot}$ \citep{Natta2004,Tannirkulam2008a,Mannings1997,Isella2007}.

Recent near-infrared studies of the inner disc of HD 163296 indicate an inner dust rim feature enclosing a bright emission region extending inwards towards the star. This emission inwards of the dust rim has been attributed to an optically thin inner disc, although there is uncertainty regarding the size and composition of such a disc. Derived radii for the dust rim lie in the range (0.2 -- 0.55)\,AU \citep{Renard2010,Eisner2009,Tannirkulam2008,Benisty2010,Monnier2006}. An increase in opacity at the inner rim due to sublimation of grains could cause the rim to puff up, and it has been suggested that this leads to time-variable self-shadowing of the outer disc \citep{Sitko2008, Wisniewski2008}. Mid-infrared imaging of warm dust in the disc surface layers seems to indicate little flaring \citep{Doucet2006}, and scattered light brightness profiles are consistent with a non-flared outer disc \citep{Wisniewski2008}. Fits to the ISO-SWS spectra of HAeBe discs led \citet{Meeus2001} to classify HD 163296 as one of their group II objects, in which an optically thick inner disc shadows the outer disc from stellar radiation, resulting in a non-flared geometry.   

In this study we aim to fit a wide variety of observed emission lines and the dust spectral energy distribution (SED) simultaneously with a disc with this observed geometry. 

\section{Herschel/PACS Observations}

\begin{figure}
\centering
\includegraphics[width=85mm]{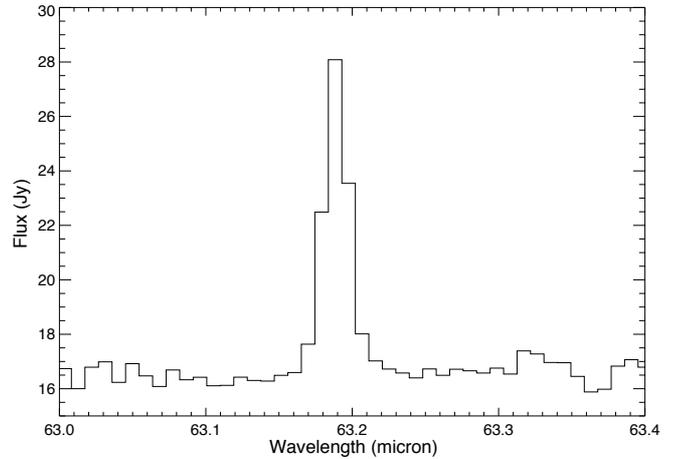}\\
\caption{Observed [{\sc Oi}]63$\mu$m emission line with Herschel/PACS.}
\label{fig:PACS}
\end{figure}

We have obtained observations of HD 163296 in the far-infrared using Herschel/PACS, consisting of spectroscopic line observations and photometry. These include a detection of the [{\sc Oi}]\,63\,$\mu$m fine structure line, and upper limits for a number of other atomic, ionic and molecular far-IR lines. We observed HD 163296 with the PACS spectrometer in line
spectroscopy mode (obsid 1342192161, 1252 sec, 3 repetitions)
and range scan mode (obsid 1342192160, 5141 sec, 3 repetitions),
while chopping and nodding.

The spectra were reduced using HIPE version 7.0, with the standard pipeline scripts for a chopped line scan of a point source. The reduction steps include division by the spectral response function and MedianOffsetCorrection, as the new flat fielding task is not yet robust for short range scans such as the present observations (PACS instrument team, private communication). For this paper, we only make use of
the data contained in the central spaxel, to which we apply a
wavelength-dependent correction factor for the loss of flux
due to diffraction losses. With these additional steps, the
absolute flux calibration accuracy is 30\% (see PACS Spectroscopy
Performance and Calibration Document, http://herschel.esac.esa.int/AOTsReleaseStatus.shtml).

%\begin{figure}
%\centering
%\includegraphics[width=85mm]{HD_163296_allblue_lines.eps}\\
%\vspace*{2mm}
%\includegraphics[width=85mm]{HD_163296_allred_lines.eps}\\
%\caption{Observed spectra with Herschel/PACS.}
%\label{fig:PACS}
%\end{figure}

\begin{figure*}
\centering
\hspace*{-5mm}
\includegraphics[width=185mm]{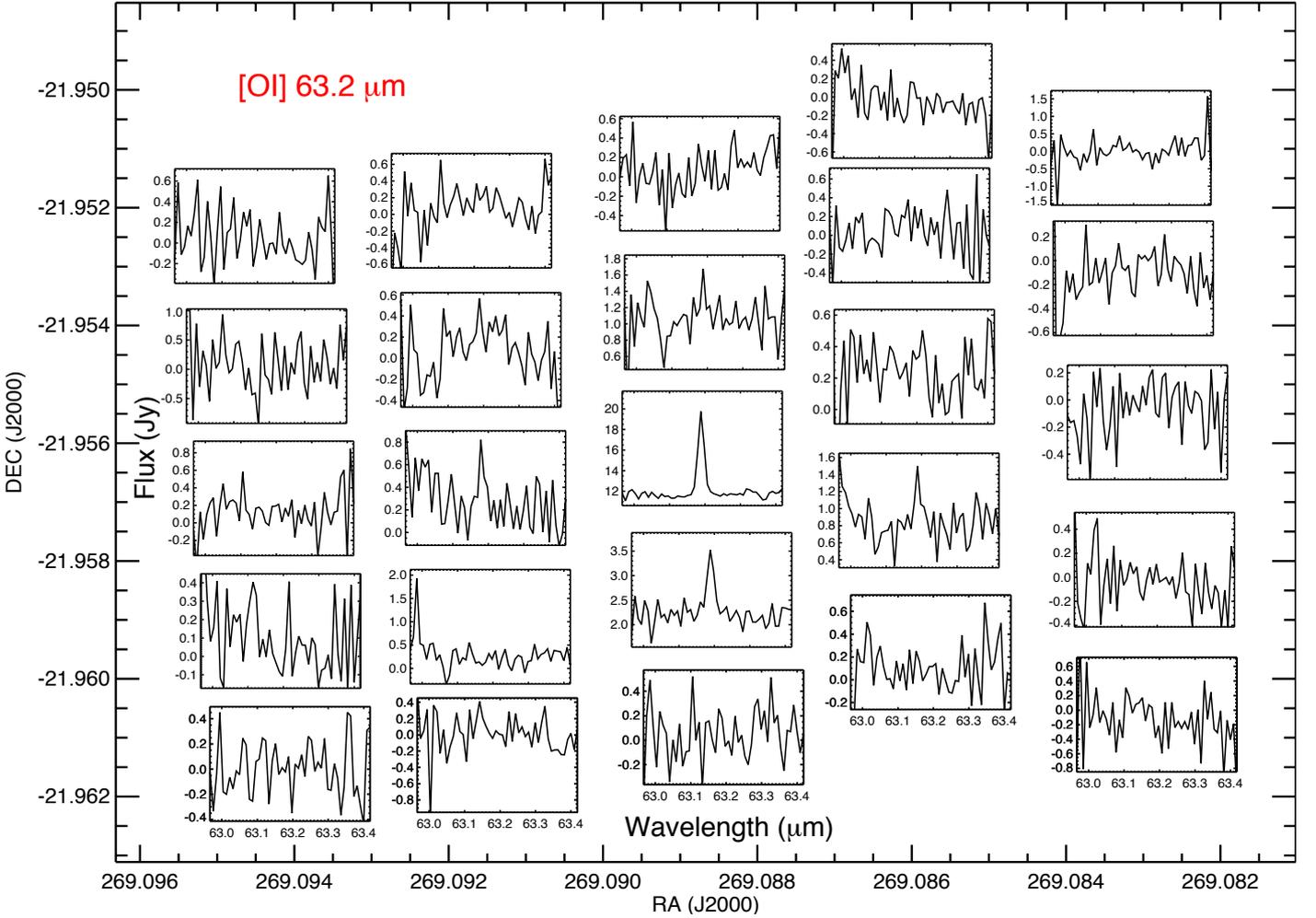}\\
\caption{[{\sc Oi}]63$\mu$m spectra observed by Herschel/PACS towards HD 163296. Each spectrum corresponds to a 9.4$\times$9.4 arcsec pixel centered at the given coordinates. Emission is dominated by the central pixel, corresponding to a maximum outer radius of $\sim$\,560AU, assuming a distance of 118.6 parsecs to this object.}
\label{fig:OI63spaxels}
\end{figure*}

The line fluxes and upper limits are given with continuum data in Table~\ref{tab:obsflux}. The PACS photometry data are given in Table~\ref{tab:PACSphot}. Comparison of the azimuthally averaged radial profiles with the observed point spread function (PSF), using Vesta, show that the disc is unresolved at both 70$\mu$m and 160$\mu$m. The spatial extent of the observed [{\sc Oi}]\,63\,$\mu$m emission is illustrated in Fig.~\ref{fig:OI63spaxels}, where it can be seen that the emission is dominated by the central 9.4$\times$9.4 arcsecond spaxel. This corresponds to a maximum radial extent of ~\,560\,AU for the emission.

The pipeline reduction of the [{\sc Cii}]\,158\,$\mu$m range shows this
line in absorption, which is unexpected. Therefore, we also
looked at the unchopped spectra in the different chop positions.
It is clear that the region around HD 163296 is full of [{\sc Cii}]\,158\,$\mu$m
emission, and that this causes the emission feature at the position of the star
to cancel out (and even become an absorption feature). This is likely due to cloud material along the line-of-sight, since HD 163296 is located close to the Galactic plane, and is consistent with the presence of CO emission at radial velocities offset from this object, but unresolvable with PACS \citep{Thi2001}. We also inspected all the other
lines for their unchopped appearance, but did not find contamination
through the offset positions in those cases.

\begin{table}
\centering
\caption{Photometric fluxes observed with Herschel/PACS. These observations (AORS 1342228401 and 1342228401) are done with the standard miniscan.}
\begin{tabular}{lcc}
\hline
$\lambda$\,[$\mu$m] & 70               & 160\\
\hline
Flux [Jy]           & $18.5\!\pm\!3.3$ & $20.2\!\pm\!4.0$\\
\hline
\end{tabular}
\label{tab:PACSphot}
\end{table}

\section{Properties of HD 163296}
\label{Section:StellarProperties}

The study of the disc of HD 163296 is directly linked to the knowledge of the
absolute parameters of the star itself and the behaviour of the spectral
energy distribution (SED) in the ultraviolet (UV), optical and near infrared
(near-IR), which is mostly dominated by stellar radiation; in particular, the
energy emitted at wavelengths shorter than 2500 \AA{} is a fundamental piece of
information used to model the disc, due to its impact on the photo-chemistry and gas heating \citep{Woitke2010}. This section describes the stellar
properties and studies the spectral energy distribution at the range
mentioned above.

\subsection{Stellar parameters}
\label{Subsection:StellarParameters}

The determination of the stellar parameters of HD 163296 was done in an
extensive work devoted to the study of a sample of Herbig Ae/Be stars
\citep{Montesinos2009}. Details of the methodology followed and the
observations used can be found in that paper. We outline in this section the
relevant steps of the process.

An iterative distance-independent algorithm based on the analysis of the SED
and optical high-resolution spectra (for the estimation of the effective
temperature and metallicity), and mid-resolution spectra of the Balmer lines
H$\beta$, H$\gamma$ and H$\delta$ (for the estimation of the gravity), was
applied.

The knowledge of the effective temperature, gravity and metal abundance allows
us to place the star in a $\log g_* - \log T_{\rm eff}$ diagram and
superimpose the appropriate set of evolutionary tracks and isochrones for that
specific metallicity \citep{Yi2001}. This gives directly --or after a simple interpolation
between the tracks and isochrones enclosing the position of the star-- the
stellar mass and the age. Since there is a one-to-one correspondence between a
pair $(T_{\rm eff},\,g_*)$ on a given track and a pair $(T_{\rm
eff},\,L_*/L_\odot)$, the stellar luminosity can also be estimated.

\begin{table}
\centering
\caption[]{Parameters for HD 163296.}
\begin{tabular}{ll} \hline\hline
Temperature           & $9250\pm 150$\,K    \\
Gravity ($\log g_*$)     & $4.07\pm 0.09$   \\
Metallicity ([Fe/H])     & $+0.20\pm 0.10$  \\
Mass                     & $2.47\pm 0.10$\,$M_\odot$   \\ 
Luminosity               & $37.7\pm 7.0$\,$L_\odot$    \\ 
Age                      & $4.2\pm 0.4$\,Myr     \\ \hline
\end{tabular}
\label{Table:StellarParameters}
\end{table}

Table \ref{Table:StellarParameters} gives the stellar parameters of HD 163296
after a slight refinement of the results presented in \citet{Montesinos2009}.

\subsection{The SED: UV, optical and near-IR observations}
\label{sec:TheSED}

Simultaneous optical UBVRI and near-infrared JHK photometry obtained as part of the EXPORT collaboration \citep{Eiroa2001,Oudmaijer2001}, were used to build the optical and near-IR
photospheric spectral energy distribution of HD 163296. Table \ref{Table:Photometry} shows the wavelengths, magnitudes and
corresponding fluxes with their uncertainties. The calibration from magnitudes
to fluxes has been done using the zero points given by \citet{Bessell1979} for the
optical and \citet{Allen1999} for the near-IR magnitudes.

\begin{table}
\caption[]{Photometry for HD 163296.}
\begin{tabular}{llcc}\hline
Band  & $\lambda$ ($\mu$m) & Magnitude and error & Flux and error (Jy)\\\hline
U & 0.360 & $6.95\pm 0.03$ & $3.00\pm 0.08$ \\ 
B & 0.440 & $6.92\pm 0.04$ & $7.27\pm 0.27$ \\ 
V & 0.550 & $6.86\pm 0.03$ & $6.56\pm 0.18$ \\ 
R & 0.640 & $6.73\pm 0.05$ & $6.26\pm 0.29$ \\ 
I & 0.790 & $6.67\pm 0.10$ & $5.48\pm 0.51$ \\ 
J & 1.215 & $6.17\pm 0.05$ & $5.55\pm 0.26$ \\ 
H & 1.654 & $5.49\pm 0.05$ & $6.69\pm 0.31$ \\ 
K & 2.179 & $4.71\pm 0.05$ & $8.56\pm 0.39$ \\  \hline
\end{tabular}
\label{Table:Photometry}
\end{table}

Ultraviolet IUE spectra of HD 163296 were extracted from the INES
archive\footnote{http://sdc.cab.inta-csic.es/ines/}.  From the collection of
68 SW and LW (for ``short'' and ``long wavelength'') spectra in low
resolution, obtained through the large aperture of the spectrograph, all those
that looked clean (few or none bad pixels) and with the correct exposure
classification codes were selected. The spectra (SW+LW) cover the range
1250--3000 \AA{}.

A high-resolution, far-UV spectrum of HD 163296 was constructed by combining data taken with HST STIS and FUSE. 
The STIS spectra, covering 1150~\AA\ to 3000~\AA, were obtained from the HST STIS Echelle Spectral Catalog of Stars (StarCat\footnote{http://casa.colorado.edu/$\sim$ayres/StarCAT/}).
The FUSE spectrum of HD 163296 is extremely noisy below 970~\AA; therefore, we used only the portion between that wavelength and 1150~\AA\ in our high-resolution composite UV spectrum for HD 163296.
FUSE spectra contain strong terrestrial airglow emission lines, due to the large size of the observing aperture.  
We removed these lines from the HD 163296 spectrum before stitching it to the STIS spectrum. 

A comparison of the IUE and FUSE+STIS spectra shows that they differ
in the level of their continuum. The IUE spectra with the largest flux
levels are a factor $\sim\!1.6$ higher than the FUSE+STIS spectrum,
therefore, regarding the modelling, the ultraviolet contribution of
the star will be represented by two different UV input spectra, namely
a ``low UV'' state, corresponding to the FUSE+STIS spectrum, and a
``high UV'' state, corresponding to the average of the highest IUE
spectra\footnote{The average IUE spectrum was build with the following
  individual spectra: SWP 28805, SWP 28811, SWP 28813 and LWR 02065.}
(see section \ref{Subsection:InputSpectrum}).

\begin{figure} 
\centering
\includegraphics[width=9.0cm]{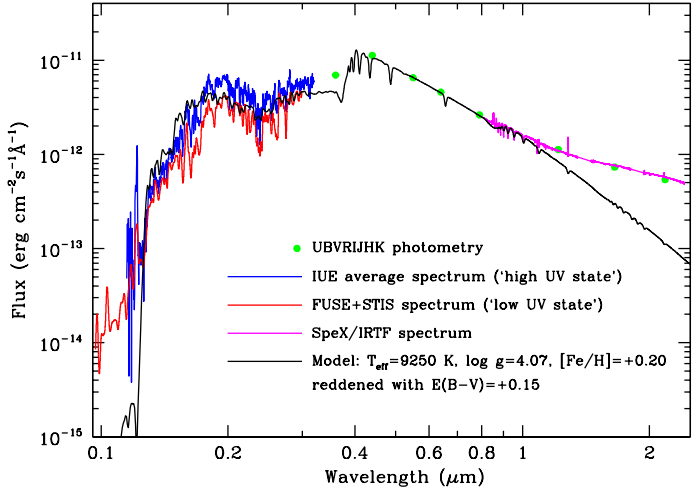}
\caption{Spectral energy distribution for HD 163296. The green dots represent
  the fluxes corresponding to the UBVRIJHK photometry. The size
of the dots is of the order or larger than the uncertainties. The IUE average
  spectrum and the FUSE+STIS spectrum are plotted in blue and red,
  respectively; the latter has been slightly smoothed to reduce the noise. The SpeX/IRTF spectrum is plotted in magenta. The
  black solid line is a photospheric model computed for the specific stellar
  parameters given in Table \ref{Table:StellarParameters}, reddened with
  E(B--V)=+0.15 and normalized at the flux in V. See text for details.}
\label{Figure:SED_UV_optical_nIR}
\end{figure}

In addition, a near-infrared spectrum of HD 163296 was obtained at the NASA
Infrared Telescope Facility (IRTF) on Mauna Kea on 2010 July 8 with SpeX
\citep{Rayner2003}. Ten individual exposures of the target, each lasting 32
s, were taken using the short-wavelength cross-dispersed (SXD) mode of
SpeX. This mode yields spectra spanning the wavelength range 0.8--2.4 $\mu$m
divided into six spectral orders. The slit width was set to 0$\farcs$3, which
yields a nominal resolving power of 2000 for the SXD spectra. Observations of
HD 156717, an A0 V star, used as a ``telluric standard'' to correct for
absorption due to the Earth's atmosphere and to flux calibrate the target
spectra, were obtained immediately preceding the observations of HD
163296. The airmass difference between the observations of the object and the
standard was 0.06. The seeing was estimated to be $\sim 0\farcs3-0\farcs5$ at
2.2 $\mu$m during the observations and conditions were clear. The statistical
S/N varies across the spectral range but is of the order of several hundred
over the entire SXD spectrum. The SpeX data were reduced with Spextool \citep{Cushing2004} and the telluric correction was performed using the procedures and software described by \citet{Vacca2003}.

Figure \ref{Figure:SED_UV_optical_nIR} shows the SED of HD 163296. The
solid black line is a photospheric model computed for $T_{\rm eff}\!=9250$ K,
$\log g_*\!=4.07$ and [Fe/H]=+0.20 from the GAIA grid of models created with
the PHOENIX code \citep{Brott2005}. The model has been reddened 
with E(B--V)=+0.15 ($R_{\rm V}\!=\!3.1$); this will be discussed in  
section \ref{Subsection:TheExtinction}.

\subsection{Spectral type from the ultraviolet and near-infrared spectra}

Despite the fact that the precise determination of the stellar parameters
has been done through a careful analysis of a variety of observations, as we
have mentioned in section \ref{Subsection:StellarParameters}, an attempt to
determine the spectral type of HD 163296 in a more qualitative way has been
done using the ultraviolet and near-infrared spectra.

A comparison of the ultraviolet data with IUE spectra of stars classified as
``spectral-type standards'' was done . The spectra were taken from the ``IUE
Ultraviolet Spectral Atlas of Standard
Stars''\footnote{http://www-int.stsci.edu/\~{}jinger/iweb/proj/project.html}
\citep{Wu1983,Wu1992}. It is interesting to note that below 2000
\AA, for a given spectral type, the ultraviolet spectra of the ``spectral-type
standards'' themselves are very different, in particular for A0 V and A1 V. If
we consider only the region above 2000 \AA, the comparison suggests that HD
163296 is closer to A3 V (and possibly slightly later), however, if a mild
extinction is present, the UV spectra would be also close to that of A1 V
stars. A larger extinction would push the spectral type determination towards
earlier spectral types.

A similar exercise of spectral typing was done by a comparison of the
SpeX spectrum in the YJ bands (0.85 -- 1.35 $\mu$m) with Kurucz
synthetic models (see \citet{Montesinos2009} for references regarding
the synthesis codes and models used). The spectrum is complex, the
range of $T_{\rm eff}$'s explored with the Kurucz models (8750--9250
K, corresponding to A3V and A1V) does not produce significant Paschen
line variations. The Paschen line wings of HD 163296 are narrower than
in any of the models, which would suggest temperatures of 9250 K or
higher. The metallic lines are faint, their typical intensity being
lower than 5\% that of the continuum.  Some lines are deeper than in
any photospheric spectrum, which suggests the presence of some
circumstellar absorption. Some lines are much fainter, which suggests
an spectral type A1V or earlier, or dust emission veiling. Good
agreement is rarely found. Although we fail to find a unique combination of $T_{\rm eff}$ and $\log g$ that provides an adequate fit to both the UV and NIR spectra of HD 163296, all data are consistent with an A1-A3 star, seen through mild foreground extinction.

\subsection{The extinction}
\label{Subsection:TheExtinction}

The determination of the extinction is an intricate problem. The classical
method of comparing the observed colours with those of standard stars of
similar properties does not lead to any conclusive result, given the slight
variability of the star and the photometric (and spectral type) uncertainties. For normal
stars, the UV part of the SED can be used to estimate this parameter, but in
the case of HD 163296 the ultraviolet spectrum is variable and has
contributions both from the photosphere and the accretion shock that are
difficult to quantify. The extinction seems to be mostly circumstellar, therefore the exact value of the constant $R_{\rm V}$ is not known. The observations available do not allow us to disentangle completely this problem.

The strategy followed to assign a value of E(B--V) to this star has four
steps: 1) introduce extinctions from E(B--V)=0.0 to 0.2 to the photospheric
model, 2) normalize the reddened model photosphere to the flux at V, 3)
deredden the normalized model and estimate the integrated stellar flux, $F_*$;
note that $F_*$ is the flux that would be observed in the complete absence of
extinction. Obviously, different extinctions imply different values of
$F_*$. A value of $R_{\rm V}\!=\!3.1$ has been used.

The fourth step is based on the following argument: since the stellar
luminosity $L_*/L_\odot$ is known through the distance-independent
method outlined in section \ref{Subsection:StellarParameters}, an
estimation of the distance, $d$, can be obtained simply from the
expression $L_*=4\pi d^2 F_*$, where it is assumed that the whole star
is visible, i.e. and no part of the stellar flux is completely blocked
by the disc of other circumstellar material (see a complete discussion
of this issue in \citet{Montesinos2009}). The value E(B--V)=+0.15
allows us to recover a distance of 128 pc, which matches the Hipparcos
distance to within the uncertainties, therefore it has been adopted as
the optimum extinction for consistency with the whole set of
parameters. We would like to point out that that value is a simple
parameterization of a very complex problem and that a deeper study,
outside the scope of this paper, would be needed to study all its
peculiarities.

In Fig. \ref{Figure:SED_UV_optical_nIR} the photospheric model reddened with
E(B--V)=+0.15 has been plotted. It goes through the two ultraviolet spectra
but falls below both of them at wavelengths shorter than $\sim\!1260$ \AA, and
it also underestimates the flux at U. Note that the model is purely {\it
photospheric}, therefore the excess fluxes inferred from the observed data
when compared with the model might be due to emission from the accretion
shock, i.e. with a non-photospheric origin. The observed SED seems to depart
from the photospheric SED at around $\sim\!1$ $\mu$m, where the contribution
from the disc starts to become noticeable.

\subsection{Observational constraints}
\label{sec:obscon}

In addition to the new Herschel data we have further constrained our disc models using SMA interferometric observations of the J=3-2 transition in $^{12}$CO and PBI interferometry of the $^{12}$CO 2-1 and $^{13}$CO 1-0 transitions \citep{Isella2007}, an upper limit for the S(1) transition in H$_2$ obtained by VLT/VISIR \citep{Martin2010}, the ISO-SWS spectrum in the infrared, and a wealth of photometric data, including SCUBA photometry in the sub-mm \citep{Sandell2011}. See Fig.~\ref{fig:SED} for a full list of the photometry sources.

\begin{table}
\centering
\caption{Fixed disc model parameters.}
\label{tab:Para_fix}
\begin{tabular}{lcc}
\\[-4.5ex]
\hline
 Quantity & Symbol & Value\\
\hline 
\hline 
stellar mass                      & $M_{\star}$   & $2.47\,M_\odot$\\
effective temperature             & $T_{\rm eff}$ & $9250\,$K\\
stellar luminosity                & $L_{\star}$   & $37.7\,L_\odot$\\
inner disc radius                 & $\Rin$       & 0.45\,AU\\
\hline
dust material mass density        & $\rho_{\rm gr}$  & 3.36\,g\,cm$^{-3}$\\
strength of incident ISM UV       & $\chi^{\rm ISM}$ & 1\\
cosmic ray H$_2$ ionisation rate  & $\zeta_{\rm CR}$   
                                             & $1.7\times 10^{-17}$~s$^{-1}$\\
turbulent Doppler width           & $v_{\rm turb}$ & 0.15\,km\,s$^{-1}$\\
$\alpha$ viscosity parameter      & $\alpha$       & 0\\
\hline 
&&\\[-2.2ex]
disc inclination                  & $i$ & $50\degr$\\
distance                          & $d$ & 118.6\,pc\\
\hline
\end{tabular}
\end{table}

There is a wealth of observational data available for HD 163296 which constrains to varying extents our choice of model parameters. As well as fitting to the data mentioned in the previous paragraph, we have taken care not to contradict current understanding of the disc geometry. There is at present some uncertainty regarding the radial extent of the disc. Scattered light imaging indicates an outer edge for the dusty disc of $\sim$\,540\,AU, consistent with that derived for the gas disc from millimetre CO emission by \citet{Isella2007}. However, the resolved mm continuum emission seems to indicate a smaller outer disc radius $\sim\,200$\,AU \citep{Isella2007}. \citet{Hughes2008} propose that the millimetre continuum and CO 3-2 emission can be fit simultaneously by a disc with an exponentially-tapered outer edge. For the purposes of our modelling, we adopt the surface density profile derived by \citet{Hughes2008} for this object, and attempt to fit simultaneously the remaining observational data (SED, line fluxes, CO line profiles) by varying the remaining disc parameters. As an additional test, we compute a model with a power-law density profile, varying in addition the column density power index to fit the same data (SED, line fluxes, CO line profiles). For the purposes of this power-law model the disc outer edge is fixed at the scattered light radius of 540\,AU. These adopted disc outer radii are consistent with our far-infrared photometry measurements, in which the disc is unresolved, corresponding to the emission at these wavelengths being dominated by the inner $\sim$\,200--300\,AU of the disc.

The disc is also resolved at the inner edge, and while there is some uncertainty regarding the precise structure in this region \citep{Renard2010, Benisty2010, Eisner2009,Tannirkulam2008}, we adopt the median value of 0.45AU for the inner disc radius. The disc is assumed to have constant flaring with radius, i.e. the assumed inner rim structure does not cause shadowing. The reference scale heights referred to in our results (see Table~\ref{tab:Para_fit}) are defined at the inner rim, i.e. $R\!=$\,0.45\,AU. The disc inclination is fairly well-constrained by imaging at various wavelengths \citep{Benisty2010,Wassell2006,Isella2007,Tannirkulam2008a,Grady2000}, and is fixed at $50\degr$ throughout the modelling in this paper. See Table~\ref{tab:Para_fix} for a full list of the parameters which are fixed during the model-fitting process.

While we have not used the spatial CO emission maps as a constraint for our modelling efforts, we do use the extracted line profiles for the three CO lines \citep{Isella2007} as a constraint for the radial origin of the emission. The interferometry allows us to isolate the emission from the disc, eliminating contamination from cloud material. The bulk of the modelling uses a surface density profile derived from fitting to the spatial CO 3-2 emission \citep{Hughes2008}.

%======================================================================

\section{Modelling}

The disc modelling was carried out using the radiation thermo-chemical disc code \PRODIMO~\citep{Woitke2009a,Kamp2010}. The code solves in turn the radiative transfer problem, chemical network, and gas heating and cooling balance. Finally, the level populations are calculated, followed by line transfer calculations to give the predicted line emission.

The frequency-dependent 2D radiative transfer solver calculates the dust temperature structure and internal continuous radiation field for a given disc and stellar spectrum. For the purposes of this paper it was used to determine an SED-fitting dust model for HD 163296, by varying the total dust mass, grain size parameters and the spatial distribution of dust in the disc, including dust settling as a function of grain size (see Section~\ref{sec:SED}).

Once the dust temperature structure $T_{\rm dust}(r,z)$ and internal radiation field $J_\nu(r,z)$ have been computed, \PRODIMO~solves the chemical network assuming kinetic equilibrium (such that the net concentration of each species does not change over time), and the gas energy balance. This gives the chemical composition of the disc, the gas temperature $T_{\rm gas}(r,z)$ and the gas emission lines, and depends on the gas-to-dust ratio and PAH abundance. The chemical network includes 973 reactions between 76 gas phase and solid ice species composed of 10 elements \citep{Woodall2007,Schoier2005}. There is a detailed treatment of UV-photoreactions \citep[see][]{Kamp2010}, as well as H$_2$ formation on grain surfaces, vibrationally excited H$_2^\star$ chemistry, and ice formation (adsorption, thermal desorption, photo-desorption, and cosmic-ray desorption) for a limited number of ice species \citep[see][for details]{Woitke2009a}.

The level populations of the various atoms, molecules and ions are calculated using an escape probability method, and used to solve the line transfer for selected spectral lines, in this case those of CO \citep{Flower2001,Jankowski2005,Wernli2006,Yang2010}, H$_2$O \citep{Barber2006,Dubernet2002,Faure2007,Green1993,Tennyson2001},{\sc Oi} \citep{Abrahamsson2007,Bell1998,Chambaud1980,Jaquet1992,Launay1977} and {\sc Cii} \citep{Flower1977,Launay1977,Wilson2002}. See \citet{Woitke2011} for a description of the line transfer calculations, as well as recent improvements to the chemistry and gas heating/cooling balance in \PRODIMO.

\subsection{Input spectrum}
\label{Subsection:InputSpectrum}

The stellar input spectrum used for the modelling is a composite of
the available observed UV data with a model photosphere spectrum in
the wavelength range for which detailed spectral data are not
available. We have accounted for the observed ultraviolet variability
of this object (see Section.~\ref{Section:StellarProperties}) by
running two sets of disc models with different UV input spectra,
namely, one representing the ``low UV'' state and the other a ``high
UV'' state, as described in Section~\ref{sec:TheSED}. These
were provided by the FUSE+STIS and the IUE average spectra,
respectively. At the upper wavelength boundary we switch in each case
to the GAIA PHOENIX model photosphere computed with $T_{\rm
  eff}\!=\!9250$ K, $\log g_*\!=\!4.07$ and [Fe/H]=+0.20. The
observations were de-reddened using the Fitzpatrick parameterization
(Fitzpatrick, 1999) with E(B--V)=+0.15 and $R_{\rm V}\!=\!3.1$.

The UV spectrum is of central importance to the gas modelling, both in the chemical network and in the gas heating/cooling balance. This requires UV input for wavelengths of 912\AA{} and above (the wavelength at which atomic hydrogen is ionised). The FUSE spectrum in this case only extends down to 970\AA, and the IUE spectrum down to 1150\AA. In order to give a full spectrum for modelling purposes, the ``high UV'' input is extended down to 970\AA{} by scaling the STIS+FUSE (low UV) spectrum accordingly. In both cases a power-law fit to the data is employed for the remaining shortfall. The total photon particle fluxes are 60\% greater for the high UV input spectrum than for the low UV state. 

%\begin{figure}
%\centering
%\includegraphics[width=90mm]{UV.eps}
%\caption{The UV spectra used for disc modelling. Plotted are the ``low UV'' input spectrum (solid black line), ``high UV'' input spectrum (dotted line), STIS+FUSE spectrum (cyan) and the average IUE spectrum (orange). Blue circles denote simultaneous UBV EXPORT photometry. The green and yellow bars indicate the wavelength ranges of UV band 1 (912 - 1110 \AA) and UV band 2 (912 - 2050 \AA) respectively, over which the input spectra are integrated for the gas modelling.}  
%\label{fig:UV}
%\end{figure}

\subsection{Fitting procedure}
\label{sec:fit}
An evolutionary $\chi^2$-minimisation strategy was employed to simultaneously fit the observed SED, ISO-SWS spectrum and gas line observations (see Section~\ref{sec:obscon}). In the case of the line observations the models were fitted to the observed fluxes, and the line widths in cases where profiles were available. The upper limits were not used as a fitting constraint, but were checked against the predicted best-fit model fluxes for any contradictions.

We have utilised a Monte-Carlo evolutionary method, varying the 11 parameters listed in Table~\ref{tab:Para_fit} until the minimum $\chi^2$ value was reached (see \citet{Woitke2011} for more information regarding the evolutionary strategy and method of $\chi^2$ calculation). We note that the best-fit parameters depend in general on the weighting scheme employed for the three groups of observations, and in practice it was found that weighting slightly in favour of the line emission and ISO-SWS observations at the expense of the photometry gave the best fit overall. We also note the possibility that the converged parameter values correspond to a local $\chi^2$ minimum, and not a single global minimum, and we do not claim a unique fit to the observations. Indeed, it is clear from our modelling efforts that there exist a lot of degeneracies between the various parameters. However, each of the models discussed in the following sections are the result of several hundred generations of $\chi^2$-minimisation.

\subsection{Dust properties}
\label{sec:dust}

The rich solid-state spectrum of HD 163296 was observed by \citet{vandenAncker2000}, using the ISO-SWS and ISO-LWS spectrometers. This study found the dust in this object to consist of amorphous silicates, iron oxide, water ice and a small fraction of crystalline silicates, with the presence of large millimetre-sized grains indicated by the continuum temperature.

We assume homogeneous and spherical dust grains (Mie theory), with a power-law size-distribution defined by a minimum radius $a_{\rm min}$, maximum radius $a_{\rm max}$, and power-law index $p$. The grain composition is assumed to be constant throughout the disc, and we adopt the dust species mixture determined by \citet{Bouwman2000} for this object, averaging their fractional species abundances over the disc (see Table~\ref{table:dustcomp}). All grains have the same species composition regardless of their size, and the grain composition is not treated consistently with the physics and chemistry of grains. For instance, the assumed grain water ice fraction is constant throughout the disc, independent of the water ice abundances calculated in the solution of the chemical network. For the dust material mass density, we take the average value for this mix of 3.36\,g\,cm$^{-3}$. Optical constants for the various dust species were taken from measurements made by the studies listed in Table~\ref{table:dustcomp}.

\citet{Bouwman2000} carried out a detailed study of this object, fitting to spectral data over a large wavelength range. We note however that the dust opacity law represents a potentially large source of uncertainty when deriving the disc parameters. For instance, the two olivine species FeMgSiO$_4$ and Mg$_2$SiO$_4$, whilst having broadly similar spectra in the wavelength range observable by ISO-SWS, have a factor $>\!30$ difference in absorption opacity at $\sim1$ micron. This will inevitably affect the derived disc parameters, although the difference in opacities at millimetre wavelengths is negligible. The models referred to in this study have extinction opacities in the range (10.4-12.8) cm$^2$g$^{-1}$(dust) at 1mm.

\subsection{Gas/dust ratio}
\label{sec:gasdust}

The models considered have constant dust properties with radius. However, in general the models do allow for differential dust grain scale heights as a function of grain size, to represent the major effects of dust settling. This leads to a local gas/dust ratio which varies with height in the disc, yet at any given radius the ratio of the vertical gas and dust total column densities is constant, and equal to the global gas/dust ratio. The degree of dust settling is determined by a simple parameterisation (see Section~\ref{sec:SED}), and models in which no dust settling is present are referred to as ``fully mixed''.

%Following the evolutionary scheme, a grid of models was computed with parameter values enclosing the best-fit model, varying each of the 12 fitted parameters in turn. This allows us to estimate the strength of the parameter dependence of the observables, and the degree of uncertainty attached to each of the fitted values in Table~\ref{tab:Parameter}. A model is deemed to no longer be a good fit when its chi value reaches twice that of the best fit model, either for the combined weighted chi value, or for a particular observable (photometry, ISO-SWS spectrum, gas emission lines). For instance, in the case of the surface density power index, $\epsilon$, the photometry provides the strongest constraint to the permissible value, but in other cases the line emission is seen to dominate (see Fig.~\ref{fig:chi}). In the case of disc inclination the fit does not vary strongly with any of the observable categories (due to individual lines etc. cancelling out one another), but a value of 40 degrees gives the best fit to the observed CO line profiles (see Section~\ref{sec:gasmodel}), and marginally the best overall chi value.

%=============================================================================

\section{Results}
\label{sec:results}

\begin{table*}
\centering
\caption{Model parameters for four well-fitting disc models. The first three columns refer to models with exponentially-tapered density profiles, while model 4 is a simple power law. Models 1, 3 and 4 are for a ``low UV'' input spectrum while model 2 is for a ``high UV'' flux. Finally, models 1, 2, and 4 allow dust grains to settle towards the midplane while model 3 is fully mixed. Model 3 represents our overall ``preferred'' model, which is later discussed in detail, and is marked by an asterisk.}
\label{tab:Para_fit}
\begin{tabular}{lccccc}
\\[-4.5ex]
          &        &  (1)       &   (2)       &  ~(3)$^{*}$ & (4) \\
\hline
Quantity & Symbol & ``low UV'' & ``high UV'' & Fully mixed & Power-law density profile\\
\hline 
\hline 
disc mass                         & $M_{\rm disc}$& $2.2 \times 10^{-2}\,M_\odot$ &$ 2.0 \times 10^{-2}\,M_\odot$ & $ 7.1 \times 10^{-2}\,M_\odot$ & $ 1.1 \times 10^{-2}\,M_\odot$\\
column density power index        & $\epsilon$   & n/a & n/a & n/a & 0.085 \\
reference scale height            & $H_0$        & $0.019$\,AU & $0.019$\,AU & $0.019$\,AU & $0.027$\,AU\\
flaring power                     & $\beta$      & $1.068$ & $1.066$ & 1.066 & 1.019\\
gas-to-dust mass ratio            & $\rho/\rho_d$  & $21.8$ & $20.5$ & 101.1 & 9.1\\
minimum dust particle radius      & $\amin$        & $5.95 \times 10^{-3}\,\mu$m & $1.25 \times 10^{-3}\,\mu$m & $9.63 \times 10^{-3}\,\mu$m & $1.24 \times 10^{-2}\,\mu$m\\
maximum dust particle radius      & $\amax$        & $1134\,\mu$m & $1405\,\mu$m & $2041\,\mu$m & $1015\,\mu$m\\
dust size dist.\ power index      & $p$            & $3.61$ & $3.57$ & 3.68 & 3.75 \\
dust settling parameter           & $s$            & $0.57$ & $0.39$ & 0 (fixed) & 0.56\\
minimum dust settling radius      & $a_{\rm set}$    & $1.09\,\mu$m & $0.44\,\mu$m & n/a & 0.81\\
PAH abundance relative to ISM     & $f_{\rm PAH}$  & $4.3\times 10^{-2}$ & $3.5\times 10^{-2}$ & $6.8\times 10^{-3}$ & $2.2\times 10^{-2}$\\
\hline
fit to observed photometry        & $\chi_{\rm PHOT}$ & 1.723 & 2.061 & 1.323 & 1.279 \\
fit to ISO-SWS spectrum           & $\chi_{\rm ISO}$  & 1.326 & 1.378 & 1.780 & 0.827\\
fit to observed line data         & $\chi_{\rm LINE}$ & 0.661 & 0.741 & 0.742 & 0.607 \\
\hline
\end{tabular}
\end{table*}

\begin{figure}
\centering
\includegraphics[width=90mm]{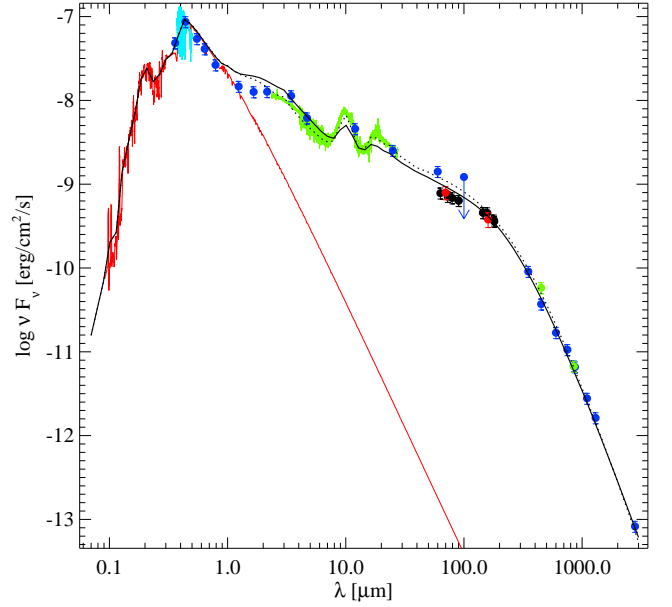}
\caption{Spectral energy distribution for the best-fit fully mixed disc model (``preferred'' model), obtained from a simultaneous fit to the observed SED, ISO-SWS spectrum and line emission from HD 163296 (solid black line). Black dotted line indicates the SED for the model with power-law density profile, which is unable to fit the mm continuum image for HD 163296. Blue circles indicate (with increasing wavelength) simultaneous UBVRIJHK photometry \citep{Eiroa2001,Oudmaijer2001}, LM photometry \citep{deWinter2001}, IRAS photometry (12-100 mic), sub-mm photometry \citep{Mannings1994},  and millimetre photometry \citep{Isella2007}. Also marked are PACS photometric observations (red circles), PACS continua derived from the spectroscopic observations (black circles), SCUBA photometry (green circles) \citep{Sandell2011}, scaled VLT/UVES spectrum (blue line, Martin-Zaidi, in preparation) and the ISO-SWS spectrum (green line). The red line shows the stellar+UV input spectrum. Downwards arrow denotes upper limit. All fluxes were corrected for interstellar reddening using the Fitzpatrick parameterisation \citep{Fitzpatrick1999} with $R_V\!=\!3.1$ and E(B--V)$\!=\!0.15$.}
\label{fig:SED}
\end{figure}

Table~\ref{tab:Para_fit} gives the derived parameter values for four runs of the evolutionary scheme outlined in the previous section. The first three columns refer to models with a fixed exponentially-tapered density profile. The fourth column refers to a model with power-law density profile, $\Sigma\!\propto\!R^{-\epsilon}$, in which the outer radius is fixed at 540\,AU but the power index, $\epsilon$, is allowed to vary.

\begin{figure*}
\begin{tabular}{ccccc}
  \hspace*{-6mm}
  \includegraphics[width=4.2cm]{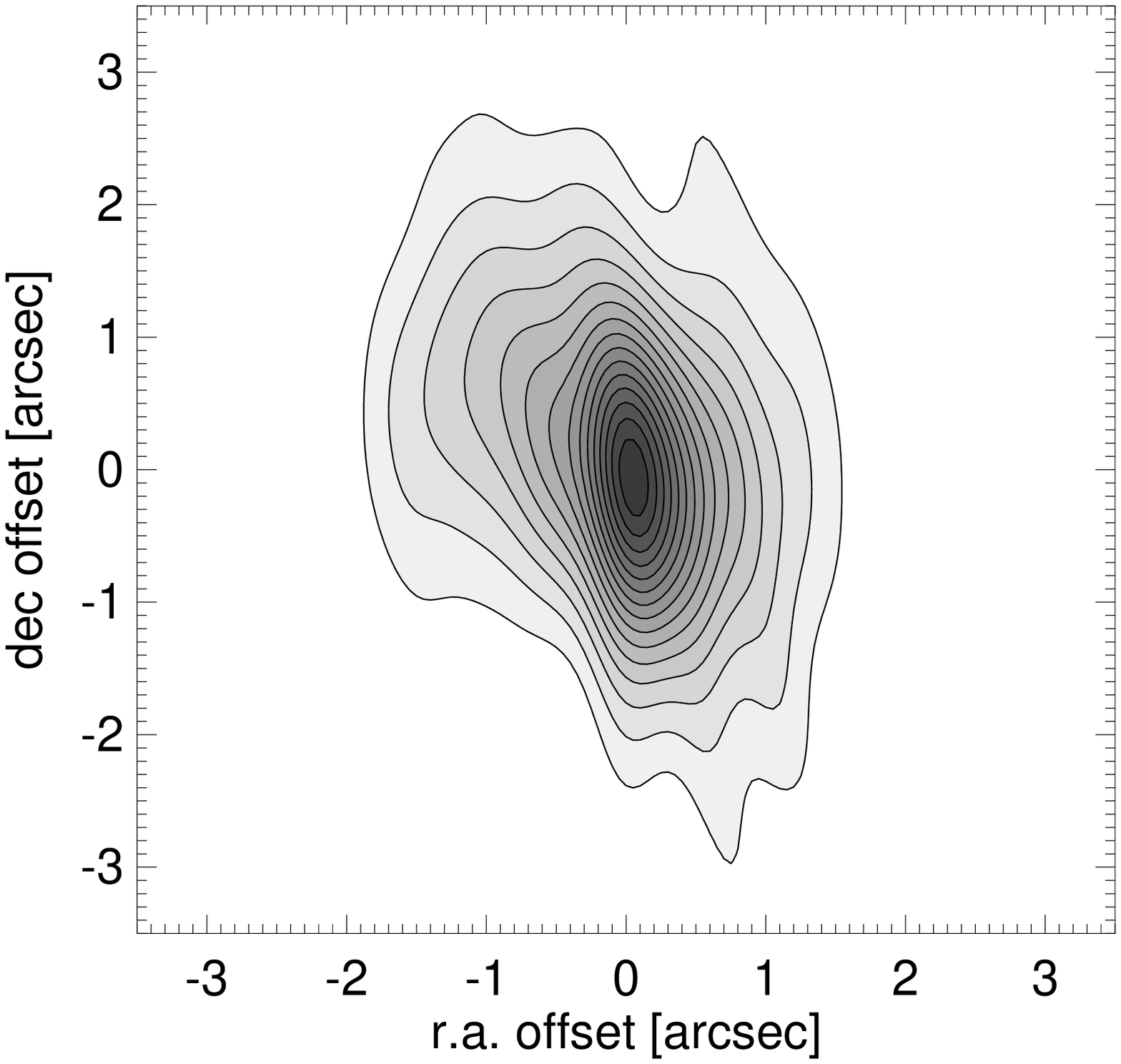} &
  \hspace*{-4mm}
  \includegraphics[width=3.5cm]{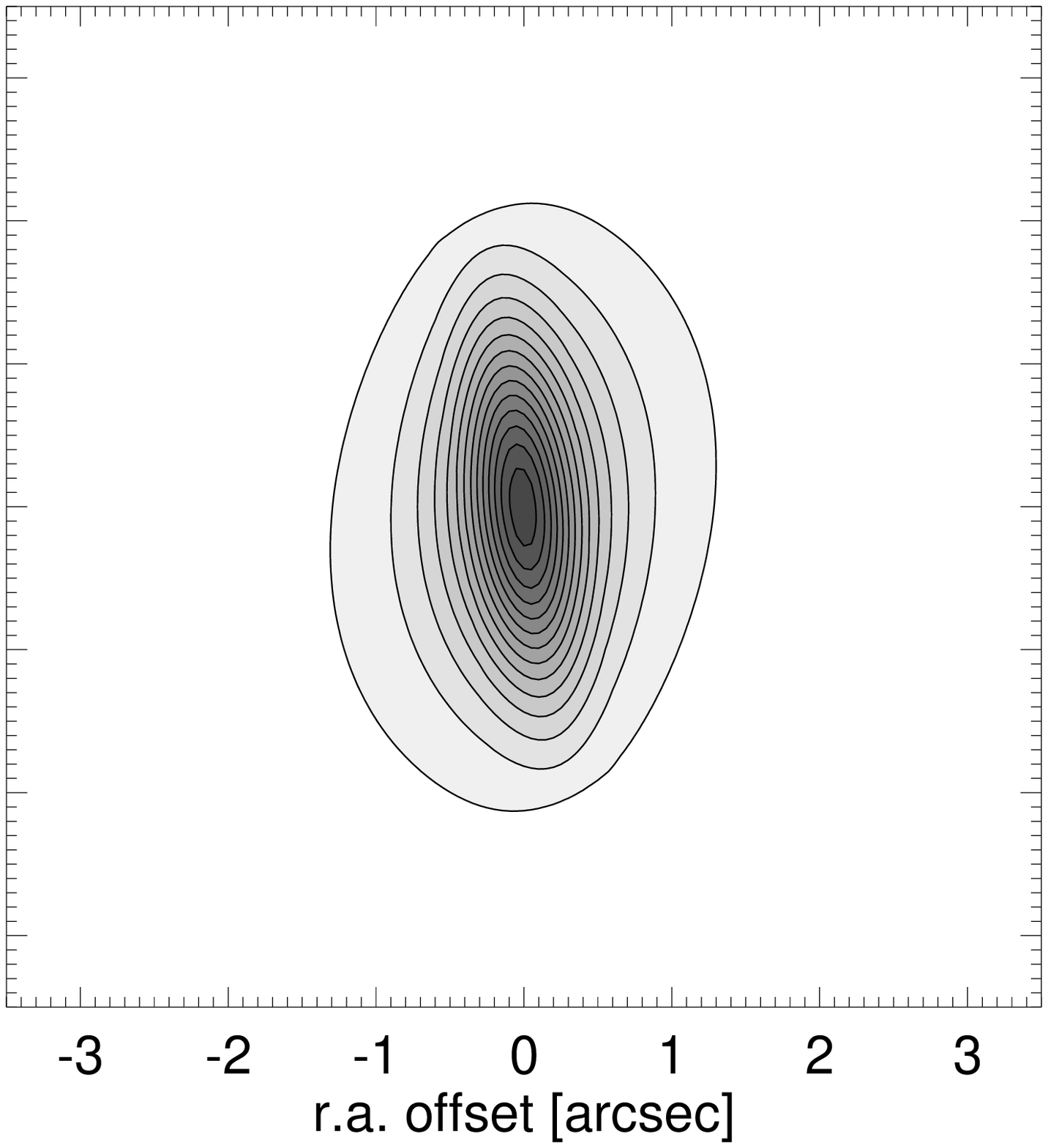} &
  \hspace*{-4mm}
  \includegraphics[width=3.5cm]{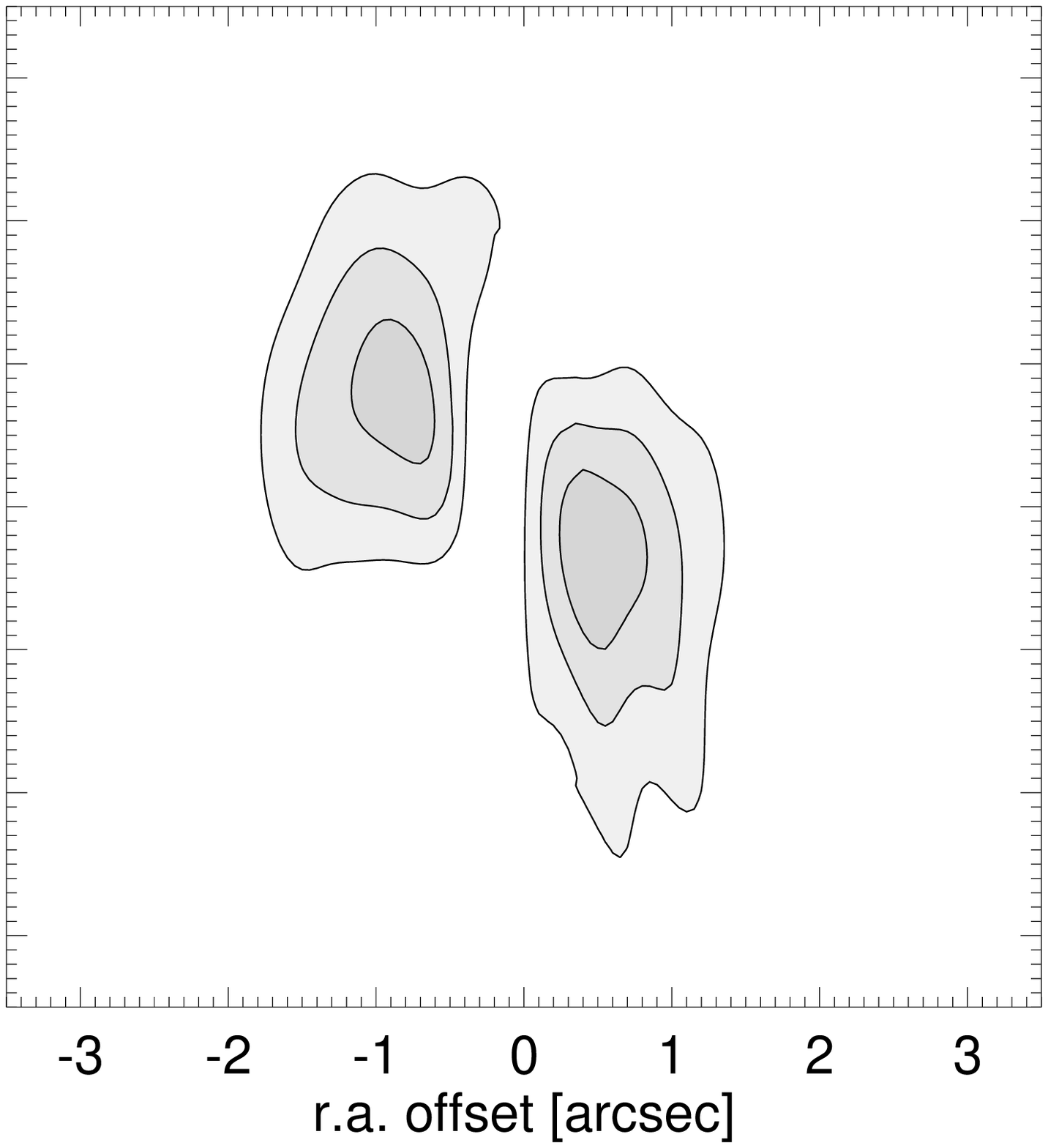} &
  \hspace*{-4mm}
  \includegraphics[width=3.5cm]{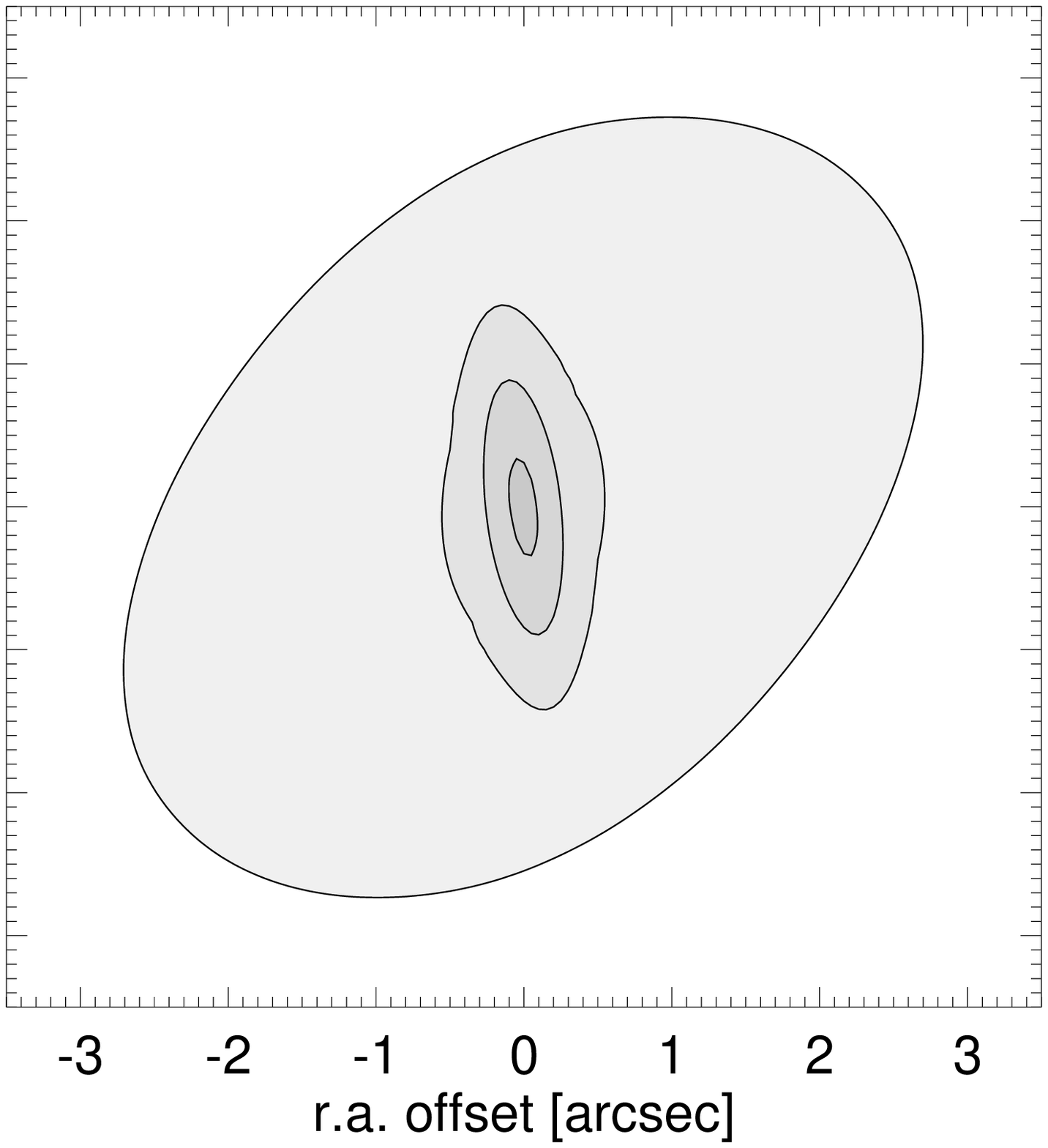} &
  \hspace*{-4mm}
  \includegraphics[width=3.5cm]{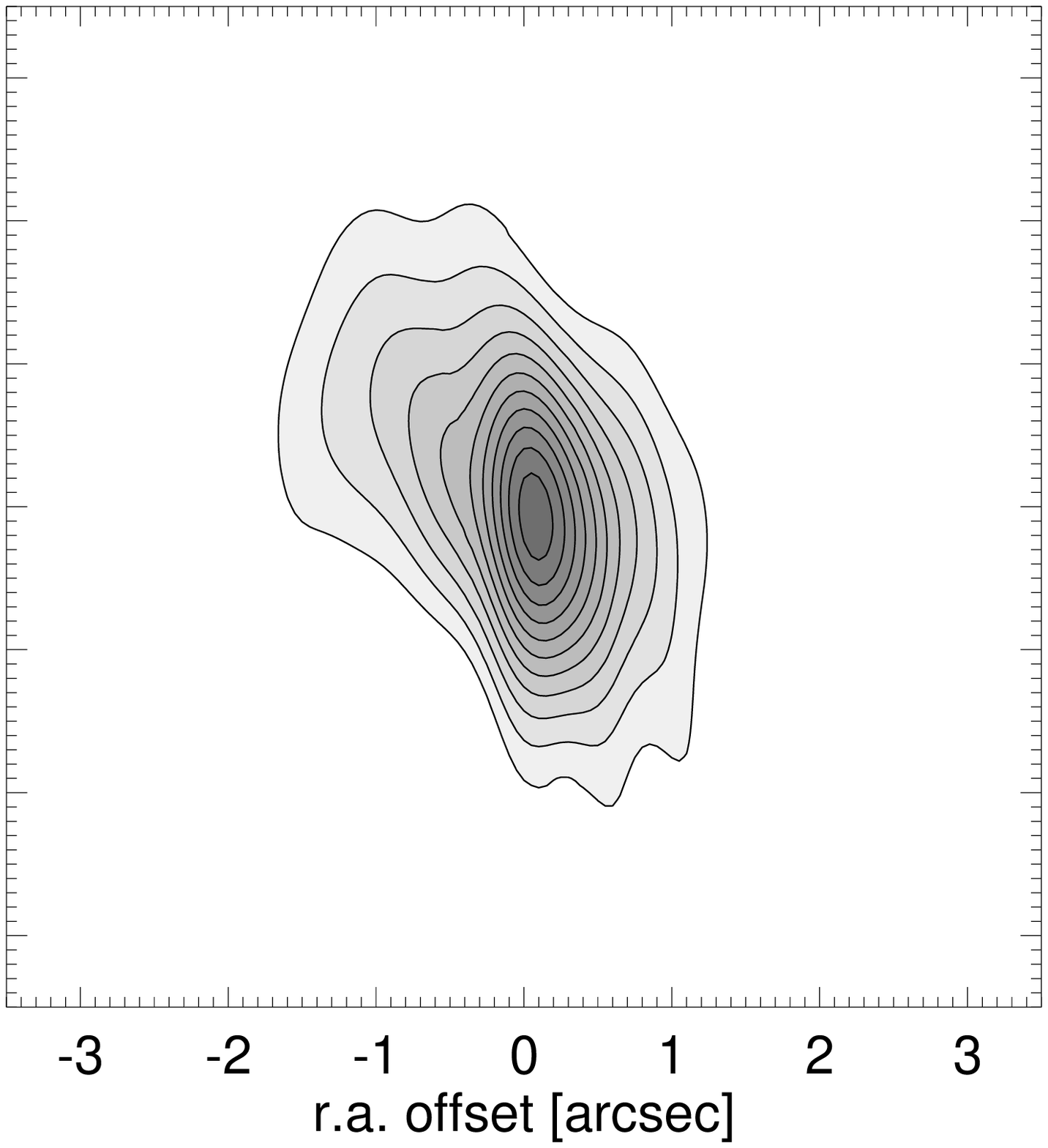} \\
\end{tabular}
\caption{1.3mm continuum emission maps for HD 163296. From left: observed map; emission map for preferred model (column 3 in Table~\ref{tab:Para_fit}); residuals for preferred model; emission map for power-law model (column 4 in Table~\ref{tab:Para_fit}); residuals for power-law model. Residuals computed by subtracting the model intensity from the observed intensity. Contours spaced at 12\,mJy intervals, corresponding to [3,6,9,12,15,18,21,24,27,30,33,36,39,42,45]$\times\,\sigma$.}
\label{fig:maps}
\end{figure*}

\begin{table}
\centering
\caption{Dust grain composition.}
\label{table:dustcomp}
\begin{tabular}{lll}
\hline
&&\\[-2.2ex]
Dust Species               & Mass Fraction & Optical constant ref.
\\
\hline
&&\\[-2.2ex]
Amorphous FeMgSiO$_4$      & 0.745           & \citet{Dorschner1995}\\      
Amorphous Carbon           & 0.15            & \citet{Jager1998}\\
Crystalline Mg$_2$SiO$_4$  & 0.035           & \citet{Servoin1973}\\
Water Ice                  & 0.035           & \citet{Warren2008}\\
Iron Oxide                 & 0.02            & \citet{Henning1995}\\
Iron                       & 0.015           & \citet{Posch2003}\\
\hline
\end{tabular}
\end{table}

\begin{figure}[h!]
\centering
\includegraphics[width=70mm]{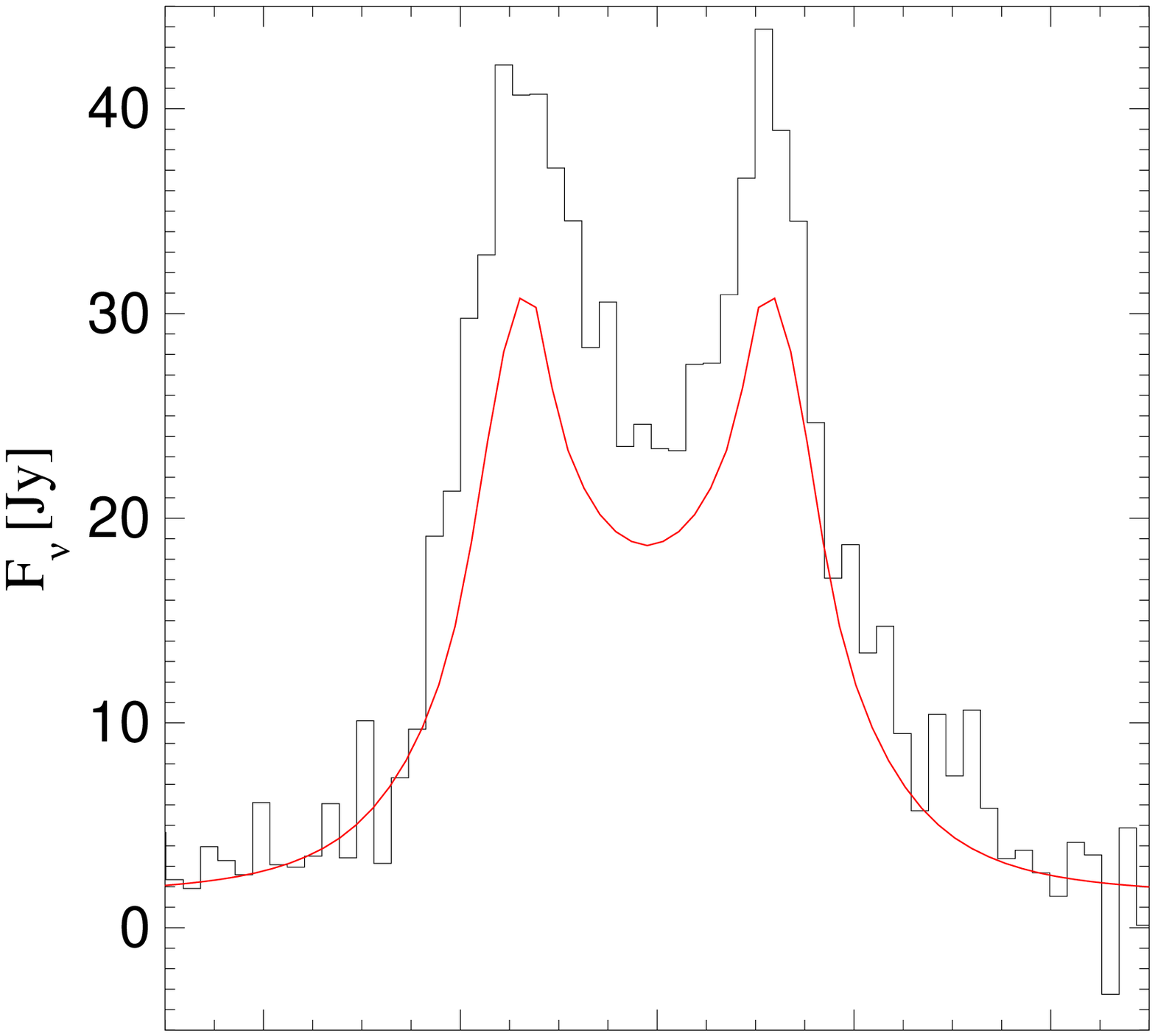} \\
\includegraphics[width=70mm]{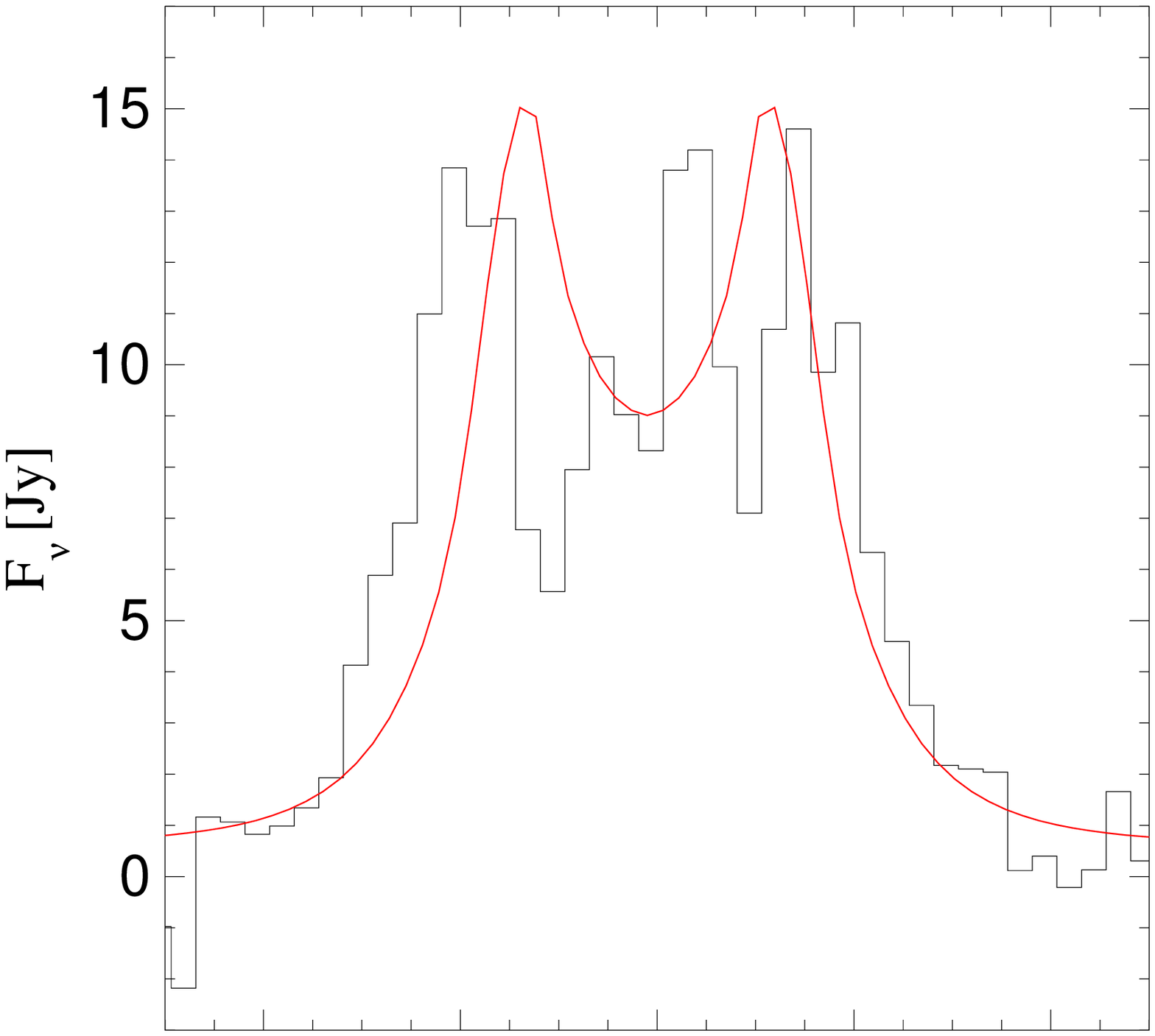} \\
\includegraphics[width=70mm]{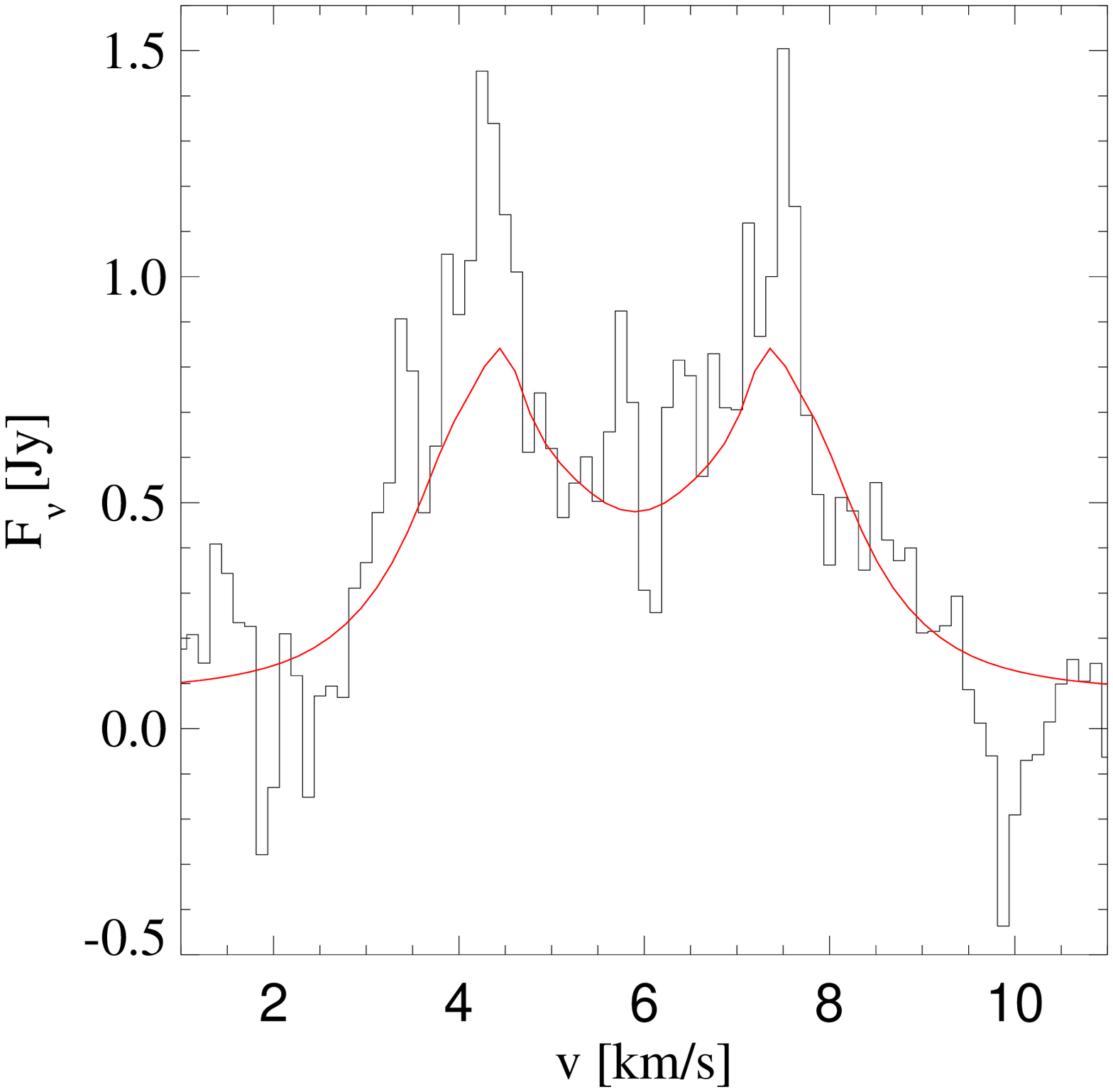} \\
\caption{Plotted in black are the line profiles observed by \citet{Isella2007} for the $^{12}$CO J=3-2 (upper panel), $^{12}$CO J=2-1 (middle panel) and $^{13}$CO J=1-0 (lower panel) transitions, with the corresponding profiles from the preferred disc model in red. This refers to a single simultaneous fit to the observed continuum and line data (column 3 in Table~\ref{tab:Para_fit}). Observed profiles are obtained by integrating over the whole disc.}
\label{fig:profiles}
\end{figure}

The surface density profile in the models with an exponential outer edge (columns 1-3) takes the form
\begin{equation}
\centering
\Sigma(R)\!\propto\!R^{-\gamma}\,\exp\,\left[-\left(\frac{R}{R_0}\right)^{2-\gamma}\right],
\end{equation} 
where $\gamma$ is analagous to the power law index, and $R_0$ is the scale length over which the disc surface density tapers exponentially. We adopt the previously derived values of $\gamma\!=\!0.9$ and $R_0\!=\!125$\,AU \citep{Hughes2008}. For the purposes of our modelling, the disc chemistry etc. was computed out to 850\,AU, at which point the column density is negligible. 

The first, ``low UV'' column uses the FUSE+STIS UV spectrum as input for the modelling, and the second ``high UV'' column uses the average IUE spectrum, as described in Section~\ref{Subsection:InputSpectrum}. Both runs produce settled discs, with variable scale heights for dust grains of different sizes. The third column in Table~\ref{tab:Para_fit} gives the results for a run with well-mixed dust grains, i.e. no settling is introduced. This run uses the FUSE+STIS (low UV) spectrum, as does the run with power-law surface density referred to in column 4.

None of the models described in Table~\ref{tab:Para_fit} represent a perfect fit to all the available data for HD 163296, but in all cases the models are able to fit almost all the observations. Considering the large number of data used for the modelling, and the range of parameters covered by the well-fitting models, is is clear that there is some degree of parameter degeneracy present.

%The ``low UV'' results represent the best overall fit to the available data, and any reference to the ``best-fit model'' (e.g. Figs.~\ref{fig:SED}\,\&\,\ref{fig:profiles}) in the rest of this paper refers to this individual model. We will go on to discuss in more detail the results of our modelling, including the motivation for introducing dust-settling, and its implications for the disc properties derived from the observations.

\subsection{Continuum emission and spectral energy distribution}
\label{sec:SED}

The SED for the best-fit fully mixed disc model (column 3 in Table~\ref{tab:Para_fit}) is shown in Fig.~\ref{fig:SED}. With the constraints outlined in previous sections, our aim was to fit the observed SED with a simple dust model, i.e. a continuous disc with constant flaring, and dust species composition constant throughout the disc. All of the models described in Table~\ref{tab:Para_fit} require the presence of mm-sized grains in order to fit the observed SED. It was not possible to fit both the 10 micron silicate feature and the millimetre tail with a well-mixed disc. This can be seen in Table~\ref{tab:Para_fit}, where the fully mixed model gives a worse fit to the ISO-SWS spectrum than for the models in which dust settling is present. Fig.~\ref{fig:SED} shows that the best-fit fully mixed model gives a smaller than observed silicate emission feature. \citet{Sitko2008} observe a spectral variability of $\sim\,10$\% in the wavelength range covered by the ISO-SWS data, and while we are unable to fit the observations to within this range with even a settled model (simultaneously with the overall SED and line data), the fit is considerably better than is possible with a well-mixed disc. This result is similar to that found for IM Lupi by \citet{Pinte2008}, in which a settled disc was needed to fit the 10\,$\mu$m silicate feature and SED millimetre tail simultaneously. 

On the other hand, the constraint of fitting to the ISO-SWS spectrum leads to models which produce too much emission in the near-infrared (J,H \& K bands), giving an overall worse fit to the observed photometry. This effect is less pronounced in the best-fit fully mixed model (column 3 c.f. columns 1\&2 in Table~\ref{tab:Para_fit}). The flux overprediction in the settled models is likely mainly due to the smaller average grain size in the strongly illuminated disc surface layer. The condition of radiative equilibrium between grains means that this gives a higher average grain temperature at the disc surface, and the re-emitted light peaks at shorter wavelength than for the fully mixed models (in the near-IR as opposed to the mid-IR). This flux overprediction is still present to a smaller extent in the fully mixed model (see Fig.~\ref{fig:SED}). This could be a consequence of our fixing the disc inner radius for modelling purposes, and a better fit in the near-IR would be possible if we had allowed the inner radius to vary. It is also probable that the structure of the inner rim is more complex than allowed by our parameterised model, with evidence for puffing-up of material, which would certainly affect the emission at these wavelengths. There is also significant variability observed for this object in this wavelength region \citep{deWinter2001,Sitko2008}.

Dust settling enhances the silicate emission feature for an optically thick disc
since it removes the large grains from the surface layers, and places them at smaller heights in the disc where they cannot be observed at mid-infrared wavelengths. The flat blackbody opacity of these larger grains is overwhelmed by the characteristic spectrally-varying opacity of smaller micron-size grains which remain at lower optical depth, dominating the observed emission.

In cases such as this where a simple parameterised disc structure is assumed, \PRODIMO\ implements a simple recipe to account for the major effects of vertical dust settling. We assume that the dust grains are distributed vertically with a scale height which decreases for large dust particles:

\begin{equation}
H'(a,r)\!=\!H(r)\cdot{\rm max}\{1,a/a_{\rm s}\}^{-s/2}
\end{equation}
where $H(r)$ is the gas scale height, and $s$ and $a_{\rm s}$ are two free parameters (see \citet{Woitke2010} for details).

Introducing two additional parameters to the modelling inevitably improves the fit to the observed data, and indeed the combined fit to the photometry, ISO-SWS spectrum and line emission does improve overall as dust settling is introduced. The fact remains, however, that it is possible to obtain a good fit to almost all the available data with a well-mixed disc, with a gas/dust ratio almost equal to the canonical value of 100, and so we adopt this model, described in column 3 of Table~\ref{tab:Para_fit}, as our ``preferred'' model. However, the inability of a well-mixed disc such as this to fit the observed emission in the mid-infrared means that there remains compelling evidence for a disc exhibiting dust-settling, able to fit the line emission with a depleted gas/dust ratio (see column 1 in Table~\ref{tab:Para_fit}). The effect of dust settling on the line emission is further explored in Section~\ref{sec:settle}.

The model with power-law density profile (column 4 in Table~\ref{tab:Para_fit}) gives the best overall fit to the SED, ISO-SWS spectrum and line emission (see Fig.~\ref{fig:SED}), with quite different disc parameters to the three models with exponential outer edges (columns 1-3). The disc requires a very flat density profile, $\epsilon\,\sim\,0.085$, in order to fit these observations (not including spatial data). The predicted millimetre continuum emission from this model is plotted with the observed maps in Fig.~\ref{fig:maps}. It is clear that the flat density profile leads to an emission deficit in the inner disc in comparison to observations, and so this cannot be considered to be a good model for the disc of HD 163296. A better power-law model could be found by using the maps as a further constraint for the modelling, but that is beyond the scope of this paper. The fully mixed ``preferred'' model gives a better match to the observed millimetre emission, as should be expected since its density profile is derived from fitting to this data \citep{Hughes2008}. The factors driving the power-law model to such a flat density profile will be discussed in Section~\ref{sec:gasmodel}.

The power-law density profile model is, however, able to fit the non-spatially-resolved data with a smaller flaring index ($\sim1.02$) than the exponentially-tapered discs ($\sim1.07$), more in keeping with the observational evidence for a disc that is not strongly flaring \citep{Meeus2001,Doucet2006,Wisniewski2008}. \citet{Meeus2001} classify HD 163296 as a group II object, in which the inner disc shadows the outer disc, as opposed to group I objects which have a flared geometry. \citet{Meijer2008} find HD 163296 to lie on the transition between flared and non-flared geometry, i.e. a flaring index close to 1. It is clear that our ``preferred'' model is unable to account for every observed property of the disc of HD 163296, and it is probable that a more complex model - with non-constant flaring and variable dust properties with radius - would be required to simultaneously fit the full set of observational data available for this object.

The models described in Table~\ref{tab:Para_fit} have disc dust masses in the range $(7-12) \times 10^{-4}$M$_{\odot}$, which is within the range of masses $(5-17) \times 10^{-4}$M$_{\odot}$ found in the literature for HD 163296. This small spread in dust mass from our fitting efforts is unsurprising given the fixed disc size and constant grain composition.

\subsection{Gas properties}
\label{sec:gasmodel}

\begin{table*}
\centering
\caption{Integrated line fluxes for the transitions observed by PACS, and additional CO and H$_2$ lines, as predicted by the models listed in Table~\ref{tab:Para_fit}. Observed fluxes for comparison. Fluxes in $[10^{-18}\rm\,W/m^2]$. Errors are a quadratic sum of the calibration error (20\% for the interferometric CO observations and 30\% for the PACS observations) and the RMS continuum noise.}
\begin{tabular}{lccccccc}
\hline
Species       &  $\lambda$\,[$\mu$m] & $\nu$\,[GHz] & ``low UV'' model & ``high UV'' model & Fully mixed model & Power-law  & Observed Flux \\
\hline\hline
OI            &  63.18             & 4745.05   & 200.6 & 222.0 & 191.4 &  170.3 & 193.1\,(58.2)\\
OI            &  145.52            & 2060.15   & 6.29  & 7.33  & 5.39  & 4.59  & $<$ 8.5\\
CII           &  157.74            & 1900.55   & 8.32  & 8.40  & 11.2  & 9.98  & -- \\ 
p-H$_2$O $3_{22} \rightarrow 2_{11}$     &  89.99             & 3331.40   & 1.22  & 1.68  & 1.20  & 0.353 & $<$ 9.4\\
o-H$_2$O $2_{12} \rightarrow 1_{01}$      &  179.52            & 1669.97   & 4.40  & 4.19  & 2.43  & 3.94  & $<$ 14.5\\
o-H$_2$O  $2_{21} \rightarrow 2_{12}$    &  180.49            & 1661.64   & 0.847
 & 0.993 & 0.618 & 0.243 & $<$ 16.2\\
o-H$_2$O  $4_{32} \rightarrow 3_{12}$    &  78.74            & 3810.01   & 2.33
 & 2.94 & 2.20 & 0.840 & $<$ 15.0\\
o-H$_2$ S(1)  &  17.03             & 17603.78  & 1.15  & 1.59  & 1.15  & 0.572 & $<$ 28\\
OH            &  79.11            & 3792.19    & 4.59  & 6.31  & 5.40  & 2.71 & $<$ 17.0\\
OH            &  79.18            & 3788.84    & 4.78  & 6.54  & 5.59  & 2.82 & $<$ 17.0\\
CO J=36-35    &  72.85             & 4115.20   & 0.335 & 0.489 & 0.423 & 0.132 & $<$ 11.6\\
CO J=33-32    &  79.36             & 3777.63   & 0.497 & 0.704 & 0.578 & 0.197 & $<$ 22.8\\
CO J=29-28    &  90.16             & 3325.12   & 0.763 & 1.03  & 0.732 & 0.311 & $<$ 11.1\\
CO J=18-17    &  144.78            & 2070.68   & 1.91  & 2.31  & 1.49  & 0.851 & $<$ 13.1\\
\hline
CO J=3-2      &  866.96            & 345.80    & 1.26  & 1.27  & 1.22  & 1.49  & 1.65\,(0.39) \\
CO J=2-1      &  1300.40           & 230.54    & 0.401 & 0.401 & 0.397 & 0.464 & 0.379\,(0.118)\\
$^{13}$CO J=1-0 &  2720.41          & 110.20    & 0.0122 & 0.0115 & 0.0120 & 0.0127 & 0.0124\,(0.007) \\
\hline
\end{tabular}
\label{modelflux}
\end{table*}

The predicted line fluxes for the various models are given in Table~\ref{modelflux}. We have ignored the [{\sc Cii}]\,158\,$\mu$m result during the model-fitting process, due to systematic uncertainties arising from strong emission at the offset positions.

\begin{figure*}
\begin{tabular}{cc}
  \includegraphics[width=8.0cm]{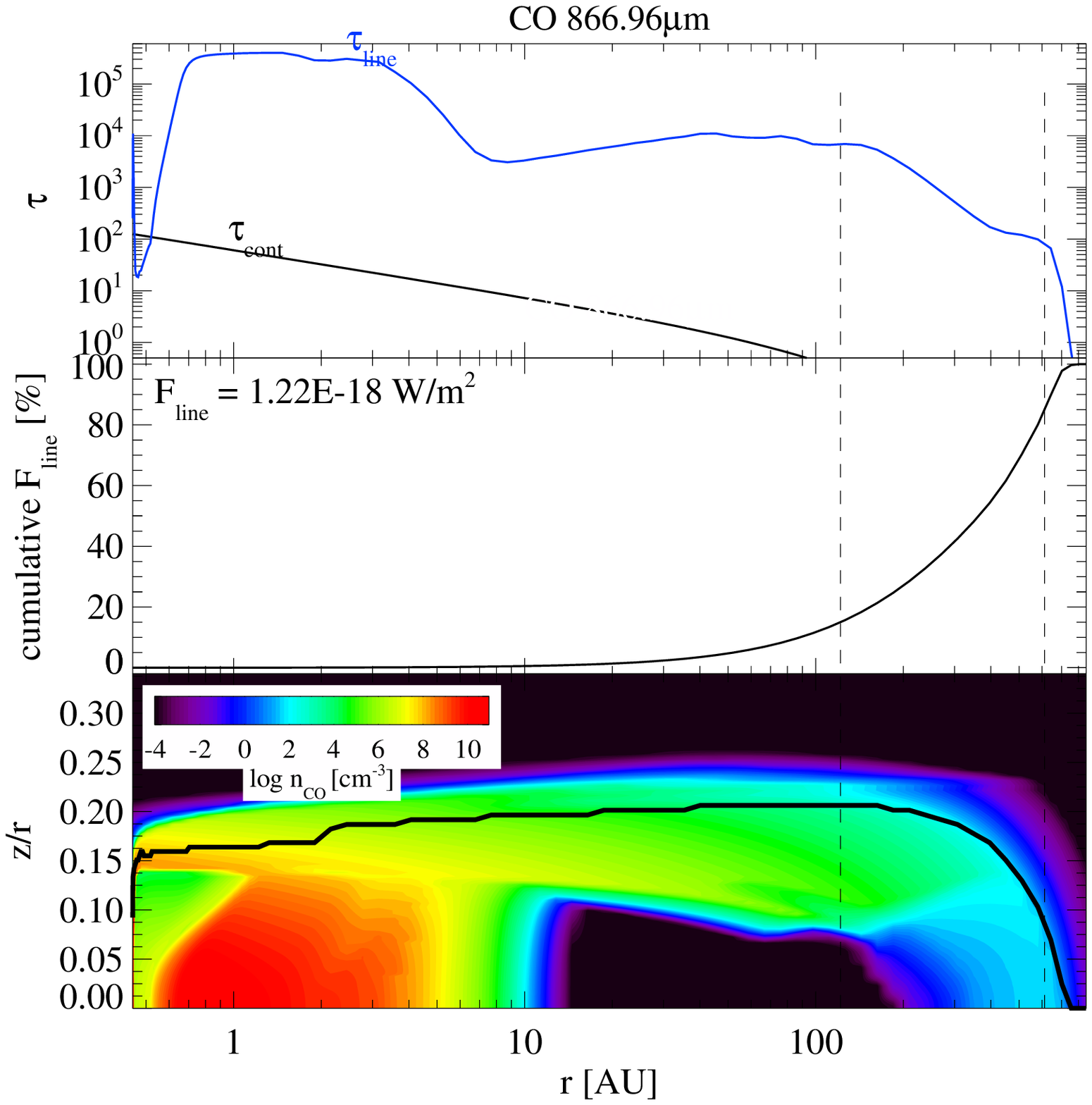} &
  \includegraphics[width=8.0cm]{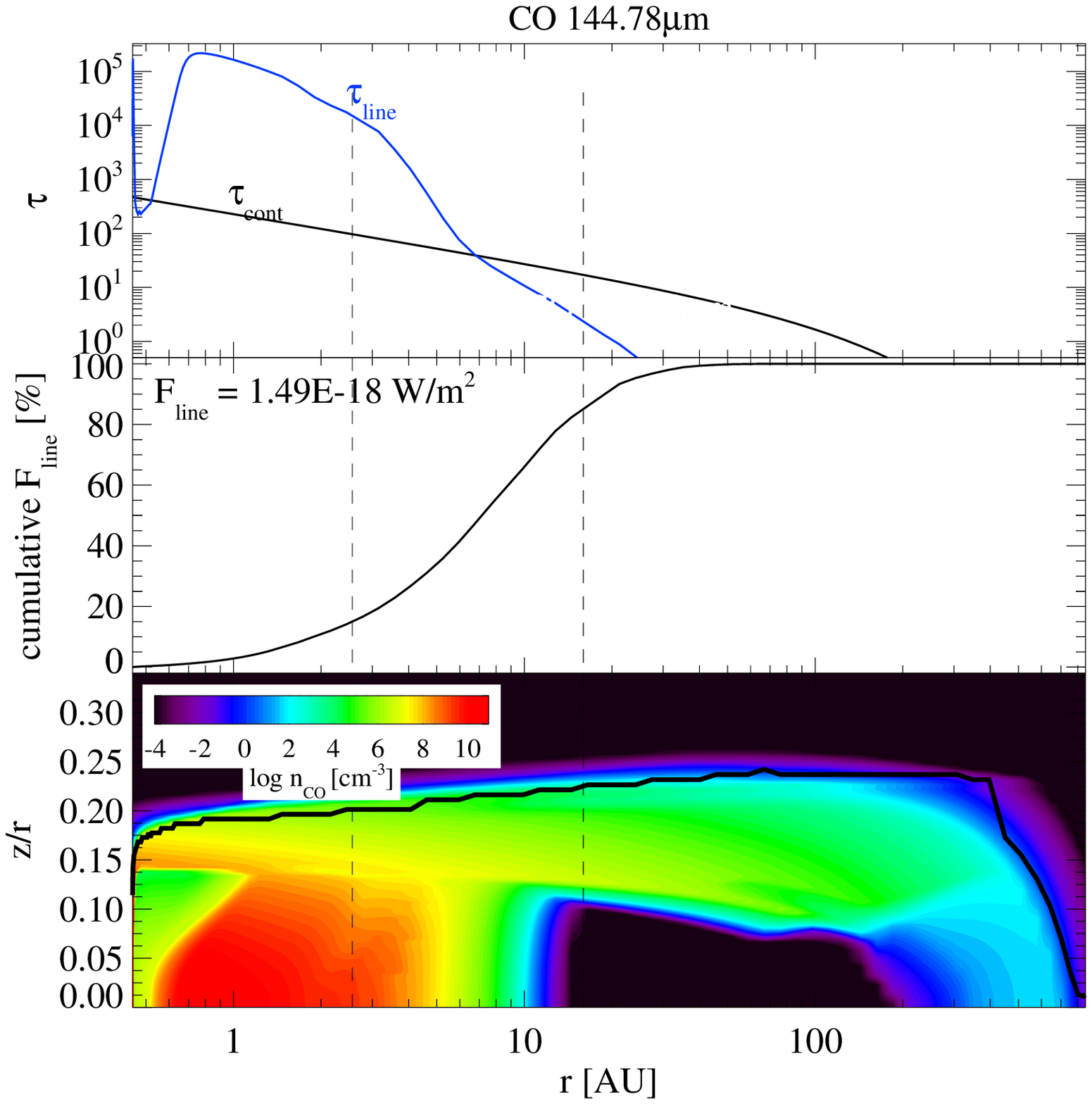}\\
  \includegraphics[width=8.0cm]{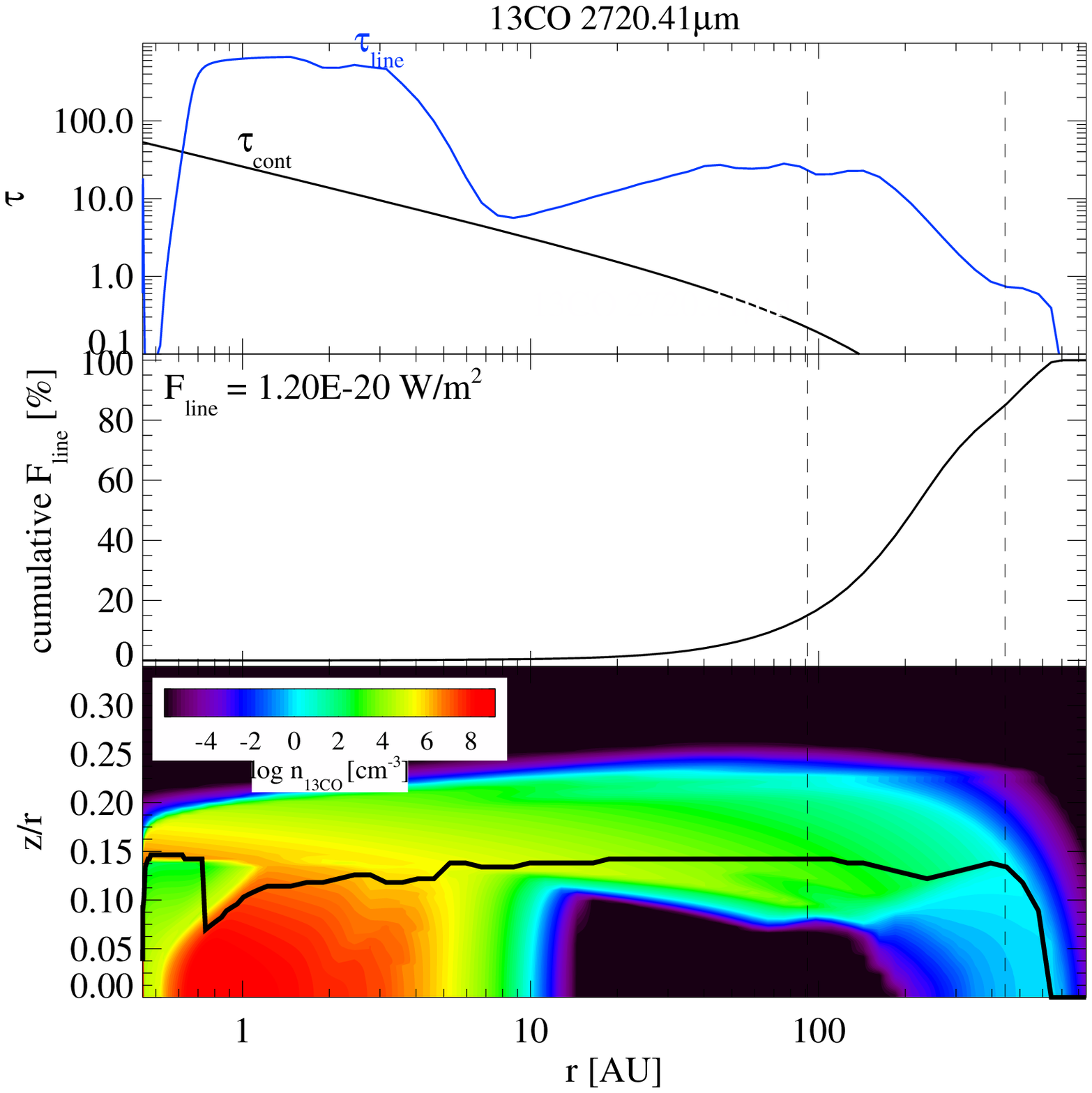} &
  \includegraphics[width=8.0cm]{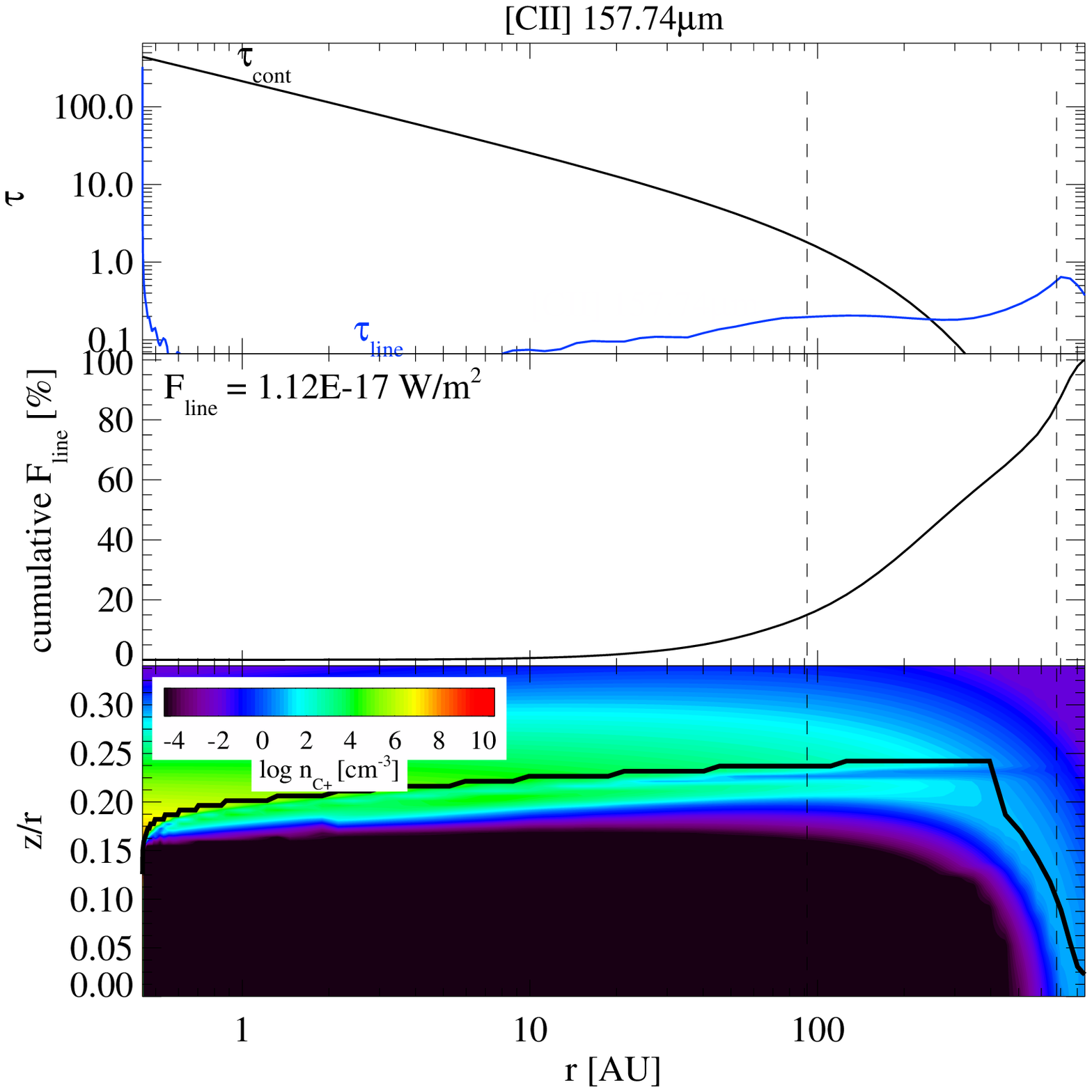}\\
  \includegraphics[width=8.0cm]{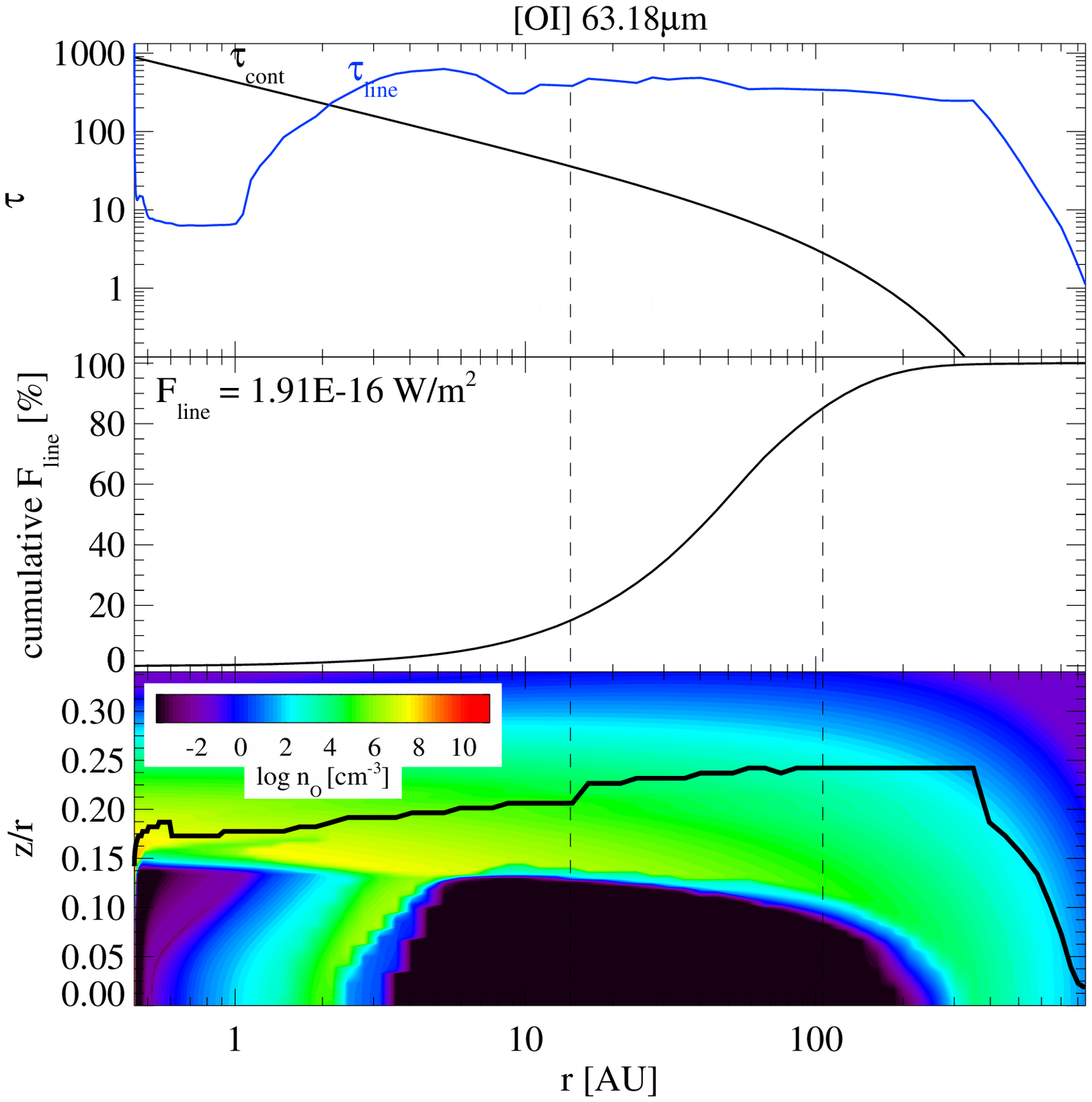} &
  \includegraphics[width=8.0cm]{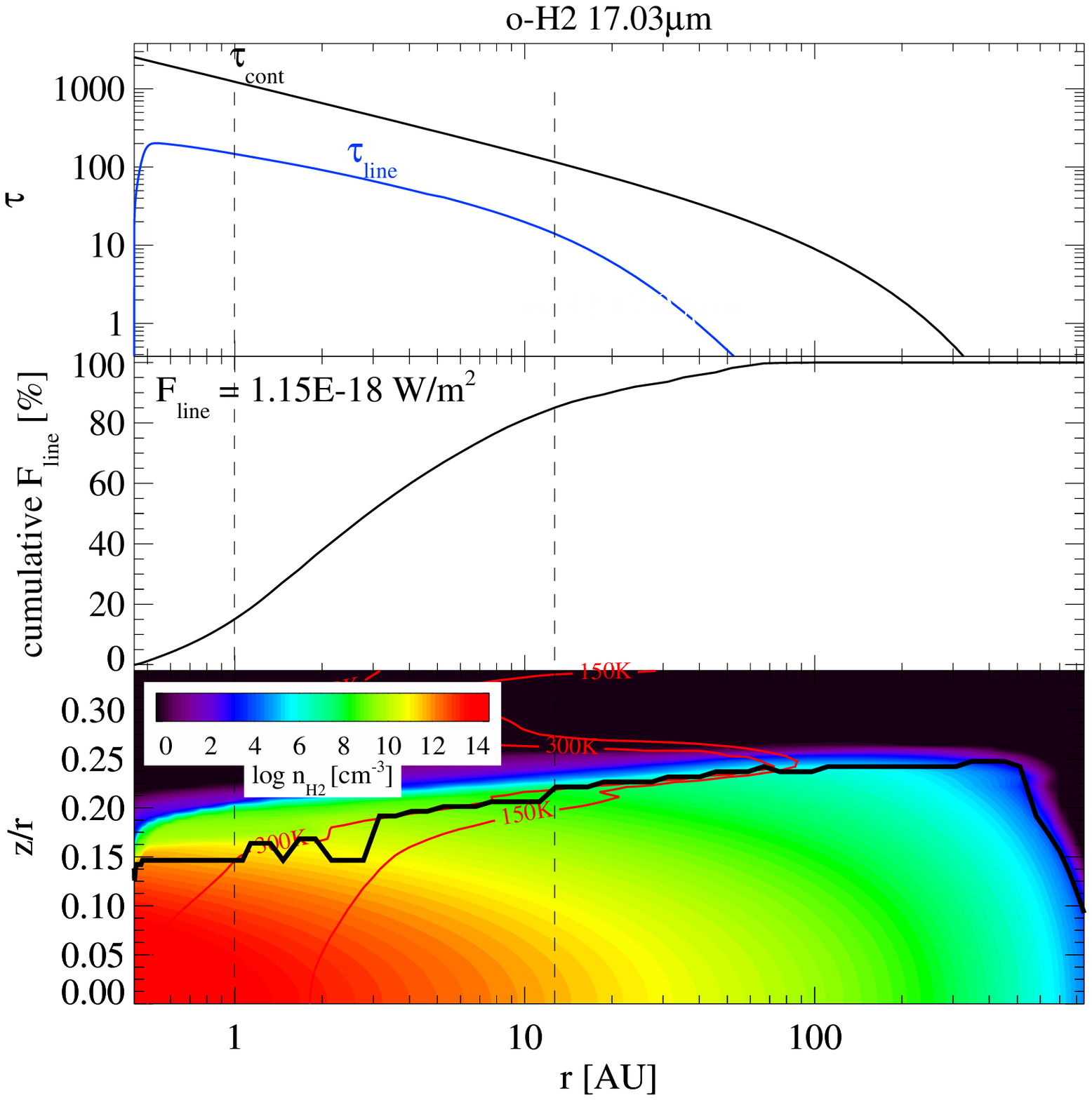}
\end{tabular}
\caption{Spatial origin of the various gas emission lines in the preferred model (column 3 in Table~\ref{tab:Para_fit}). From top-left clockwise: CO J=3-2, CO J=18-17, [CII] 157.74$\mu$m, o-H$_2$ S(1) (with red $T_g$ contours), [OI] 63.18$\mu$m, $^{13}$CO J=1-0. In each panel, the upper plot shows the line optical depth as a function of radius (blue line) and the continuum optical depth at the corresponding wavelength (black line). The middle plot shows the cumulative contribution to the total line flux with increasing radius. The lower plot shows the gas species number density, and the black line marks the cells that contribute the most to the line flux in their vertical column. The two vertical dashed lines indicate 15\% and 85\% of the radially cumulative face-on line flux respectively, i.e. 70\% of the line flux originates from within the two dashed lines.}
\label{fig:analysis}
\end{figure*}

The observed line profiles for the $^{12}$CO J=3-2, $^{12}$CO J=2-1  and $^{13}$CO J=1-0 transitions \citep{Isella2007} are plotted with the preferred model profiles in Fig.~\ref{fig:profiles}. The observed profiles are double-peaked, consistent with a disc in Keplerian rotation. The CO 3-2 line is observed to be brighter than predicted by the model. This behaviour is repeated across the three exponentially-tapered models, but is less pronounced in the power-law model, whose flat surface density profile gives a larger CO 3-2 flux. Discs with flatter density profiles allow stellar radiation to penetrate deeper into the inner disc, resulting in a slight increase in temperature throughout the disc, including its outer regions. This leads to brighter CO 3-2 lines since this transition is optically thick in all of our models, and is largely dependent on the disc outer radius (which is fixed) and the outer disc temperature at intermediate height (see Fig.~\ref{fig:analysis}). The CO 2-1 line also follows this behaviour, with the line flux ratio in our models staying roughly constant at a value of CO 3-2/1-0 $\sim$\,3, as expected for lines formed under optically thick LTE conditions \citep{Kamp2010}. This is slightly lower than the observed ratio of CO 3-2/2-1 = 4.3. We note that recent observations of this object give a peak intensity of 25\,Jy for the 3-2 line \citep{Hughes2011}\footnote{This CO 3-2 data also seems to indicate the presence of turbulence of order 0.3\,km/s in the disc of HD 163296 \citep{Hughes2011}. We have not yet explored the effect of this possible stronger turbulence on our results.}, compared with 43\,Jy in the \citet{Isella2007} data and the 30\,Jy predicted by our preferred model. Our predicted flux is within the calibration error margin for both observations, but fitting to the lower value would in general tend to lead to power-law models with steeper density profiles, in contrast to the flat power-law profile we obtain from fitting to the brighter observation. Another factor driving the power-law models to flat density profiles is the overprediction in near-IR continuum emission by our models. The inner radius is fixed in accordance with high resolution imaging, and so the power-law models tend to reduce their near-IR emission by removing material from the inner disc, giving a better fit to the photometry.

PAHs are one of the main sources of gas heating in the disc, as UV photons cause them to emit excited electrons via the photoelectric effect, which thermalise in the gas. For the purposes of our modelling we consider a typical size of PAH molecule (circumcoronene, $N_{\rm C}\!=\!54$ carbon atoms and $N_{\rm H}\!=\!18$ hydrogen atoms) and include PAH$^{-}$, PAH, PAH$^{+}$, PAH$^{2+}$ and PAH$^{3+}$ as additional species in the chemical network (see \citet{Woitke2010} for further details). The fractional PAH abundance, $f_{\rm PAH}$, is defined relative to the standard ISM particle abundance with respect to hydrogen nuclei, $X^{\rm ISM}_{\rm PAH}\!=\!3 \times 10^{-7}$ \citep{Tielens2008}, so that the PAH abundance, $\epsilon({\rm PAH})\!=\!f_{\rm PAH}\,X^{\rm ISM}_{\rm PAH}\,\frac{50}{N_{\rm C}}$. Small $f_{\rm PAH}$ values (0.007-0.04 relative to ISM abundances) are required in our models in order to fit the observed line data. This is consistent with the findings of \citet{Geers2006}, who require PAH abundances $f_{\rm PAH}\!<\!0.1$ in order to fit the observed PAH emission from Herbig discs. PAH emission was not observed in HD 163296 by \citet{Acke2010}, and we have used the Monte Carlo radiative transfer code {\sc Mcfost} \citep{Pinte2006} to check that our results are consistent with this non-detection. A disc model with parameters and PAH abundance identical to our preferred model gives an infrared spectrum in which no PAH emission is seen. This is due in part to the low flaring in the disc, $\beta\!=\!1.066$, and the large fraction of small dust grains in the disc, with significant opacity in the optical-UV range over which PAH molecules absorb. This results in the PAHs being ``hidden'' from direct stellar illumination, absorbing a fraction $\sim\!10^{-4}$ of the energy absorbed by the disc. In our preferred model 0.3\% of the total carbon mass is tied up in PAH molecules.

\begin{figure*}
\begin{tabular}{cc}
  \includegraphics[width=8.5cm,height=7.5cm]{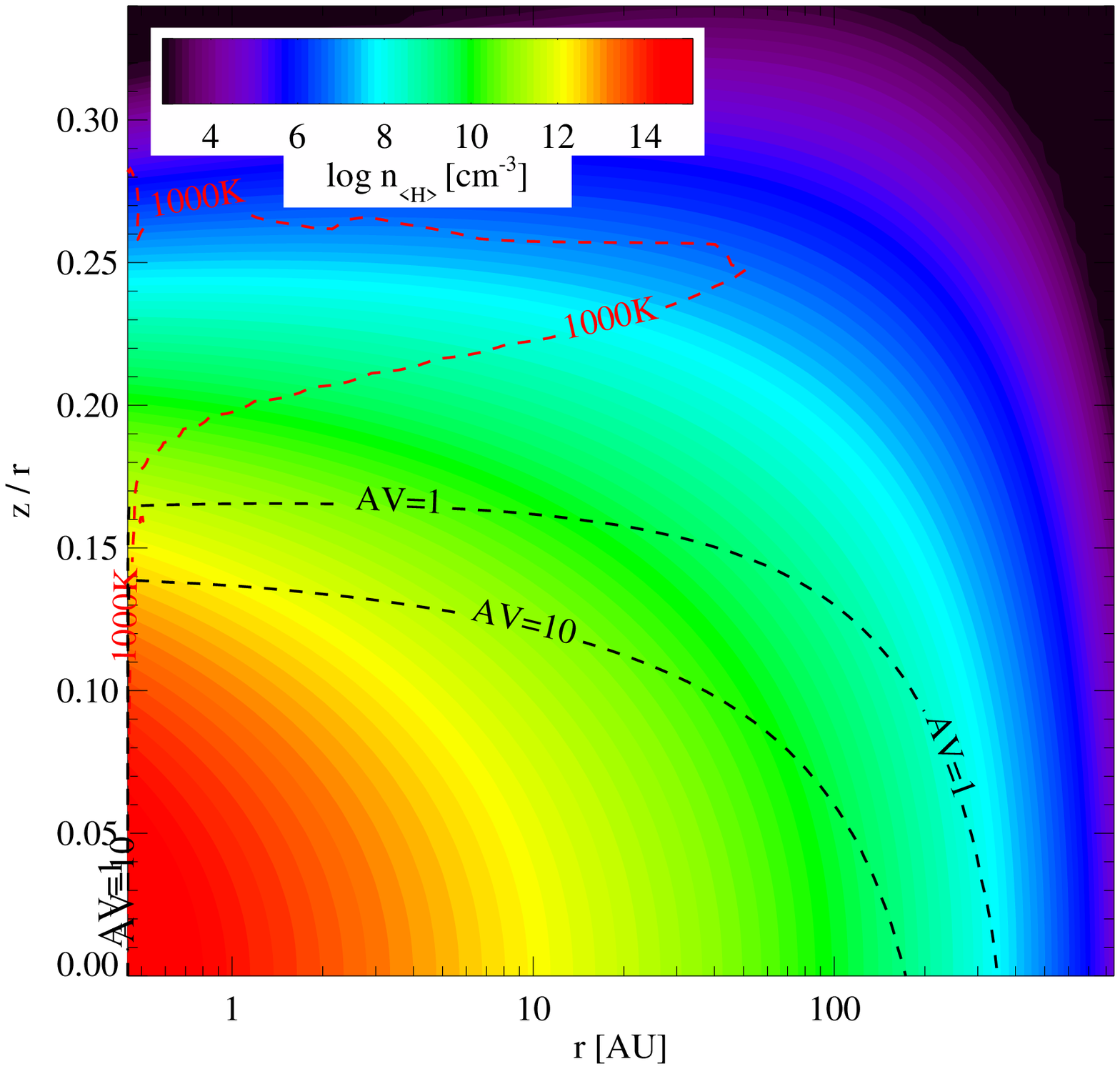} &
  \includegraphics[width=8.5cm,height=7.5cm]{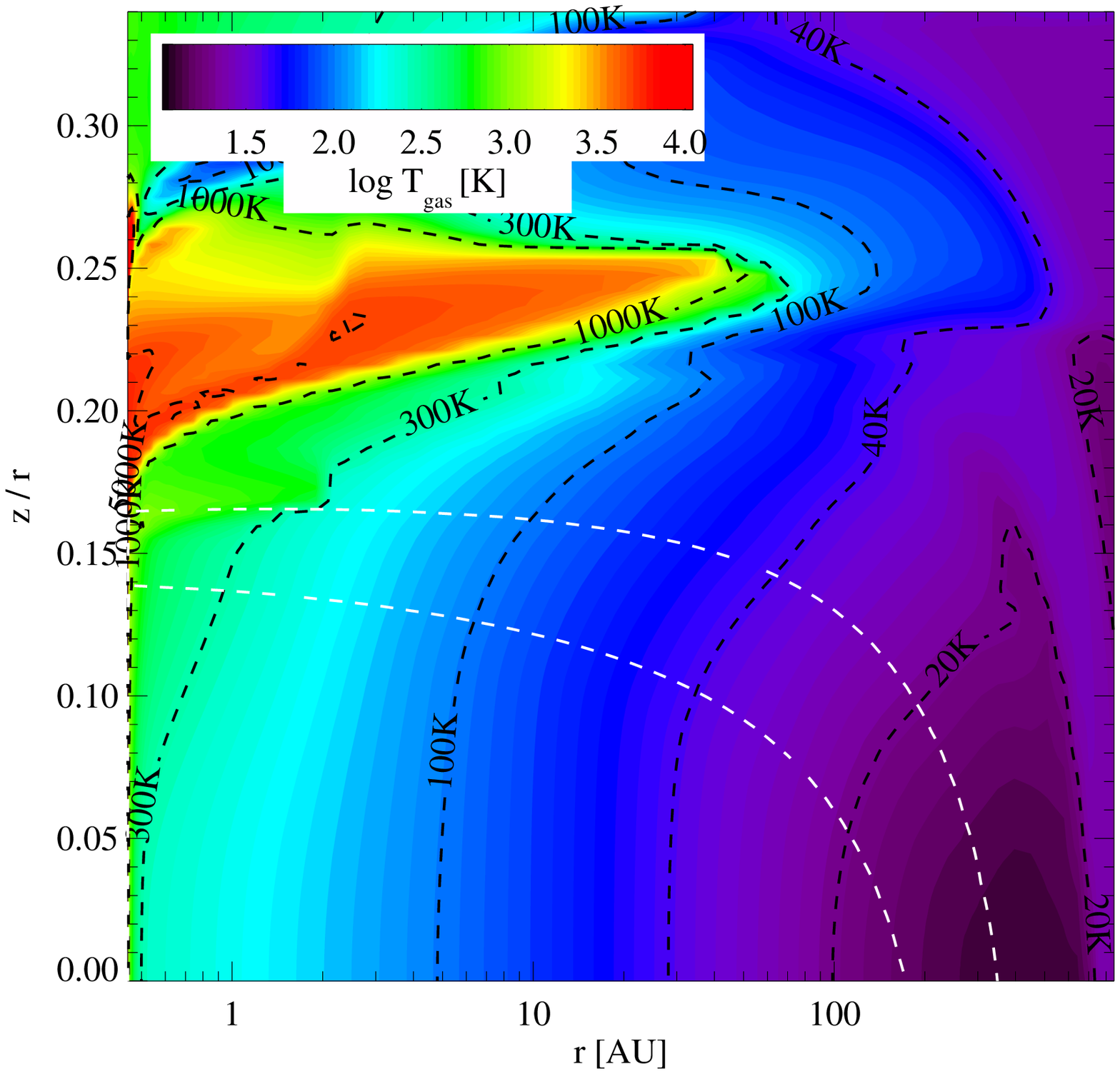}\\
\end{tabular}
\caption{The total hydrogen number density (left panel) and gas temperature (right panel) are plotted for a vertical cross-section through the preferred disc model (column 3 in Table~\ref{tab:Para_fit}). Dashed lines indicate contours of gas temperature and visual extinction $A_{\rm V}$, as marked.}
\label{fig:disc}
\end{figure*}

The total hydrogen number density and gas temperature within the disc are plotted in Fig.~\ref{fig:disc}, and the spatial origin of the various emission lines is visualised in Fig.~\ref{fig:analysis}. These plots all refer to the ``preferred'' disc model, i.e. one in which no dust settling is present. The CO J=3-2 and [{\sc Oi}]\,63\,$\mu$m lines are optically thick throughout the disc, and cannot be used alone to trace the gas mass. Even the $^{13}$CO line is optically thick throughout much of the disc, only becoming optically thin outside of $\sim\,400$\,AU. Also, there is evidence that this line (and most others) can be affected by the degree of dust settling in the disc (see Section~\ref{sec:settle}). The [CII] 157.74$\mu$m line is optically thin, but traces only the warm ionised gas in the disc surface. Clearly care must be taken when attempting to use individual emission lines as a tracer of gas mass. Note the contrast in disc radii at which line and continuum become optically thin in the case of CO 3-2, reflecting the conflicting derived radii for the gas and dust discs in HD 163296. The line optical depths in Fig.~\ref{fig:analysis} were computed using an escape probability formalism, and do not take into account the effects of Keplerian shear in the disc. 

\begin{figure*}
  \begin{tabular}{cc}
    \includegraphics[width=8.5cm]{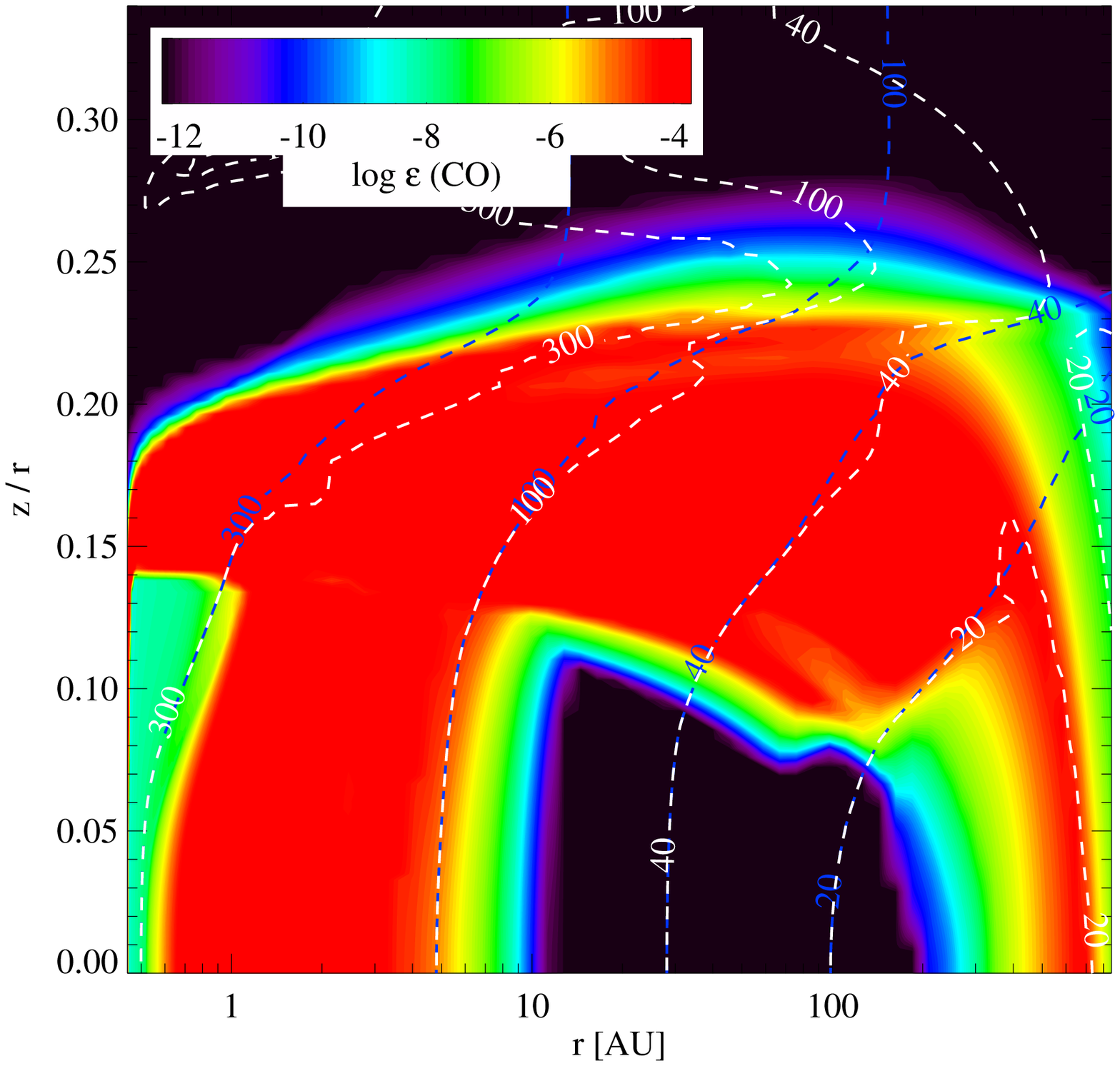} &
    \includegraphics[width=8.5cm]{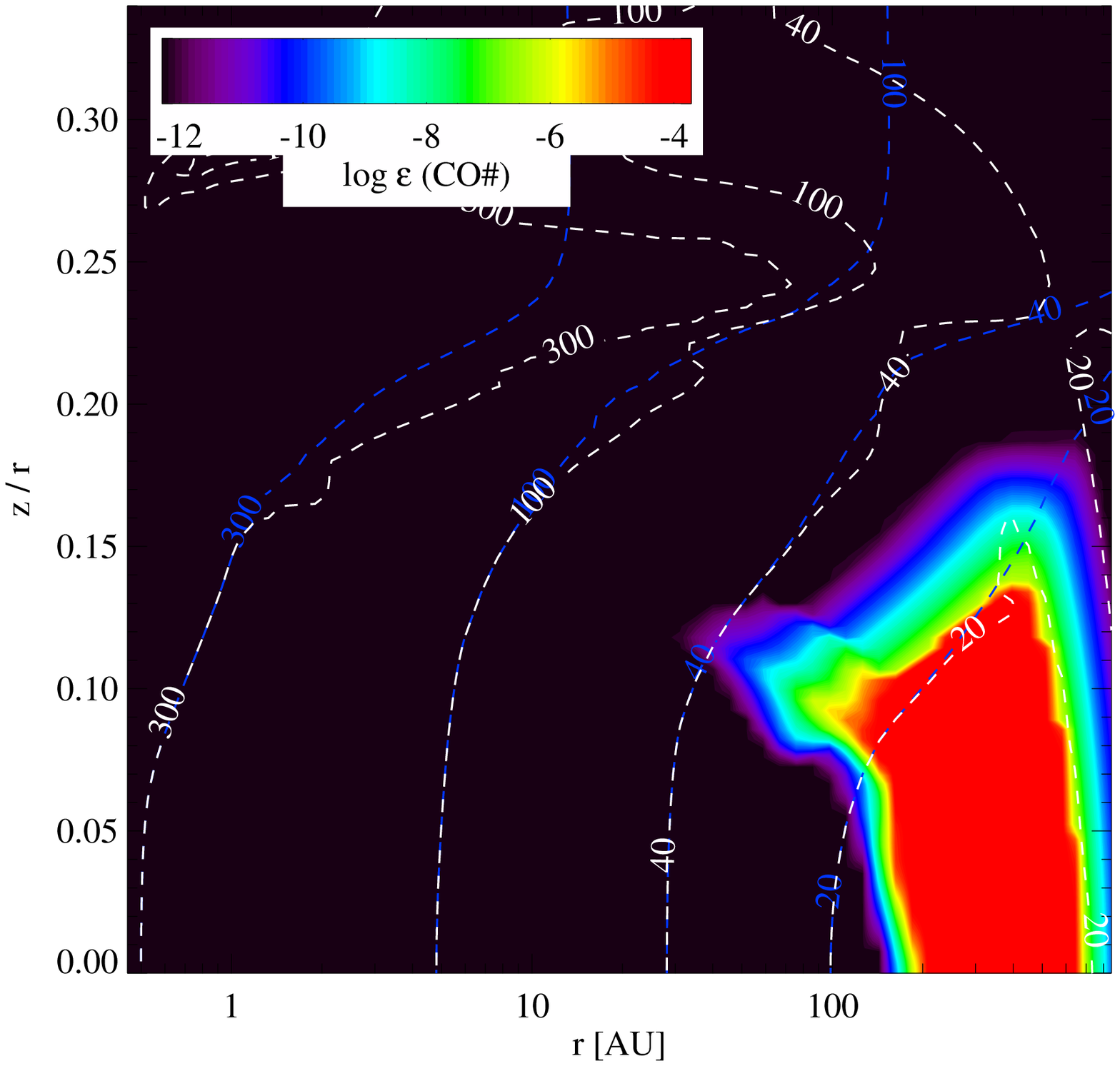}\\
  \end{tabular}
  \caption{CO abundances in the preferred model (column 3 in Table~\ref{tab:Para_fit}). Left panel: gas-phase CO abundances. White and blue dashed lines plot the gas and dust temperature contours respectively. Right panel: CO ice abundance.}
  \label{fig:freeze}
\end{figure*}

As well as optical depth effects, disc gas mass estimates from CO emission alone can be affected by the extent of CO freeze-out on grain surfaces. \PRODIMO\ calculates the grain adsorption and desorption as part of the solution of the chemical network, and the spatial abundances of gas-phase and ice CO in our best-fit model are plotted in Fig.~\ref{fig:freeze}. The total mass of gas-phase CO in the disc is $1.51\,\times\,10^{-5}$M$_{\odot}$, and the CO ice mass is $4.32\,\times\,10^{-5}$M$_{\odot}$. We would expect a smaller fraction of CO to be frozen out on grain surfaces in discs with flatter density profiles, due to weaker dust shielding in the inner disc leading to warmer temperatures throughout the disc, and less CO freeze-out.

The derived chemical abundances and gas properties in our preferred model have been checked by re-computing the chemistry using a time-dependant solver, and comparing the results to those obtained with the assumption of kinetic equilibrium. We assume molecular cloud initial abundances, and after running the solver for 4\,Myr there is no major departure from the equilibrium chemistry. The biggest change in line flux is a reduction of 20\% in the 180.4\,$\mu$m water line, while the [{\sc Oi}]\,63\,$\mu$m line decreases by 4\% and the CO lines by $<$\,1\%. The assumption of a constant $^{13}$CO/$^{12}$CO ratio is also valid since dust shielding dominates over CO self-shielding in the models, with negligible change to the results when CO self-shielding is swtiched off. We would therefore expect there to be no fractionation effect present in this disc.

All of the disc models considered are passive discs, i.e. the viscous ``$\alpha$'' heating parameter is set to zero. This is not strictly consistent with the presence of accreting gas in this object, since this implies some form of viscosity to transport angular momentum through the disc. However, viscous heating is likely to be unimportant in the case of Herbig Ae discs such as this, which are thought to be dominated by radiative heating by the central star \citep{DAlessio1998}. 

\subsubsection{Effect of UV variability}

\begin{table*}
\centering
\caption{Effects of UV variability. Comparison between the preferred model, and an identical model illuminated by an input spectrum typical of a ``high UV'' state for this object. Shown are total species masses, average overall CO temperature, average CO temperature for the warm gas regions in the inner disc, and line fluxes. $L_{\rm UV}$ refers to the energy emitted in the wavelength range 912-2500\AA.}
\begin{tabular}{lcc}
\hline
                    & Best-fit model & Test model (high UV)\\ 
\hline\hline
$L_{\rm UV}/L_{\star}$ & 0.097                   & 0.155\\
$M_{\rm H}$          & $2.67\times10^{-5}$M$_{\odot}$   & $2.82\times10^{-5}$M$_{\odot}$\\ 
$M_{\rm C^{+}}$       & $8.67\times10^{-8}$M$_{\odot}$   & $9.42\times10^{-8}$M$_{\odot}$\\
$M_{\rm CO}$         & $1.54\times10^{-5}$M$_{\odot}$   & $1.66\times10^{-5}$M$_{\odot}$\\
\hline
$\langle T_{\rm CO}\rangle$ & 55.3\,K                     & 55.0\,K\\
$\langle T_{\rm CO}\rangle\,(z/R\!>\!0.2,\,R\!<\!20$\,AU) & 262.8\,K & 332.6\,K\\
\hline
$F_{\rm [OI]\,63\,\mu{\rm m}}$ & $1.91\times10^{-16}$W\,m$^{-2}$ & $2.49\times10^{-16}$W\,m$^{-2}$\\
$F_{\rm [CII]\,58\,\mu{\rm m}}$ & $1.12\times10^{-17}$W\,m$^{-2}$ & $1.41\times10^{-17}$W\,m$^{-2}$\\
$F_{\rm CO~3-2}$             & $1.22\times10^{-18}$W\,m$^{-2}$ & $1.31\times10^{-18}$W\,m$^{-2}$\\
$F_{\rm CO~36-35}$           & $4.23\times10^{-19}$W\,m$^{-2}$ & $6.22\times10^{-19}$W\,m$^{-2}$\\
\hline
\end{tabular}
\label{tab:UVtest}
\end{table*}

We have investigated the effects of UV variability on the gas in the disc in two ways. Firstly, by computing a test model with parameters identical to our ``preferred'' model, but with the high UV state spectrum used as input. This allows us to isolate the effect of increasing the UV intensity on the line emission, since all other model parameters remain the same. The results of this test are summarised in Table~\ref{tab:UVtest}. The second approach consists of a separate evolutionary run of models using the high UV state as input, and has been covered in Section~\ref{sec:results}. This method allows us to assess how the disc model parameters change in order to fit the same line data with an increased UV intensity (see Table~\ref{tab:Para_fit}). By examining firstly the physical effect of the UV on the gas in the disc, and secondly the effect of the UV on the model parameter fit, we can estimate the degree of uncertainty introduced by UV variability in this object.

An increased UV strength affects the gas chemistry by promoting photochemical reactions, and the photodissociation of H$_2$ and CO. It also leads to increased desorption of ice species from grain surfaces in regions where the UV is able to penetrate. It also strongly affects the gas heating, via the photoelectric effect in dust grains and PAHs.

The ``high UV'' input spectrum represents a 60\% increase in UV intensity over the low UV input. Table~\ref{tab:UVtest} summarises the effects of this increase on the gas in the preferred model. There is a slight overall increase in the disc gas temperature, and an increase in the atomic hydrogen and C$^{+}$ masses, due to photoelectric effects and PAH heating, and increased photodissociation of gas molecules. The total gas-phase CO mass actually increases slightly, since the increased rate of photodissociation is balanced by an increase in the rate of ice desorption. The J=36-35 line in particular sees the biggest increase in flux, $\sim$\,50\%. By considering only the gas in the warm layers of the inner disc, where the high J CO lines form, it can be seen that this increase in line flux is caused by a substantial increase in the gas temperature in this region. The [{\sc Oi}]\,63\,$\mu$m flux increases by $\sim$\,30\%, roughly equal to the calibration uncertainty margin, and the low J CO lines increase by $\sim$\,5\%. All non-detected lines stay within the observed upper limits. In summary, it would not be possible to distinguish between the high and low UV states from these disc observations, since all fluxes are within the observational error margins. The general influence of the UV intensity on the disc gas is discussed by \citet{Woitke2010}, in the context of a large grid of models.

The derived parameters for the evolutionary run using the ``high UV'' spectrum as input are listed in Table~\ref{tab:Para_fit}. They are largely similar to those derived from the low UV run (columns 2 c.f. 1), as would be expected from the small fractional change in UV intensity. The slight decrease in gas mass and flaring will tend to counteract any increase in line emission. The PAH fractional abundance decreases slightly as might be expected. The main change lies in the grain size distribution, with a reduction in the minimum grain size. This makes it harder for the UV to penetrate the disc, balancing the increase in UV intensity. This also produces even more continuum emission in the near-IR, worsening the SED fit still further in comparison to the ``low UV'' model.

\subsubsection{Effects of dust settling}
\label{sec:settle}

\begin{figure}
\centering
\hspace{-6mm}
\includegraphics[width=8.5cm,height=7.0cm]{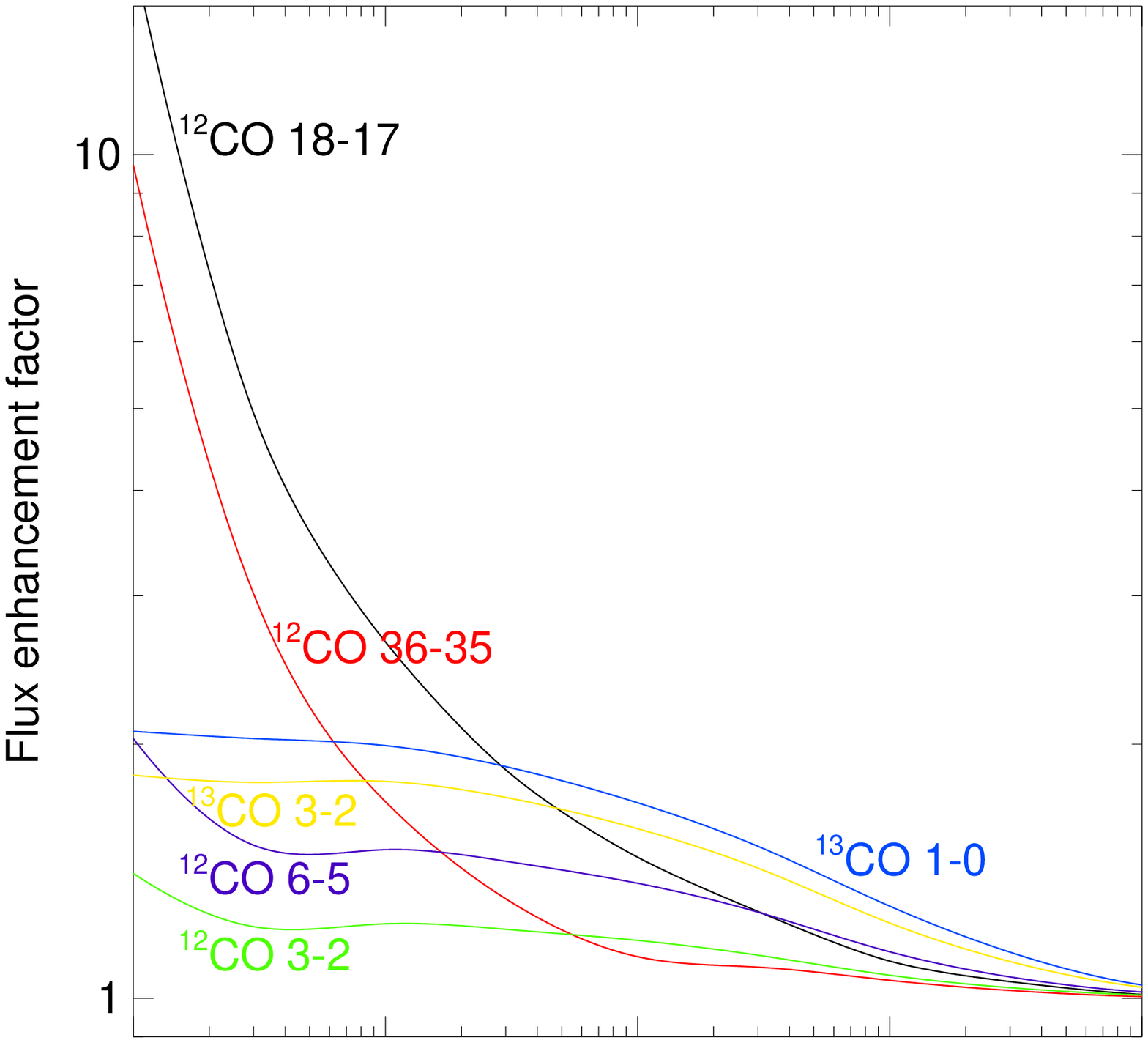}\\
\hspace{-6mm}
\includegraphics[width=8.5cm,height=7.0cm]{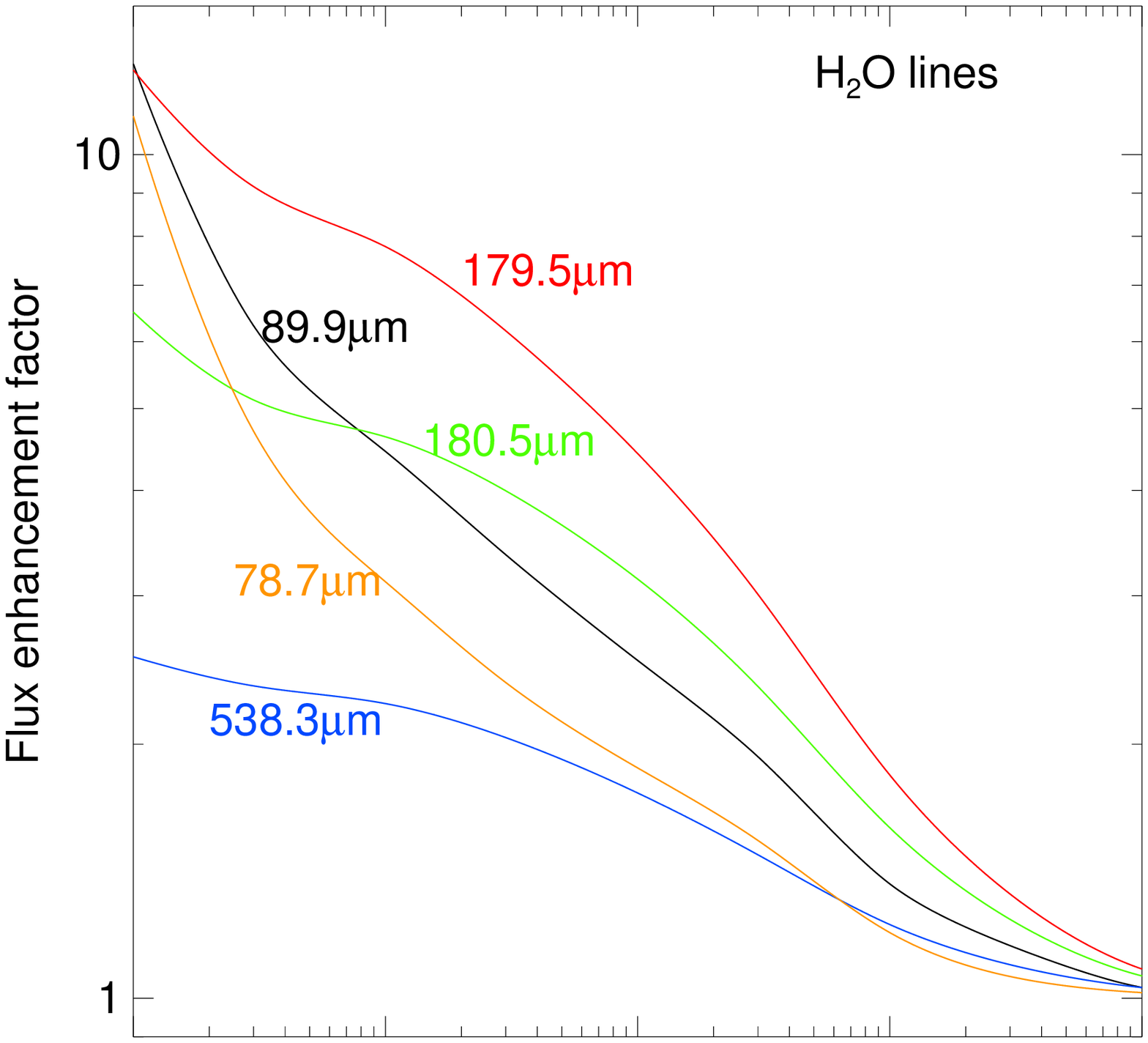}\\
\hspace{-6mm}
\includegraphics[width=8.5cm,height=8.0cm]{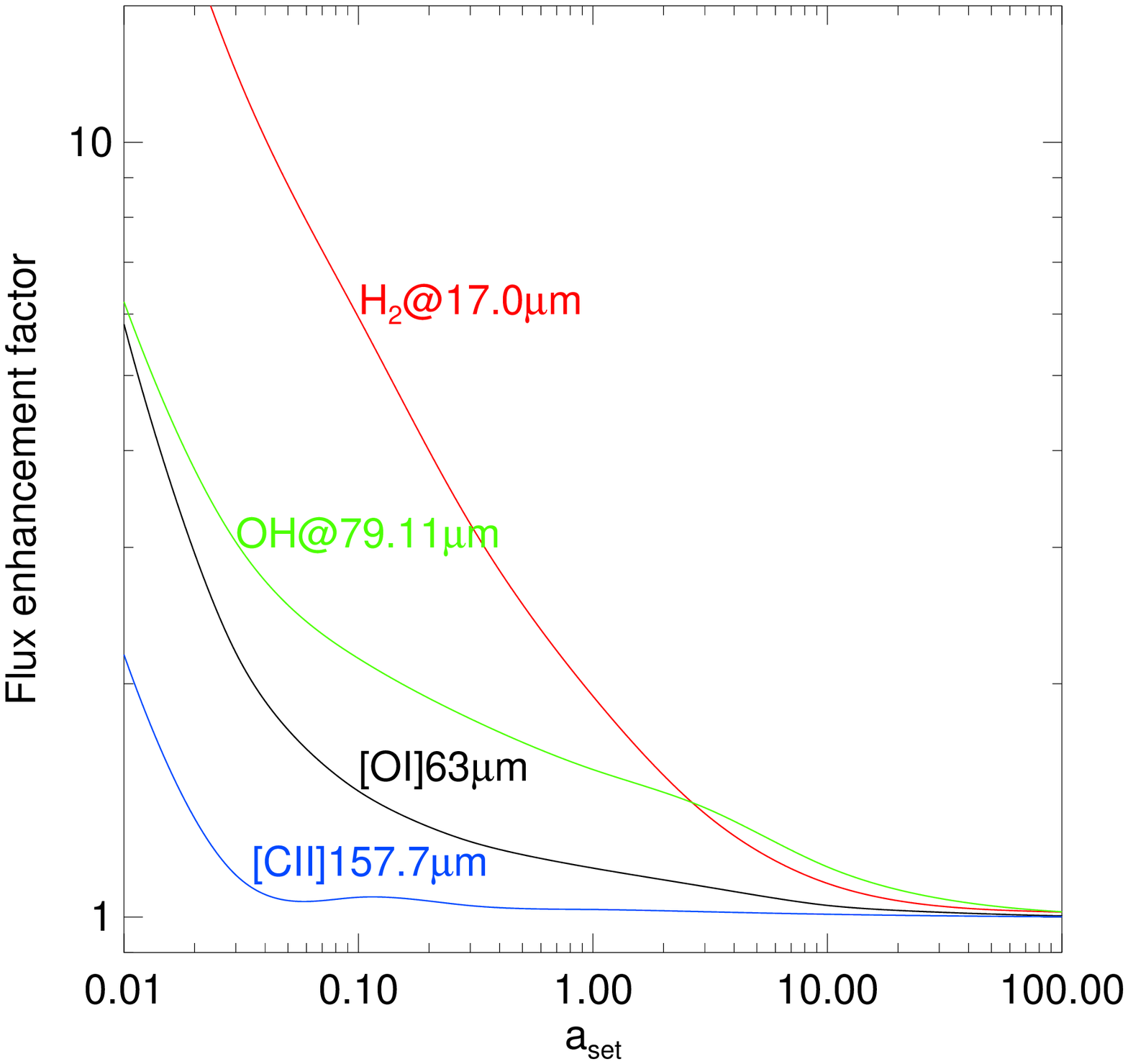}\\
\caption{The fractional enhancement in line flux is plotted against the minimum grain size affected by dust settling, relative to the preferred (fully mixed) model ($a_{\rm s}\!=\!a_{\rm max}$). All models have a settling parameter of 0.5. Top panel: CO lines; Middle panel: water lines; Lower panel: [{\sc Oi}]63$\mu$m, [{\sc Cii}]158$\mu$m, OH 79.11\,$\mu$m and H$_2$ S(1) lines.}
\label{fig:settle}
\end{figure}

The effect of dust settling on various line fluxes can be seen in Fig.~\ref{fig:settle}. This shows the increase in line flux across the various transitions as smaller and smaller grains are allowed to settle towards the midplane. The models are otherwise identical to the preferred (fully mixed) model, but with a settling parameter of 0.5. The models run from strongly settled discs through to entirely well-mixed, i.e. identical to the preferred model. For the purposes of this exercise we have computed some extra line fluxes in addition to those observed in HD 163296, namely the $^{12}$CO 6-5 and 1-0 transitions, the $^{13}$CO 3-2 transition, and the o-H$_2$O 538.3$\mu$m transition.  

\begin{figure*}
\begin{centering}
\begin{tabular}{cc}
  \includegraphics[width=7.5cm]{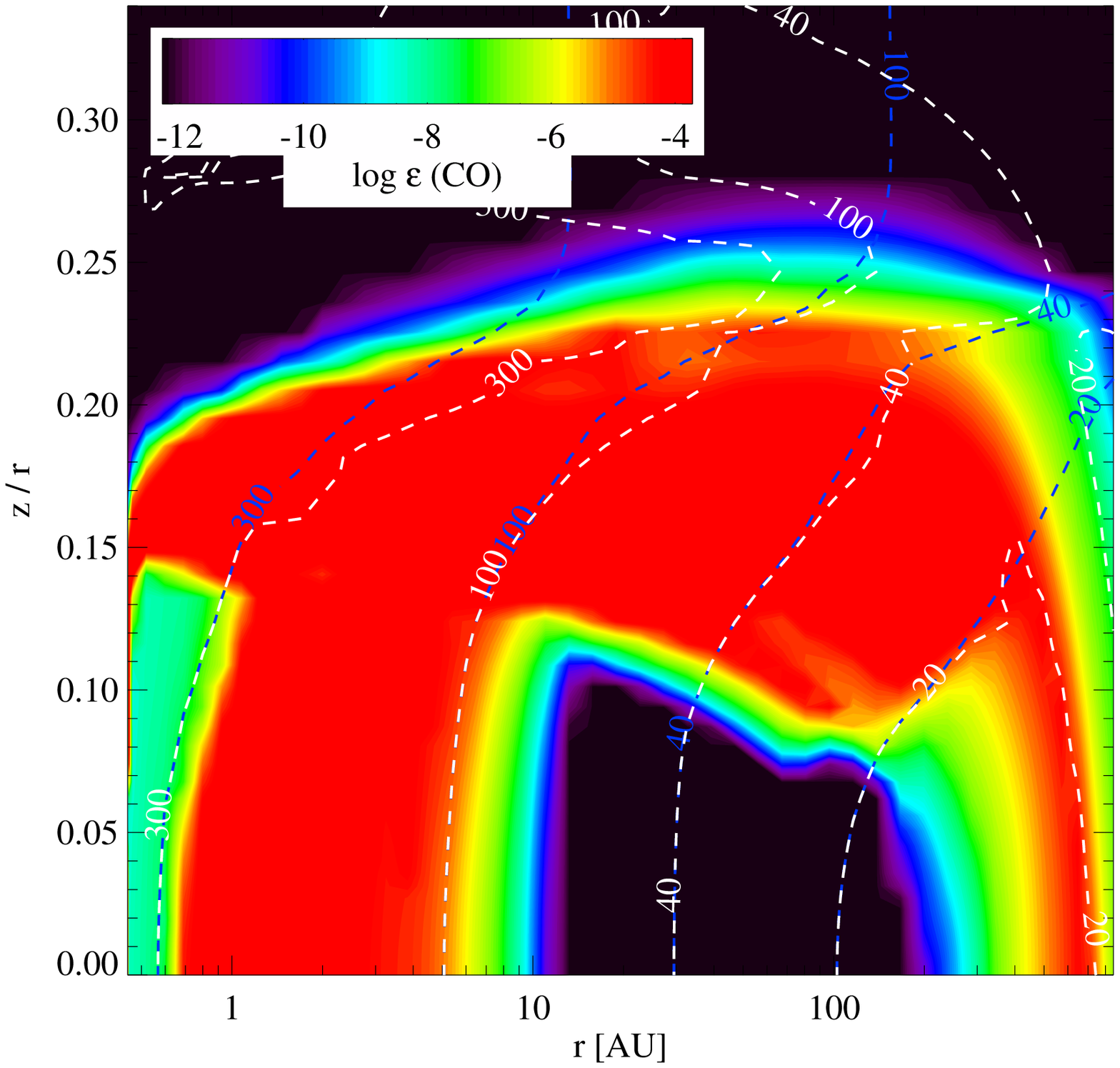} &
  \includegraphics[width=7.5cm]{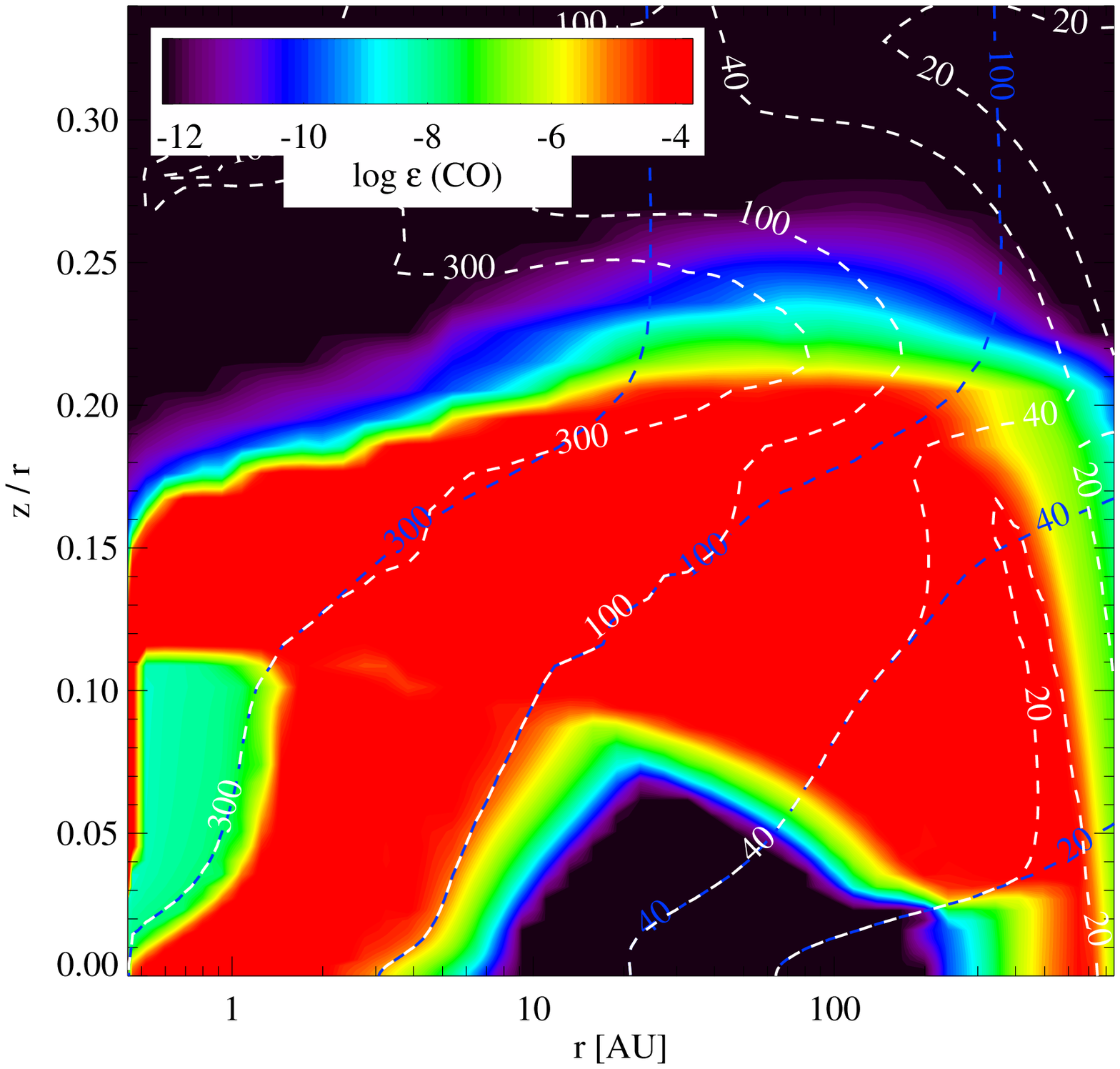}\\
  \includegraphics[width=7.5cm]{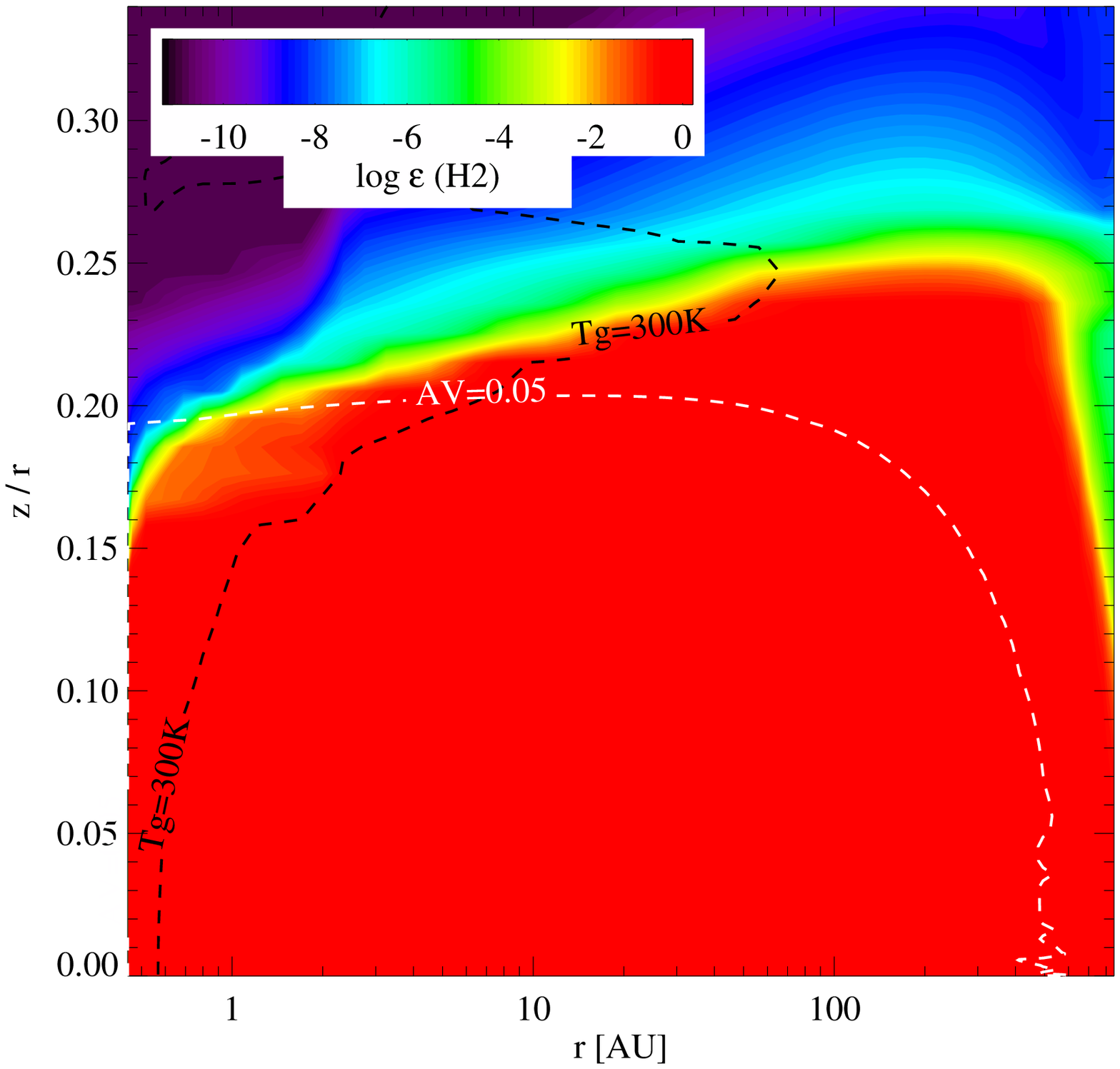} &
  \includegraphics[width=7.5cm]{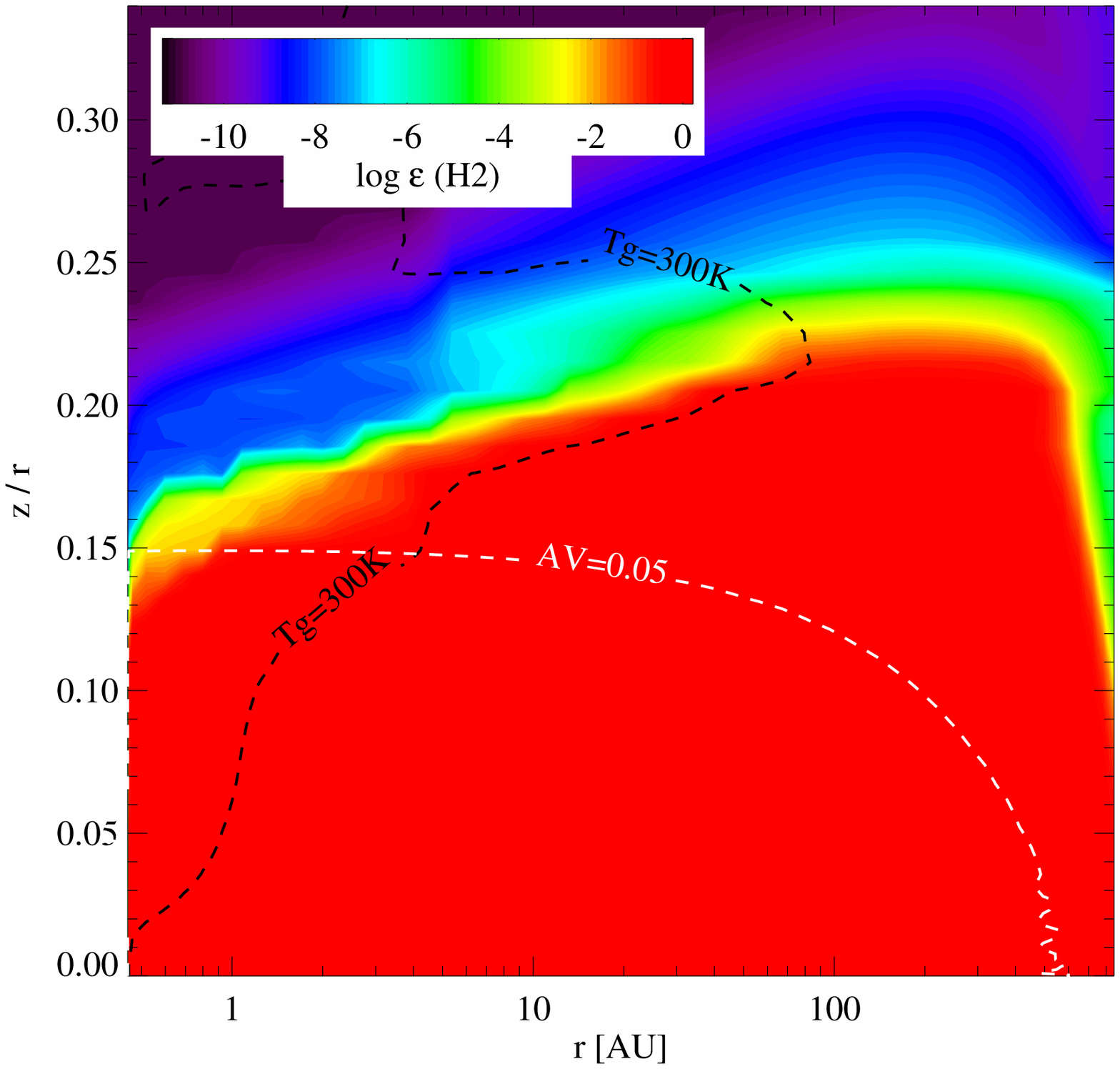}\\
  \includegraphics[width=7.5cm]{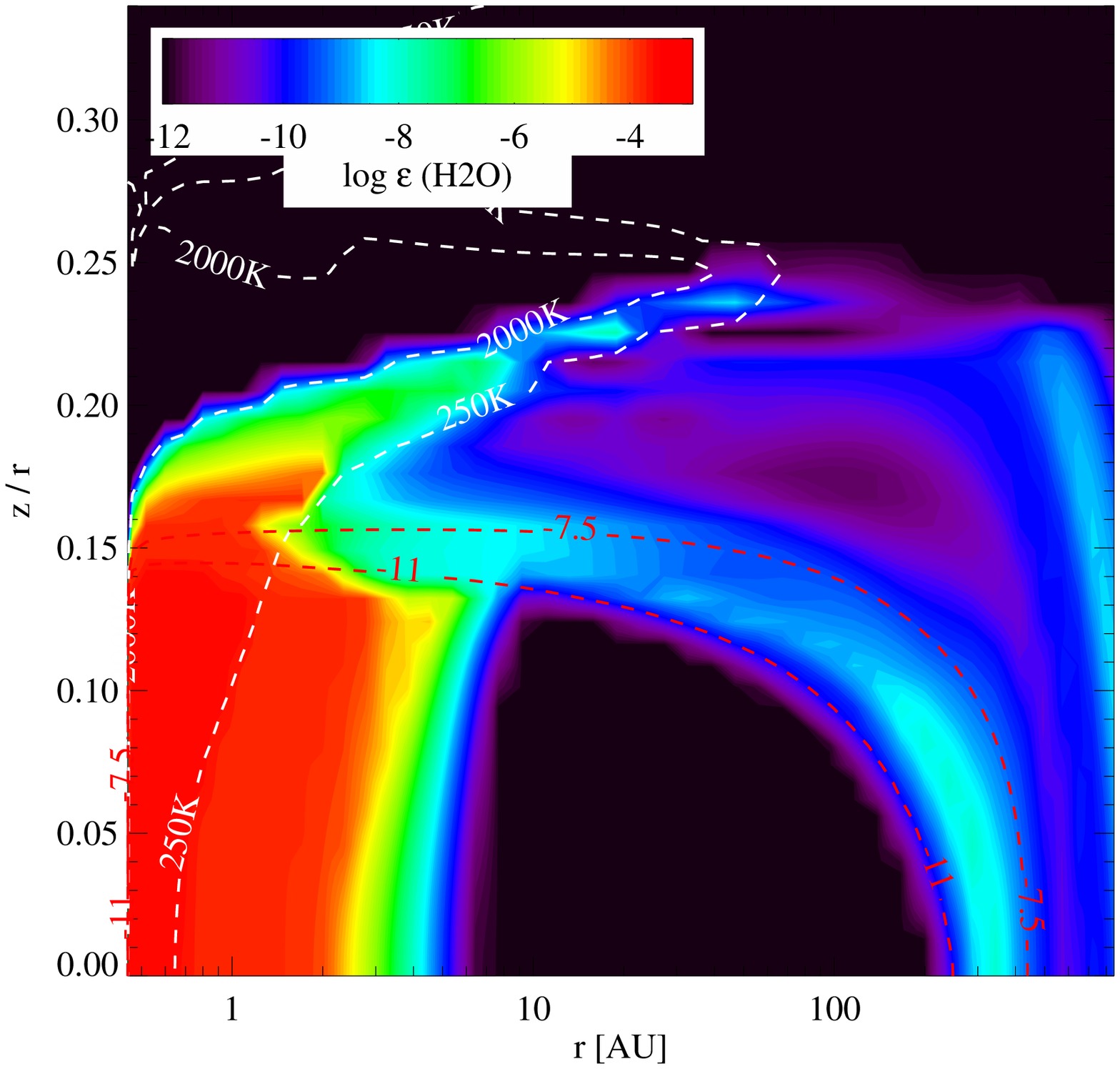} &
  \includegraphics[width=7.5cm]{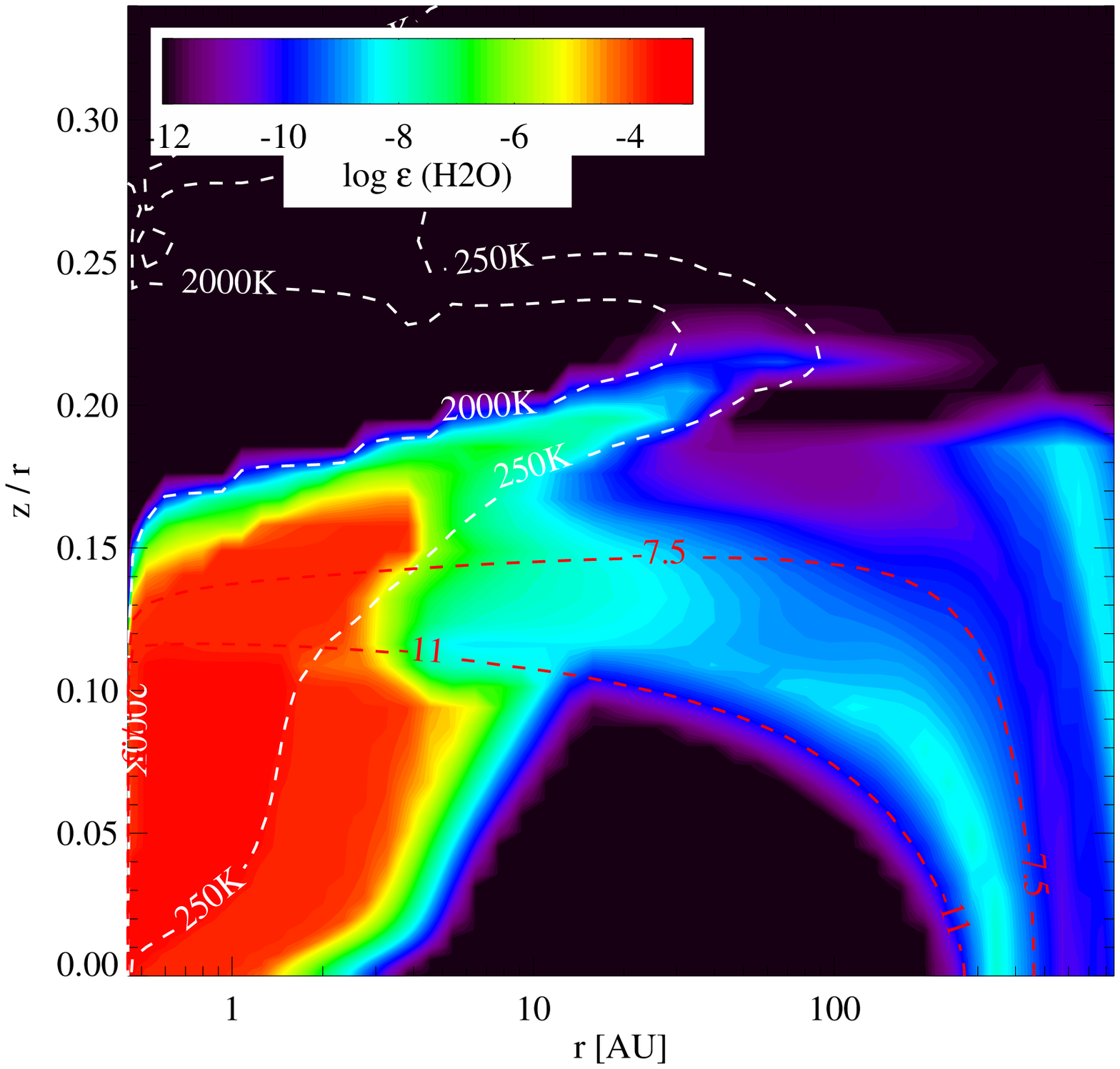}
\end{tabular}
\caption{Effect of dust settling on the disc chemical abundances. Left column indicates abundances in a fully mixed model, and right column represents a strongly-settled model with otherwise identical parameters. Top row: CO abundance with dust (blue dashed lines) and gas (white dashed line) temperature contours. Middle row: H$_2$ abundance with 300K gas temperature contour (black dashed line) and visual extinction contour (white dashed line). Bottom row: H$_2$O abundance. White dashed lines indicate $T_{\rm gas}$ contours enclosing ``hot water'' region, red dashed lines indicate contours of UV field strength per hydrogen nucleus, $\log(\chi/n_{\rm H})$ as defined in \citet{Draine1996}, enclosing the cool water belt.}
\label{fig:gassettle}
\end{centering}
\end{figure*}

\citet{Kamp2011} computed a large grid of disc models using \PRODIMO, in tandem with {\sc Mcfost} \citep{Pinte2006}. This study noted a trend of increasing disc dust temperature with settling, due to a combination of reduced emissivity at long wavelengths in the disc surface, and increased illumination of the dust as the stellar radiation is able to penetrate further into the disc. Our models of HD 163296 follow this same behaviour, albeit with an accompanying increase in gas temperature. The effect of dust settling on the gas and dust temperature contours in our models is demonstrated in Fig.~\ref{fig:gassettle}, where the chemical abundances of various chemical species are plotted. The chemical structure is seen to ``follow'' the settled dust grains towards the midplane, since the UV penetrates further into the disc, causing the characteristic layering of the various gas species to move closer to the midplane. This is despite the gas scale heights being identical in both models. In the settled models the warm gas in the disc resides at lower heights where the gas density is higher, leading to a general brightening of the emission lines. This seems to be a firm result for this object, with serious implications for the derivation of a precise disc gas mass from the emission lines.

The brightening of the line fluxes seems to be a general result for all the transitions considered. The exact behaviour of the various line fluxes with variable dust settling, and the extent to which they are affected, depends on the height and radial position in the disc from which they originate. This is well-illustrated by the CO lines in Fig.~\ref{fig:settle}. As the largest grains begin to settle, the $^{13}$CO 1-0 line is the first to show an increase in flux. This is due to an increase in thermal desorption of CO ice  from grain surfaces in the outer disc midplane, which this line traces (see Fig.~\ref{fig:analysis}). This behaviour is echoed in the other $^{13}$CO line, the J=3-2 transition. In contrast, the $^{12}$CO 3-2 line, which is also formed in the outer disc, shows a smaller increase in flux. This is because this line is optically thick, and formed higher in the disc where it remains unaffected by the thermal desorption of ice in the midplane. The $^{12}$CO 2-1 and 1-0 lines follow this same behaviour (not plotted). As smaller grains begin to settle, the $^{13}$CO lines increase less rapidly than for the lines formed in the warm gas higher in the disc. This is most pronounced in the high J CO lines, which form in the inner disc where the temperature and density gradients are more extreme, and the lines more sensitive to the downwards shift in chemical structure. The $^{12}$CO 6-5 line forms at intermediate distance in the disc, and this is reflected in the level of flux enhancement (Fig.~\ref{fig:settle}; top panel).

The behaviour of the water lines (middle panel in Fig.~\ref{fig:settle}) is less straightforward. The o-H$_2$O 179.5$\mu$m line shows a stronger increase in flux than both the 180.5$\mu$m line, which is at higher excitation, and the 538.3$\mu$m line, which is at lower excitation. The p-H$_2$O 89.99$\mu$m and o-H$_2$O 78.7$\mu$m lines are at higher excitation, forming in the inner disc, and these also show a large increase in flux as the dust grains settle. It is likely that the relative behaviour of the various water lines is a complex function of the changing temperature structure and H$_2$O chemical abundance structure in the disc, affecting the excitation of the various levels.

The largest increase in line flux occurs in the H$_2$ S(1) transition (Fig.~\ref{fig:settle}; lower panel). The emission region for this line is seen to follow the 300K gas temperature contour (see Fig.~\ref{fig:analysis}), which intersects a greater molecular hydrogen mass in the settled models. This leads to a dramatic enhancement in flux in the strongly settled models, a factor of $\sim\,30$ increase on the fully mixed model. The [{\sc Oi}]\,63\,$\mu$m line also starts to increase as the smallest grains are allowed to settle, coinciding with an increase in atomic gas, but the enhancement is less extreme than since this line is formed further out in the disc (see Fig.~\ref{fig:analysis}). The OH 79.11\,$\mu$m line is formed at intermediate distance between H$_2$ S(1) and [{\sc Oi}]\,63\,$\mu$m, and its behaviour with increasing dust settling reflects this. The [{\sc Cii}]\,158\,$\mu$m line is less strongly affected by settling. This line is formed in a thin layer at the disc surface, with emission dominated by the outer disc, and so it is less sensitive to changes in the internal disc temperature structure.

The settling of dust grains might be expected to drive line formation conditions closer to LTE, with the various species residing lower in the disc, in regions of higher density, with more frequent collisions between particles. However, there is no evidence for this in our models. The majority of lines show only minor departures from LTE, in both settled and well-mixed models. The largest departure from the line fluxes assuming LTE level populations occurs in the water lines, but the LTE line fluxes increase at roughly the same rate as those computed using escape probability, so the ratio $F_{\rm NLTE}/F_{\rm LTE}$ remains roughly constant. The critical density for the water lines, $n_{\rm crit}\!\sim10^8 - 10^{10}$cm$^{-3}$. These densities are reached only in the inner disc in our models, with each of the lines arising in regions with $n\!<\!n_{\rm crit}$, even in the settled models. The water line which is closest to LTE in our models is the 538.3$\mu$m line, which has the lowest level of excitation and the smallest critical density. In general, the critical densities decrease rapidly at large optical depths, driving most of the transitions towards LTE.

%The local gas/dust ratio in the best-fit model is plotted in Fig.~\ref{fig:dustgas}, where it is seen to vary by a factor $\sim$\,30 at various heights in the disc. Indeed, the local gas/dust ratio has a value $\sim$\,100 throughout much of the disc, dropping to $\sim$\,3 in the midplane due to the build-up of dust grains here. The best-fit model has an overall gas/dust ratio of 9.

These results are in contrast to the findings of \citet{Jonkheid2007}, who find a trend of decreasing gas temperature and line emission as the dust grains are allowed to settle towards the midplane. However, a direct comparison is not appropriate, since their models assumed a simultaneous reduction in the dust/gas ratio and PAH abundances with dust settling. The reduction in gas temperature can largely be attributed then to the drop in photoelectric heating from dust grains and PAH molecules. \citet{Jonkheid2007} also note that the dust temperatures in their models are high enough to prevent CO freeze-out, whereas considerable freeze-out is present here. \citet{Meijerink2009} also predict an increase in line emission with dust settling, from models in which the effect of settling is simulated by simply increasing the global gas/dust ratio, as opposed to considering the vertical distribution of grains in the disc.

Our models indicate that the $^{13}$CO 1-0 line flux can change by a factor $\sim$\,2 in strongly settled models. This is in a high mass disc for which the line is optically thick throughout most of the disc, and so this settling flux enhancement might be expected to be even stronger in a lower mass disc. In any case, variable grain freeze-out would seem to limit the ability of this line to trace the total disc gas mass. In general, the effect of dust settling on the vertical thermal structure of the gas in discs is seen to introduce new degeneracies when attempting to fit the disc parameters to the observed line emission. While our preferred model fits the observed data with a well-mixed disc with the canonical gas/dust ratio, we cannot rule out the possibility of a gas-depleted disc in which dust settling gives an enhancement in the various line fluxes. Indeed, such a disc with gas/dust $\sim\,20$ allows a better fit to the line data and the observed 10 micron silicate emission (column 1 in Table~\ref{tab:Para_fit}).

\subsection{Effect of X-rays}

It is currently unclear how important a role X-rays play in determining the gas chemistry and temperature structure in discs. \citet{Aresu2011} studied the effects of stellar X-ray emission on models of T Tauri discs computed with \PRODIMO. This study found that while Coulomb heating by X-rays introduced an extended hot gas surface layer to the discs, the [{\sc Oi}] and [{\sc Cii}] fine structure lines were only affected for X-ray luminosities $L_{\rm X}\!>\!10^{30}$\,erg\,s$^{-1}$. One would expect this to hold for the warmer gas in Herbig discs, and \citet{Gunther2009} derive an X-ray luminosity of $L_{\rm X}\!=\!10^{29.6}$\,erg\,s$^{-1}$ for HD 163296, lower than but close to the threshold value noted by \citet{Aresu2011} for T Tauri discs. HD 163296 represents an interesting test case for the influence of X-rays in Herbig discs.

\begin{figure}
\includegraphics[width=8.5cm]{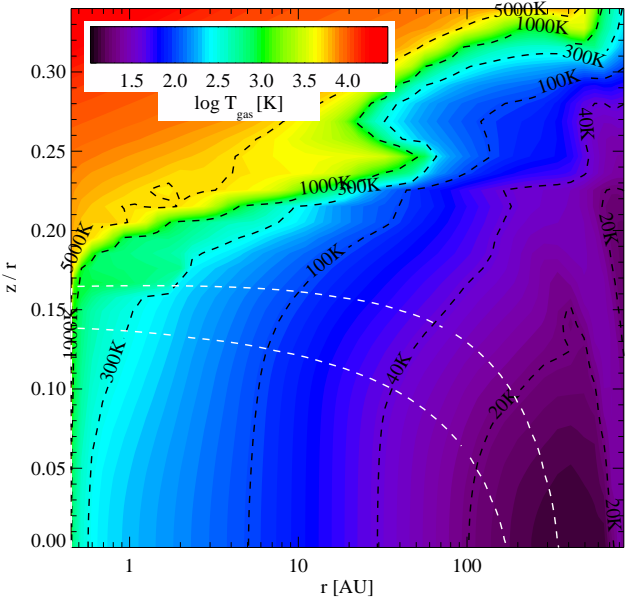}
\caption{Gas temperature structure for the preferred model (column 3 in Table~\ref{tab:Para_fit}), irradiated by an additional X-ray spectral component with $L_{\rm X}\!=\!10^{29.6}$\,erg\,s$^{-1}$, as observed by \citet{Gunther2009} in HD 163296. This is in comparison to the right hand panel of Fig.~\ref{fig:disc}, in which no X-ray spectral component is present.}
\label{fig:xrays}
\end{figure}

We have computed a disc model with parameters identical to our preferred model, but with an additional X-ray component in the input spectrum. The X-ray luminosity was set equal to the value of $L_{\rm X}\!=\!10^{29.6}$\,erg\,s$^{-1}$ observed by \citet{Gunther2009} for HD 163296. The effect of this on the gas temperature is illustrated by Fig.~\ref{fig:xrays} (c.f. Fig.~\ref{fig:disc}). The X-ray heating processes produce an extended hot surface gas layer, as noted by \citet{Aresu2011}, with gas temperatures $>$5000K extending out to the outer disc.

There is little discernible effect on the disc chemistry, with none of the chemical species masses or predicted line fluxes changing by more than a few percent.In all cases the effect of the additional X-rays is less than produced by switching to the ``high UV'' input spectrum . We would therefore expect the UV to dominate the gas chemistry in a Herbig disc of the sort considered here.

%\begin{figure}
%\includegraphics[width=8.5cm]{gasdust.eps}
%\caption{The local dust/gas ratio in the best-fit model. This varies locally due to settling of dust grains towards the midplane.}
%\label{fig:dustgas}
%\end{figure}

%The only emission line to show a large deviation from the non-X-ray-irradiated model is the o-H$_2$O 180.49\,$\mu$m line, which increases by a factor of 2. This appears to be caused by a slight extension outwards of the ``hot water band'' noted by \citet{Woitke2009b}, where this line originates, and which closely follows the gas temperature contours in the disc (see Fig.~\ref{fig:gassettle} c.f. the gas contours in Fig.~\ref{fig:xrays} produced by X-ray Coulomb heating). There is no corresponding jump in the other water lines, with an increase of $\sim$10\% in the p-H$_2$O 89.99\,$\mu$m and o-H$_2$O 179.53\,$\mu$m lines (with higher and lower excitation energies respectively). There are no discernible changes to any of the other emission lines.

\section{Summary and Conclusions}

This paper presents new observations of the far-IR lines of the Herbig Ae star HD 163296, obtained using Herschel/PACS as part of the GASPS open time key program. These consist of a detection of the [{\sc Oi}]\,63.18\,$\mu$m line, as well as upper limits for eight additional lines. We have supplemented these observations with additional line and continuum data, as well as high resolution observations of the variable optical spectra obtained as part of the EXPORT collaboration.

We have computed radiation thermo-chemical disc models using the disc code \PRODIMO, and employed an evolutionary $\chi^2$-minimisation strategy to find the best simultaneous model fit to the continuum and line data. The stellar parameters and UV input spectrum for the modelling were determined through detailed analysis in the UV, optical and near-infrared.

We obtain reasonable fits to the observed photometry, ISO-SWS spectrum, far-IR line fluxes and millimetre CO profiles for a variety of discs, and note that parameter degeneracies preclude the precise derivation of the disc properties. In particular, the effects of dust settling on the vertical thermal structure of the gas in discs strongly influence the line emission, placing further limits on the derivation of a disc gas mass from the line data. We are able to fit the photometry and line data with a well-mixed disc with gas/dust $\sim$\,100, but an equally good fit is possible with a disc with gas/dust $\sim$\,22, in which dust settling is present. The main advantage of the settled model is that the fit to the observed ISO-SWS spectrum is greatly improved, and it seems probable that some degree of dust settling and gas-depletion is present in the disc. These discs have a radial surface density profile as derived by \citet{Hughes2008}, with an exponentially-tapered outer edge. 

A model fit to the same data with power-law radial density profile gives a gas/dust ratio of $\sim$\,9. This power-law model gives the best fit to the observed SED data, although it does very badly when it comes to fitting the radial intensity distribution of the resolved millimetre continuum. This is due to the flat density power index required to fit the SED and the observed CO 3-2 emission, and presents a challenge for future modelling efforts.

The emitted line fluxes are in general sensitive to the degree of dust settling in the disc. This is a firm result in our models, and has serious implications for attempts to derive the disc gas mass and other properties from line observations. This settling flux enhancement arises from changes to the vertical temperature and chemical structure, as settled dust grains allow stellar UV to penetrate deeper into the disc. The effect is strongest in lines which are formed in the warm gas in the inner disc (e.g. a factor $\sim$\,30 increase in the H$_2$ S(1) line), but the low excitation molecular lines are also affected, e.g. a factor $\sim$\,2 increase in the $^{13}$CO 1-0 line.

We explore the effects of the observed UV variability in this object on the gas chemistry in our models, and conclude that this effect is not large enough to affect the observable line fluxes beyond the current range of instrumental uncertainty.

We also examine the effect of X-rays on the gas chemistry of our models, and find that while X-rays present a significant source of gas-heating in the disc surface layers, the observed X-ray luminosity of $L_{\rm X}\!=\!10^{29.6}$\,erg\,s$^{-1}$ does not significantly alter the gas chemistry or line emission. Any effects are smaller than those expected as a result of the observed UV variability in this object.

It is difficult to reach any firm conclusions regarding the evolutionary state of the disc of HD 163296. There is some evidence to suggest that the disc is gas-depleted. This is in contrast to the result found for the Herbig Ae star HD 169142 by \citet{Meeus2010}, where despite indications that the disc is transitional, it was found to be gas-rich. We note that there are uncertainties associated with our gas/dust values arising from uncertainties in the disc dust composition, and the associated opacity law. This is reflected by the factor of $\sim$\,3 spread in the range of derived dust masses found in the literature for this object \citep{Natta2004,Tannirkulam2008a,Mannings1997,Isella2007}. All of our derived dust masses are within the range $(5-17) \times 10^{-4}$M$_{\odot}$ from the literature. We also note the restrictions placed on our conclusions by the assumption of constant dust grain properties throughout the disc, and it has been suggested that the dust properties in discs should in general be variable with radius \citep{Birnstiel2010,Guilloteau2011}.

It would appear that the modelling of HD 163296 is a far from straightforward task, and it is difficult to fit it into the standard evolutionary picture developed over the past decade. We are unable to find a single disc model which fits perfectly the entire wealth of observational data for this object. As well as the gas/dust ratio, there is uncertainty regarding the disc flaring, which itself is strongly tied in to the disc gas heating and line emission, as well as being indicative of the disc's evolutionary state. Evidence of in-falling material and the presence of a bipolar outflow in HD 163296 seem to indicate that the star is actively accreting material, typical of a young object with a gas-rich disc. However, it has been suggested that the observed disc geometry and flaring could indicate a star at a later stage of its evolution. This would be consistent with the evidence for grain growth from our modelling, with all the models requiring large grains to fit the observations. There is also possible further evidence for particle growth from high resolution optical spectra of this object (see Appendix). This is also consistent with possible evidence for dust settling and gas-depletion from this study, but is hard to reconcile with evidence for substantial ongoing activity in the inner disc. It is clear that this object is at a fascinating stage in its evolution.

\section*{Appendix: \'echelle spectra analysis}

Sect.~\ref{sec:results} provides possible evidence of a low gas-to-dust ratio in the HD 163296
circumstellar disc as compared to interstellar abundances. This result
is also  supported in this Appendix from the analysis  of visible high
resolution \'echelle spectra.

The EXPORT collaboration \citep{Eiroa2001} obtained high resolution
\'echelle  spectra ($R =  \lambda /  \Delta \lambda  = 48000$)  in the
range 3800-5900 \AA\ during five nights in May and July 1998 using the
Utrecht Echelle Spectrograph (UES)  on the William Herschel Telescope,
La Palma.

Significant  spectral  variability  was  found  in  the  optical
spectrum.  The EXPORT  data monitored those changes on  time scales of
hours,  days  and months \citep{Mend2011}.   Almost  any  strong  photospheric line  is
variable  to   some  extent.   In   particular,  Transient  Absorption
Components (TACs)  are observed in  many hydrogen (Balmer  series) and
metallic lines, such as \ion{Ca}{ii} K  \& H, \ion{Na}{i} D2 \& D1 and
the  \ion{Fe}{ii}  4924,  5018  and  5169 \AA\  triplet.  The Appendix summarizes the main  results on the dynamics of  the circumstellar gas
obtained from the TACS.

Many  TACs observed  are  similar to  those  studied by \citet{Natta2000} and \citet{Mora2002,Mora2004} in similar  age Herbig Ae stars.
These authors  propose magnetospheric  accretion of primordial  gas as
the most likely origin. That is, the TACs in Herbig Ae stars typically
trace the  kinematics of material with  gas to dust  ratios similar to
that of the interstellar medium.

\begin{figure}
\centering
\includegraphics[width=85mm]{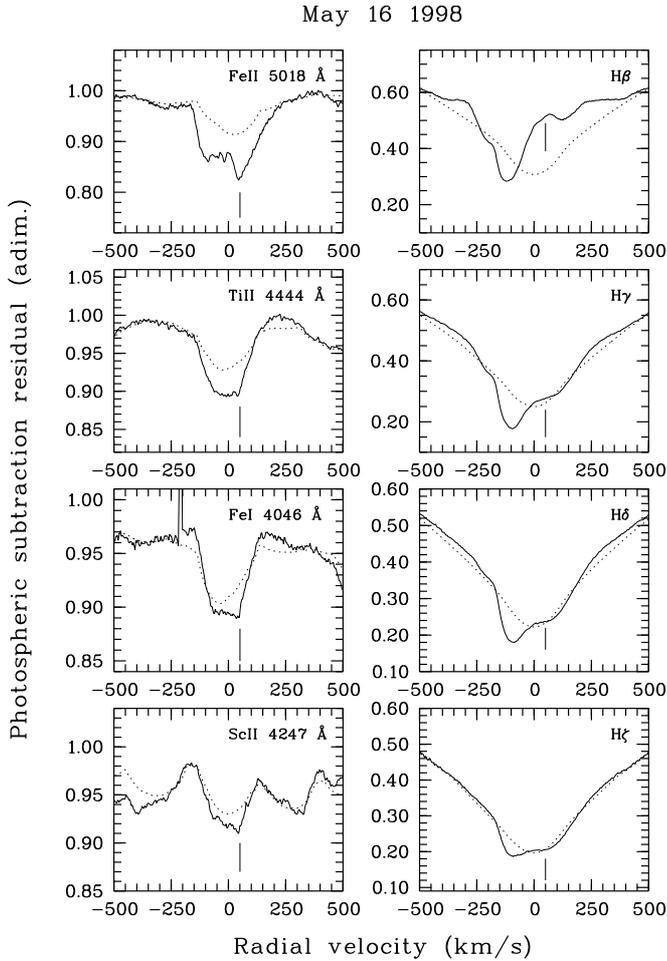}
\caption{Transient  absorption components  in  16 May  1998. The  high
  resolution  observed  spectrum  of  several lines  (solid  lines)  is
  plotted  along with the synthetic photospheric  spectrum (dashed  lines; \citet{Kurucz1993}). The short vertical dashes mark the approximate location
of the  metallic-only transient absorptions  at +50 km/s. A
  transient  red-shifted  component of  velocity  $v\,\sim\,50$\,km/s  is
  clearly  seen  in  the  metallic  but not  the  hydrogen  lines.  In
  addition a blue-shifted  absorption is identified in all  lines for $v\,\sim\,-100$\,km/s.}
\label{Figure:metallicEventsMa16}
\end{figure}

\begin{figure}
\centering
\includegraphics[width=85mm]{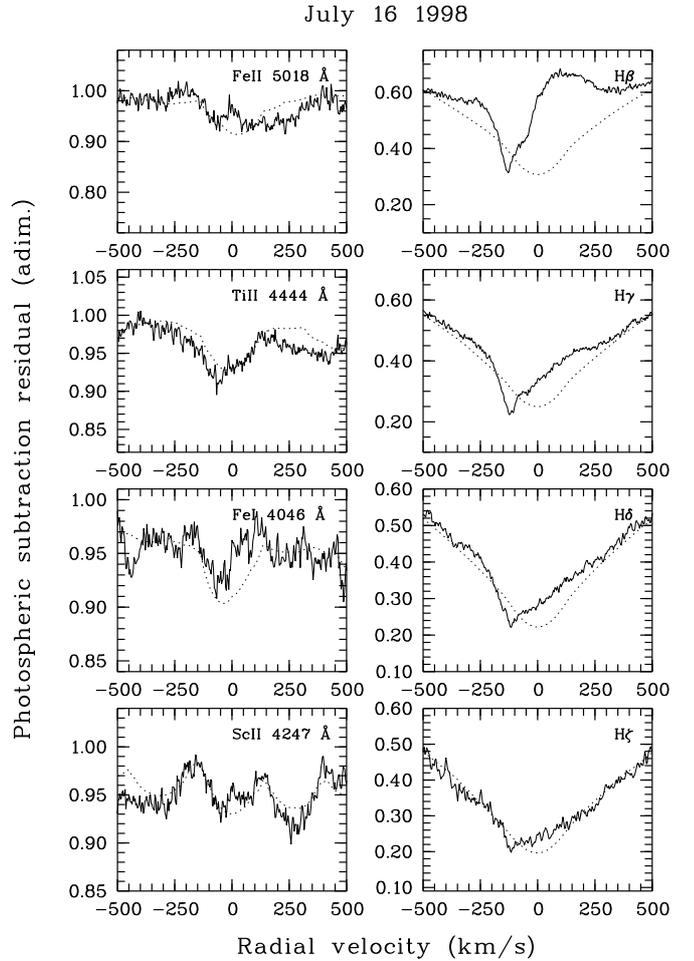}
\caption{Transient absorption components in
  16 July 1998. The high resolution observed spectrum of several lines
  (solid  lines)  is plotted along with the photospheric  spectrum
  (dashed  lines). No  clear transient  absorption is  detected  in the
  metallic  lines.   On   the  other  hand a significant  blue-shifted
  absorption is detected in all Balmer lines at $v\,\sim\,-100$\,km/s.}
\label{Figure:metallicEventsJu28}
\end{figure}

However, a  distinctive feature  of HD 163296  is the presence  of low
velocity  ($v\,\sim\,50$\,~km/s)  red-shifted  TACs   only  observed  in
refractory  elements:   \ion{Fe}{i},  \ion{Fe}{ii},  \ion{Ti}{ii}  and
\ion{Sc}{ii}.   These  ``metallic''  TACs  have no  clear  counterpart
neither  in  H nor  \ion{Ca}{ii}  and  \ion{Na}{i}.  They  were
observed  during  all nights  except  28  July  1998. These  transient
absorptions could be better explained  in terms of the model developed
by \citet{LH1988} for the evaporation of solid bodies in
the $\beta$~Pic  debris disc and subsequently postulated  by \citet{Grinin1991,Grinin1994} for accretion discs around Herbig Ae/Be stars.

Fig.~\ref{Figure:metallicEventsMa16}  shows   selected  line  profiles
observed in  16 May  1998. Wavelengths have  been converted  to radial
velocities in the stellar  reference system.  The approximate location
of the  metallic-only transient absorptions  at +50 km/s  is displayed
with a short vertical dash. The underlying photosphere is represented by a synthetic Kurucz model computed with $T_{\rm eff}\!=\!9250$ K, $\log g_*\!=\!4.07$ and [Fe/H]=+0.20, using the ATLAS9 and SYNTHE codes by \citet{Kurucz1993}.  In this work we have used the GNU Linux version of the codes available online\footnote{http://wwwuser.oat.ts.astro.it/atmos/} \citep{Sbordone2004}.
It  is  clear  that  no  Balmer line  absorption  counterpart  can  be
identified, even  for the higher  H$\delta$ and H$\zeta$ lines  in the
series for  which no significant  core line emission is  apparent.  On
the  other hand, blue-shifted  absorption with  $v \sim  -100$~km/s is
clearly seen  both in metallic  and hydrogen lines.  The  metallic only
transient  events are not  an artifact  generated by  poor photosphere
characterisation, as  revealed by Fig.~\ref{Figure:metallicEventsJu28}
(electronic edition only), which shows the spectra obtained in 28 July
1998, the only night where no metallic only events where identified.

Similar results were  found for WW Vul by \citet{Mora2004}, but not
for the other four stars  studied. In addition, both hydrogen-rich and
hydrogen-less  TACs are simultaneously  detected in  HD 163296  and WW
Vul. If particle growth indeed explains the metallic-only events, that
would   suggest  the   circumstellar  disc   around  HD 163296  (and
incidentally  WW  Vul)  has  an  evolutionary status  between  a  pure
primordial  gas   rich  accretion  disc  and   a  gas-depleted  second
generation  debris disc.  That  finding is  fully consistent  with the
results obtained  in Sec.~\ref{sec:results} from  the analysis of  Herschel/PACS far-IR spectroscopy and ground-based mm CO observations.

\subsection*{Kinematics of the circumstellar gas}
\label{sec:kin}

A detailed  study of the dynamics  of the circumstellar  gas around HD
163296 using  the method presented  in \citet{Mora2004} is  out the
scope of this paper. The main results are, however, summarized here.

The  following  lines  were  studied:  the  Balmer  series  (H$\beta$,
H$\gamma$, H$\delta$, H$\epsilon$ and H$\zeta$), The \ion{Na}{i} D and
\ion{Ca}{ii} HK  doublets, the \ion{Fe}{ii}  4924, 5018 and  5169 \AA\
triplet,  \ion{Fe}{i} 4046 \AA,  \ion{Ti}{ii} 4444  and 4572  \AA\ and
\ion{Sc}{ii}  4247  \AA.   Three  types  of  variability  were  always
simultaneously  observed: core  line emission  (most prominent  in the
Balmer,  \ion{Na}{i}  and  \ion{Ca}{ii} lines),  transient  red-shifted
absorptions (associated to in-falling  material, observed in all lines)
and  transient   blue-shifted  absorptions  (associated   to  outflows,
observed  in all lines).   HD 163296  was more  active than  the stars
studied by \citet{Mora2002,Mora2004}.  In particular, the larger line
core emission made the analysis more difficult.

The in-falling material  always span a large range  of velocities, from
near  zero  to +300  km/s  for  the red  end  of  the higher  velocity
components.  The stellar winds maximum  velocities can also be high. A
maximum of -400 km/s was observed in 17 May 1998.

\begin{figure*}
\centering
\includegraphics[angle=270,width=0.75\hsize]{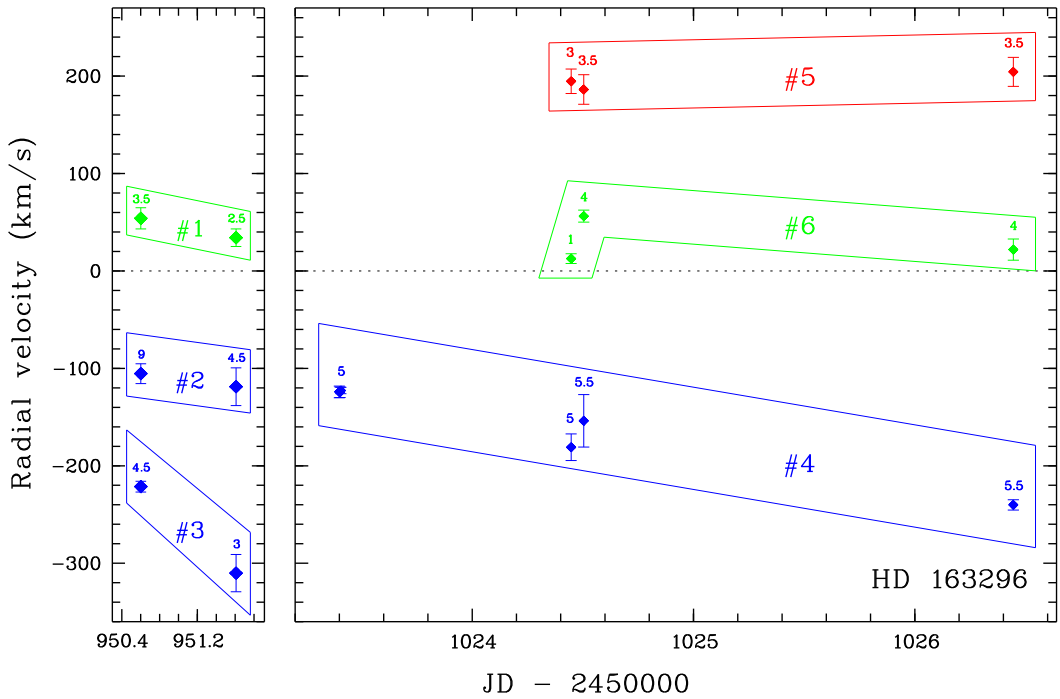}
\vspace*{4mm}
\caption{Kinematics  of the circumstellar  gas. Features  with similar
  radial velocity observed in  different lines and different dates are
  assumed to share a common origin and grouped into events. Each event
  is represented  by a box enclosing  a number of  points. Each point represents the weighted average  of the radial velocity simultaneously measured for several lines. The  error bar is the standard  deviation.  The weighted
  number of  lines is given next  to each point.   Red boxes represent
  in-falling material,  blue boxes out-flowing material  and green boxes
  in-falling material only detected in metallic lines.}
\label{Figure:eventsKinematics}
\end{figure*}

Fig.~\ref{Figure:eventsKinematics}   shows  a   summary  of   all  the
transient  absorptions  identified.    Features  with  similar  radial
velocity observed  in different  lines are assumed  to share  a common
origin.  The average radial  velocity and  the standard  deviation are
displayed as points and error bars in the figure.  The number of lines
used to compute  the average is given next to  each point. The fainter
lines are given half weight in  the average (see \citet{Mora2004} for
additional details).   Points with  similar velocity on  nearby nights
are also  assumed to  be produced by  the same clump  of circumstellar
material  and are  grouped in  "events". Events  are displayed  in the
figure with boxes  enclosing a certain number of  average points.  Red
boxes  represent in-falling  material,  blue boxes  outflows and  green
boxes   in-falling   material   only   detected  in   metallic   lines.
Metallic-only events have smaller radial velocities than hydrogen-rich
in-falling material.

One  of the most  interesting events  observed is  \#4, a  strong wind
observed in plenty of Balmer  and metallic lines during July 1998. The
most  relevant feature  is  an almost  constant  acceleration rate  of
$\sim$\,-0.4 m/s$^2$.  The average  radial velocity starts at -120 km/s on
28 July 1998 and  grows up to -235 km/s in 31  July 1998. The width of
the  absorption features  has an  almost constant  value of  120 km/s,
supporting  the hypothesis  of a  common origin  of all  the transient
absorptions in the event.

%\begin{acknowledgements}
%  We thank Catherine Dougados for fruitful discussions about the
%  properties of disc outflows and the interpretation of optical
%  emission lines.  We highly appreciated comments by Greg Herczeg
%  about the HST/COS UV-observations of RECX/,15.  W.-F.~Thi
%  acknowledges a SUPA astrobiology fellowship.  G.~Meeus, C.~Eiroa,
%  I.~Mendigut\'ia and B.~Montesinos are partly supported by Spanish
%  grant AYA 2008-01727. C.~Pinte acknowledges the funding from the EC
%  7$^{th}$ Framework Program as a Marie Curie Intra-European Fellow
%  (PIEF-GA-2008-220891). D.R.~Ardila, S.D.~Brittain, C.A.~Grady,
%  I.~Pascucci, B.~Riaz, G.~Sandell and C.D.~Howards, J.-P.~Williams,
%  G.~Matthews, A.~Roberge, W.~Danchi acknowledge NASA/JPL for funding
%  support. E.~Solano and J.M.~Alacid acknowledge the funding from the
%  Spanish MICINN through grant AYA2008-02156. The LAOG group
%  acknowledges PNPS, CNES and ANR (contract ANR-07-BLAN-0221) for
%  financial support.  I.~de~Gregorio-Monsalvo is partially supported
%  by Ministerio de Ciencia e Innovaci{\'o}n (Spain), grant AYA
%  2008-06189-C03 (including FEDER funds), and by Consejer{\'i}a de
%  Innovaci{\'o}n y Ciencia y Empresa of Junta de Andaluc{\'i}a,
%  (Spain)
%\end{acknowledgements}

\bibliography{reference}

\begin{thebibliography}{98}
\expandafter\ifx\csname natexlab\endcsname\relax\def\natexlab#1{#1}\fi

\bibitem[{{Abrahamsson} {et~al.}(2007){Abrahamsson}, {Krems}, \&
  {Dalgarno}}]{Abrahamsson2007}
{Abrahamsson}, E., {Krems}, R.~V., \& {Dalgarno}, A. 2007, \apj, 654, 1171

\bibitem[{{Acke} {et~al.}(2010){Acke}, {Bouwman}, {Juh{\'a}sz}, {Henning}, {van
  den Ancker}, {Meeus}, {Tielens}, \& {Waters}}]{Acke2010}
{Acke}, B., {Bouwman}, J., {Juh{\'a}sz}, A., {et~al.} 2010, \apj, 718, 558

\bibitem[{{Aresu} {et~al.}(2011){Aresu}, {Kamp}, {Meijerink}, {Woitke}, {Thi},
  \& {Spaans}}]{Aresu2011}
{Aresu}, G., {Kamp}, I., {Meijerink}, R., {et~al.} 2011, \aap, 526, A163+

\bibitem[{{Barber} {et~al.}(2006){Barber}, {Tennyson}, {Harris}, \&
  {Tolchenov}}]{Barber2006}
{Barber}, R.~J., {Tennyson}, J., {Harris}, G.~J., \& {Tolchenov}, R.~N. 2006,
  \mnras, 368, 1087

\bibitem[{{Bell} {et~al.}(1998){Bell}, {Berrington}, \& {Thomas}}]{Bell1998}
{Bell}, K.~L., {Berrington}, K.~A., \& {Thomas}, M.~R.~J. 1998, \mnras, 293,
  L83

\bibitem[{{Benisty} {et~al.}(2010){Benisty}, {Natta}, {Isella}, {Berger},
  {Massi}, {Le Bouquin}, {M{\'e}rand}, {Duvert}, {Kraus}, {Malbet}, {Olofsson},
  {Robbe-Dubois}, {Testi}, {Vannier}, \& {Weigelt}}]{Benisty2010}
{Benisty}, M., {Natta}, A., {Isella}, A., {et~al.} 2010, \aap, 511, A74+

\bibitem[{{Bessell}(1979)}]{Bessell1979}
{Bessell}, M.~S. 1979, \pasp, 91, 589

\bibitem[{{Birnstiel} {et~al.}(2010){Birnstiel}, {Dullemond}, \&
  {Brauer}}]{Birnstiel2010}
{Birnstiel}, T., {Dullemond}, C.~P., \& {Brauer}, F. 2010, \aap, 513, A79+

\bibitem[{{Bouwman} {et~al.}(2000){Bouwman}, {de Koter}, {van den Ancker}, \&
  {Waters}}]{Bouwman2000}
{Bouwman}, J., {de Koter}, A., {van den Ancker}, M.~E., \& {Waters},
  L.~B.~F.~M. 2000, \aap, 360, 213

\bibitem[{{Brott} \& {Hauschildt}(2005)}]{Brott2005}
{Brott}, I. \& {Hauschildt}, P.~H. 2005, in ESA Special Publication, Vol. 576,
  The Three-Dimensional Universe with Gaia, ed. {C.~Turon, K.~S.~O'Flaherty, \&
  M.~A.~C.~Perryman}, 565--+

\bibitem[{{Chambaud} {et~al.}(1980){Chambaud}, {Levy}, {Millie}, {Roueff}, \&
  {Tran Minh}}]{Chambaud1980}
{Chambaud}, G., {Levy}, B., {Millie}, P., {Roueff}, E., \& {Tran Minh}, F.
  1980, Journal of Physics B Atomic Molecular Physics, 13, 4205

\bibitem[{{Cox}(1999)}]{Allen1999}
{Cox}, A.~N., ed. 1999, {Allen's Astrophysical Quantities}, 4th edn.
  (Springer-Verlag New York)

\bibitem[{{Cushing} {et~al.}(2004){Cushing}, {Vacca}, \&
  {Rayner}}]{Cushing2004}
{Cushing}, M.~C., {Vacca}, W.~D., \& {Rayner}, J.~T. 2004, \pasp, 116, 362

\bibitem[{{D'Alessio} {et~al.}(1998){D'Alessio}, {Canto}, {Calvet}, \&
  {Lizano}}]{DAlessio1998}
{D'Alessio}, P., {Canto}, J., {Calvet}, N., \& {Lizano}, S. 1998, \apj, 500,
  411

\bibitem[{{de Winter} {et~al.}(2001){de Winter}, {van den Ancker}, {Maira},
  {Th{\'e}}, {Djie}, {Redondo}, {Eiroa}, \& {Molster}}]{deWinter2001}
{de Winter}, D., {van den Ancker}, M.~E., {Maira}, A., {et~al.} 2001, \aap,
  380, 609

\bibitem[{{Devine} {et~al.}(2000){Devine}, {Grady}, {Kimble}, {Woodgate},
  {Bruhweiler}, {Boggess}, {Linsky}, \& {Clampin}}]{Devine2000}
{Devine}, D., {Grady}, C.~A., {Kimble}, R.~A., {et~al.} 2000, \apjl, 542, L115

\bibitem[{{Dorschner} {et~al.}(1995){Dorschner}, {Begemann}, {Henning},
  {Jaeger}, \& {Mutschke}}]{Dorschner1995}
{Dorschner}, J., {Begemann}, B., {Henning}, T., {Jaeger}, C., \& {Mutschke}, H.
  1995, \aap, 300, 503

\bibitem[{{Doucet} {et~al.}(2006){Doucet}, {Pantin}, {Lagage}, \&
  {Dullemond}}]{Doucet2006}
{Doucet}, C., {Pantin}, E., {Lagage}, P.~O., \& {Dullemond}, C.~P. 2006, \aap,
  460, 117

\bibitem[{{Draine} \& {Bertoldi}(1996)}]{Draine1996}
{Draine}, B.~T. \& {Bertoldi}, F. 1996, \apj, 468, 269

\bibitem[{{Dubernet} \& {Grosjean}(2002)}]{Dubernet2002}
{Dubernet}, M.-L. \& {Grosjean}, A. 2002, \aap, 390, 793

\bibitem[{{Eiroa} {et~al.}(2001){Eiroa}, {Garz{\'o}n}, {Alberdi}, {de Winter},
  {Ferlet}, {Grady}, {Cameron}, {Davies}, {Deeg}, {Harris}, {Horne},
  {Mer{\'{\i}}n}, {Miranda}, {Montesinos}, {Mora}, {Oudmaijer}, {Palacios},
  {Penny}, {Quirrenbach}, {Rauer}, {Schneider}, {Solano}, {Tsapras}, \&
  {Wesselius}}]{Eiroa2001}
{Eiroa}, C., {Garz{\'o}n}, F., {Alberdi}, A., {et~al.} 2001, \aap, 365, 110

\bibitem[{{Eisner} {et~al.}(2009){Eisner}, {Graham}, {Akeson}, \&
  {Najita}}]{Eisner2009}
{Eisner}, J.~A., {Graham}, J.~R., {Akeson}, R.~L., \& {Najita}, J. 2009, \apj,
  692, 309

\bibitem[{{Faure} {et~al.}(2007){Faure}, {Crimier}, {Ceccarelli}, {Valiron},
  {Wiesenfeld}, \& {Dubernet}}]{Faure2007}
{Faure}, A., {Crimier}, N., {Ceccarelli}, C., {et~al.} 2007, \aap, 472, 1029

\bibitem[{{Fitzpatrick}(1999)}]{Fitzpatrick1999}
{Fitzpatrick}, E.~L. 1999, \pasp, 111, 63

\bibitem[{{Flower}(2001)}]{Flower2001}
{Flower}, D.~R. 2001, Journal of Physics B Atomic Molecular Physics, 34, 2731

\bibitem[{{Flower} \& {Launay}(1977)}]{Flower1977}
{Flower}, D.~R. \& {Launay}, J.~M. 1977, Journal of Physics B Atomic Molecular
  Physics, 10, 3673

\bibitem[{{Geers} {et~al.}(2006){Geers}, {Augereau}, {Pontoppidan},
  {Dullemond}, {Visser}, {Kessler-Silacci}, {Evans}, {van Dishoeck}, {Blake},
  {Boogert}, {Brown}, {Lahuis}, \& {Mer{\'{\i}}n}}]{Geers2006}
{Geers}, V.~C., {Augereau}, J., {Pontoppidan}, K.~M., {et~al.} 2006, \aap, 459,
  545

\bibitem[{{Grady} {et~al.}(2000){Grady}, {Devine}, {Woodgate}, {Kimble},
  {Bruhweiler}, {Boggess}, {Linsky}, {Plait}, {Clampin}, \&
  {Kalas}}]{Grady2000}
{Grady}, C.~A., {Devine}, D., {Woodgate}, B., {et~al.} 2000, \apj, 544, 895

\bibitem[{{Green} {et~al.}(1993){Green}, {Maluendes}, \& {McLean}}]{Green1993}
{Green}, S., {Maluendes}, S., \& {McLean}, A.~D. 1993, \apjs, 85, 181

\bibitem[{{Grinin} {et~al.}(1991){Grinin}, {Kiselev}, {Chernova}, {Minikulov},
  \& {Voshchinnikov}}]{Grinin1991}
{Grinin}, V.~P., {Kiselev}, N.~N., {Chernova}, G.~P., {Minikulov}, N.~K., \&
  {Voshchinnikov}, N.~V. 1991, \apss, 186, 283

\bibitem[{{Grinin} {et~al.}(1994){Grinin}, {The}, {de Winter}, {Giampapa},
  {Rostopchina}, {Tambovtseva}, \& {van den Ancker}}]{Grinin1994}
{Grinin}, V.~P., {The}, P.~S., {de Winter}, D., {et~al.} 1994, \aap, 292, 165

\bibitem[{{Guilloteau} {et~al.}(2011){Guilloteau}, {Dutrey}, {Pi{\'e}tu}, \&
  {Boehler}}]{Guilloteau2011}
{Guilloteau}, S., {Dutrey}, A., {Pi{\'e}tu}, V., \& {Boehler}, Y. 2011, \aap,
  529, A105+

\bibitem[{{G{\"u}nther} \& {Schmitt}(2009)}]{Gunther2009}
{G{\"u}nther}, H.~M. \& {Schmitt}, J.~H.~M.~M. 2009, \aap, 494, 1041

\bibitem[{{Henning} {et~al.}(1995){Henning}, {Begemann}, {Mutschke}, \&
  {Dorschner}}]{Henning1995}
{Henning}, T., {Begemann}, B., {Mutschke}, H., \& {Dorschner}, J. 1995, \aaps,
  112, 143

\bibitem[{{Hughes} {et~al.}(2011){Hughes}, {Wilner}, {Andrews}, {Qi}, \&
  {Hogerheijde}}]{Hughes2011}
{Hughes}, A.~M., {Wilner}, D.~J., {Andrews}, S.~M., {Qi}, C., \& {Hogerheijde},
  M.~R. 2011, \apj, 727, 85

\bibitem[{{Hughes} {et~al.}(2008){Hughes}, {Wilner}, {Qi}, \&
  {Hogerheijde}}]{Hughes2008}
{Hughes}, A.~M., {Wilner}, D.~J., {Qi}, C., \& {Hogerheijde}, M.~R. 2008, \apj,
  678, 1119

\bibitem[{{Isella} {et~al.}(2007){Isella}, {Testi}, {Natta}, {Neri}, {Wilner},
  \& {Qi}}]{Isella2007}
{Isella}, A., {Testi}, L., {Natta}, A., {et~al.} 2007, \aap, 469, 213

\bibitem[{{Jager} {et~al.}(1998){Jager}, {Mutschke}, \& {Henning}}]{Jager1998}
{Jager}, C., {Mutschke}, H., \& {Henning}, T. 1998, \aap, 332, 291

\bibitem[{{Jankowski} \& {Szalewicz}(2005)}]{Jankowski2005}
{Jankowski}, P. \& {Szalewicz}, K. 2005, J. Chem. Phys., 123, 104301

\bibitem[{{Jaquet} {et~al.}(1992){Jaquet}, {Staemmler}, {Smith}, \&
  {Flower}}]{Jaquet1992}
{Jaquet}, R., {Staemmler}, V., {Smith}, M.~D., \& {Flower}, D.~R. 1992, Journal
  of Physics B Atomic Molecular Physics, 25, 285

\bibitem[{{Jonkheid} {et~al.}(2007){Jonkheid}, {Dullemond}, {Hogerheijde}, \&
  {van Dishoeck}}]{Jonkheid2007}
{Jonkheid}, B., {Dullemond}, C.~P., {Hogerheijde}, M.~R., \& {van Dishoeck},
  E.~F. 2007, \aap, 463, 203

\bibitem[{{Kamp} \& {Bertoldi}(2000)}]{Kamp2000}
{Kamp}, I. \& {Bertoldi}, F. 2000, \aap, 353, 276

\bibitem[{{Kamp} {et~al.}(2010){Kamp}, {Tilling}, {Woitke}, {Thi}, \&
  {Hogerheijde}}]{Kamp2010}
{Kamp}, I., {Tilling}, I., {Woitke}, P., {Thi}, W., \& {Hogerheijde}, M. 2010,
  \aap, 510, A18+

\bibitem[{{Kamp} {et~al.}(2011){Kamp}, {Woitke}, {Pinte}, {Tilling}, {Thi},
  {Menard}, {Duchene}, \& {Augereau}}]{Kamp2011}
{Kamp}, I., {Woitke}, P., {Pinte}, C., {et~al.} 2011, \aap, 532, A85+

\bibitem[{{Kurucz}(1993)}]{Kurucz1993}
{Kurucz}, R.~L. 1993, VizieR Online Data Catalog, 6039, 0

\bibitem[{{Lagrange-Henri} {et~al.}(1988){Lagrange-Henri}, {Vidal-Madjar}, \&
  {Ferlet}}]{LH1988}
{Lagrange-Henri}, A.~M., {Vidal-Madjar}, A., \& {Ferlet}, R. 1988, \aap, 190,
  275

\bibitem[{{Launay} \& {Roueff}(1977)}]{Launay1977}
{Launay}, J.-M. \& {Roueff}, E. 1977, Journal of Physics B Atomic Molecular
  Physics, 10, 879

\bibitem[{{Mannings}(1994)}]{Mannings1994}
{Mannings}, V. 1994, \mnras, 271, 587

\bibitem[{{Mannings} \& {Sargent}(1997)}]{Mannings1997}
{Mannings}, V. \& {Sargent}, A.~I. 1997, \apj, 490, 792

\bibitem[{{Martin-Za{\"i}di} {et~al.}(2010){Martin-Za{\"i}di}, {Augereau},
  {M{\'e}nard}, {Olofsson}, {Carmona}, {Pinte}, \& {Habart}}]{Martin2010}
{Martin-Za{\"i}di}, C., {Augereau}, J., {M{\'e}nard}, F., {et~al.} 2010, \aap,
  516, A110+

\bibitem[{{Mathews} {et~al.}(2010){Mathews}, {Dent}, {Williams}, {Howard},
  {Meeus}, {Riaz}, {Roberge}, {Sandell}, {Vandenbussche}, {Duch{\^e}ne},
  {Kamp}, {M{\'e}nard}, {Montesinos}, {Pinte}, {Thi}, {Woitke}, {Alacid},
  {Andrews}, {Ardila}, {Aresu}, {Augereau}, {Barrado}, {Brittain}, {Ciardi},
  {Danchi}, {Eiroa}, {Fedele}, {Grady}, {de Gregorio-Monsalvo}, {Heras},
  {Huelamo}, {Krivov}, {Lebreton}, {Liseau}, {Martin-Zaidi},
  {Mendigut{\'{\i}}a}, {Mora}, {Morales-Calderon}, {Nomura}, {Pantin},
  {Pascucci}, {Phillips}, {Podio}, {Poelman}, {Ramsay}, {Rice},
  {Riviere-Marichalar}, {Solano}, {Tilling}, {Walker}, {White}, \&
  {Wright}}]{Mathews2010}
{Mathews}, G.~S., {Dent}, W.~R.~F., {Williams}, J.~P., {et~al.} 2010, \aap,
  518, L127+

\bibitem[{{Meeus} {et~al.}(2010){Meeus}, {Pinte}, {Woitke}, {Montesinos},
  {Mendigut{\'{\i}}a}, {Riviere-Marichalar}, {Eiroa}, {Mathews},
  {Vandenbussche}, {Howard}, {Roberge}, {Sandell}, {Duch{\^e}ne}, {M{\'e}nard},
  {Grady}, {Dent}, {Kamp}, {Augereau}, {Thi}, {Tilling}, {Alacid}, {Andrews},
  {Ardila}, {Aresu}, {Barrado}, {Brittain}, {Ciardi}, {Danchi}, {Fedele}, {de
  Gregorio-Monsalvo}, {Heras}, {Huelamo}, {Krivov}, {Lebreton}, {Liseau},
  {Martin-Zaidi}, {Mora}, {Morales-Calderon}, {Nomura}, {Pantin}, {Pascucci},
  {Phillips}, {Podio}, {Poelman}, {Ramsay}, {Riaz}, {Rice}, {Solano}, {Walker},
  {White}, {Williams}, \& {Wright}}]{Meeus2010}
{Meeus}, G., {Pinte}, C., {Woitke}, P., {et~al.} 2010, \aap, 518, L124+

\bibitem[{{Meeus} {et~al.}(2001){Meeus}, {Waters}, {Bouwman}, {van den Ancker},
  {Waelkens}, \& {Malfait}}]{Meeus2001}
{Meeus}, G., {Waters}, L.~B.~F.~M., {Bouwman}, J., {et~al.} 2001, \aap, 365,
  476

\bibitem[{{Meijer} {et~al.}(2008){Meijer}, {Dominik}, {de Koter}, {Dullemond},
  {van Boekel}, \& {Waters}}]{Meijer2008}
{Meijer}, J., {Dominik}, C., {de Koter}, A., {et~al.} 2008, \aap, 492, 451

\bibitem[{{Meijerink} {et~al.}(2009){Meijerink}, {Pontoppidan}, {Blake},
  {Poelman}, \& {Dullemond}}]{Meijerink2009}
{Meijerink}, R., {Pontoppidan}, K.~M., {Blake}, G.~A., {Poelman}, D.~R., \&
  {Dullemond}, C.~P. 2009, \apj, 704, 1471

\bibitem[{{Mendigut{\'{\i}}a} {et~al.}(2011){Mendigut{\'{\i}}a}, {Eiroa},
  {Montesinos}, {Mora}, {Oudmaijer}, {Mer{\'{\i}}n}, \& {Meeus}}]{Mend2011}
{Mendigut{\'{\i}}a}, I., {Eiroa}, C., {Montesinos}, B., {et~al.} 2011, \aap,
  529, A34+

\bibitem[{{Monnier} {et~al.}(2006){Monnier}, {Berger}, {Millan-Gabet}, {Traub},
  {Schloerb}, {Pedretti}, {Benisty}, {Carleton}, {Haguenauer}, {Kern},
  {Labeye}, {Lacasse}, {Malbet}, {Perraut}, {Pearlman}, \&
  {Zhao}}]{Monnier2006}
{Monnier}, J.~D., {Berger}, J., {Millan-Gabet}, R., {et~al.} 2006, \apj, 647,
  444

\bibitem[{{Montesinos} {et~al.}(2009){Montesinos}, {Eiroa}, {Mora}, \&
  {Mer{\'{\i}}n}}]{Montesinos2009}
{Montesinos}, B., {Eiroa}, C., {Mora}, A., \& {Mer{\'{\i}}n}, B. 2009, \aap,
  495, 901

\bibitem[{{Mora} {et~al.}(2004){Mora}, {Eiroa}, {Natta}, {Grady}, {de Winter},
  {Davies}, {Ferlet}, {Harris}, {Miranda}, {Montesinos}, {Oudmaijer},
  {Palacios}, {Quirrenbach}, {Rauer}, {Alberdi}, {Cameron}, {Deeg},
  {Garz{\'o}n}, {Horne}, {Mer{\'{\i}}n}, {Penny}, {Schneider}, {Solano},
  {Tsapras}, \& {Wesselius}}]{Mora2004}
{Mora}, A., {Eiroa}, C., {Natta}, A., {et~al.} 2004, \aap, 419, 225

\bibitem[{{Mora} {et~al.}(2002){Mora}, {Natta}, {Eiroa}, {Grady}, {de Winter},
  {Davies}, {Ferlet}, {Harris}, {Montesinos}, {Oudmaijer}, {Palacios},
  {Quirrenbach}, {Rauer}, {Alberdi}, {Cameron}, {Deeg}, {Garz{\'o}n}, {Horne},
  {Mer{\'{\i}}n}, {Penny}, {Schneider}, {Solano}, {Tsapras}, \&
  {Wesselius}}]{Mora2002}
{Mora}, A., {Natta}, A., {Eiroa}, C., {et~al.} 2002, \aap, 393, 259

\bibitem[{{Natta} {et~al.}(2000){Natta}, {Grinin}, \&
  {Tambovtseva}}]{Natta2000}
{Natta}, A., {Grinin}, V.~P., \& {Tambovtseva}, L.~V. 2000, \apj, 542, 421

\bibitem[{{Natta} {et~al.}(2004){Natta}, {Testi}, {Neri}, {Shepherd}, \&
  {Wilner}}]{Natta2004}
{Natta}, A., {Testi}, L., {Neri}, R., {Shepherd}, D.~S., \& {Wilner}, D.~J.
  2004, \aap, 416, 179

\bibitem[{{Oudmaijer} {et~al.}(2001){Oudmaijer}, {Palacios}, {Eiroa}, {Davies},
  {de Winter}, {Ferlet}, {Garz{\'o}n}, {Grady}, {Cameron}, {Deeg}, {Harris},
  {Horne}, {Mer{\'{\i}}n}, {Miranda}, {Montesinos}, {Mora}, {Penny},
  {Quirrenbach}, {Rauer}, {Schneider}, {Solano}, {Tsapras}, \&
  {Wesselius}}]{Oudmaijer2001}
{Oudmaijer}, R.~D., {Palacios}, J., {Eiroa}, C., {et~al.} 2001, \aap, 379, 564

\bibitem[{{Pani{\'c}} \& {Hogerheijde}(2009)}]{Panic2009}
{Pani{\'c}}, O. \& {Hogerheijde}, M.~R. 2009, \aap, 508, 707

\bibitem[{{Pi{\'e}tu} {et~al.}(2007){Pi{\'e}tu}, {Dutrey}, \&
  {Guilloteau}}]{Pietu2007}
{Pi{\'e}tu}, V., {Dutrey}, A., \& {Guilloteau}, S. 2007, \aap, 467, 163

\bibitem[{{Pi{\'e}tu} {et~al.}(2005){Pi{\'e}tu}, {Guilloteau}, \&
  {Dutrey}}]{Pietu2005}
{Pi{\'e}tu}, V., {Guilloteau}, S., \& {Dutrey}, A. 2005, \aap, 443, 945

\bibitem[{{Pinte} {et~al.}(2006){Pinte}, {M{\'e}nard}, {Duch{\^e}ne}, \&
  {Bastien}}]{Pinte2006}
{Pinte}, C., {M{\'e}nard}, F., {Duch{\^e}ne}, G., \& {Bastien}, P. 2006, \aap,
  459, 797

\bibitem[{{Pinte} {et~al.}(2008){Pinte}, {Padgett}, {M{\'e}nard},
  {Stapelfeldt}, {Schneider}, {Olofsson}, {Pani{\'c}}, {Augereau},
  {Duch{\^e}ne}, {Krist}, {Pontoppidan}, {Perrin}, {Grady}, {Kessler-Silacci},
  {van Dishoeck}, {Lommen}, {Silverstone}, {Hines}, {Wolf}, {Blake}, {Henning},
  \& {Stecklum}}]{Pinte2008}
{Pinte}, C., {Padgett}, D.~L., {M{\'e}nard}, F., {et~al.} 2008, \aap, 489, 633

\bibitem[{{Posch} {et~al.}(2003){Posch}, {Kerschbaum}, {Fabian}, {Mutschke},
  {Dorschner}, {Tamanai}, \& {Henning}}]{Posch2003}
{Posch}, T., {Kerschbaum}, F., {Fabian}, D., {et~al.} 2003, \apjs, 149, 437

\bibitem[{{Rayner} {et~al.}(2003){Rayner}, {Toomey}, {Onaka}, {Denault},
  {Stahlberger}, {Vacca}, {Cushing}, \& {Wang}}]{Rayner2003}
{Rayner}, J.~T., {Toomey}, D.~W., {Onaka}, P.~M., {et~al.} 2003, \pasp, 115,
  362

\bibitem[{{Renard} {et~al.}(2010){Renard}, {Malbet}, {Benisty}, {Thi{\'e}baut},
  \& {Berger}}]{Renard2010}
{Renard}, S., {Malbet}, F., {Benisty}, M., {Thi{\'e}baut}, E., \& {Berger}, J.
  2010, \aap, 519, A26+

\bibitem[{{Sandell} {et~al.}(2011){Sandell}, {Weintraub}, \&
  {Hamidouche}}]{Sandell2011}
{Sandell}, G., {Weintraub}, D.~A., \& {Hamidouche}, M. 2011, \apj, 727, 26

\bibitem[{{Sbordone} {et~al.}(2004){Sbordone}, {Bonifacio}, {Castelli}, \&
  {Kurucz}}]{Sbordone2004}
{Sbordone}, L., {Bonifacio}, P., {Castelli}, F., \& {Kurucz}, R.~L. 2004,
  Memorie della Societa Astronomica Italiana Supplementi, 5, 93

\bibitem[{{Sch{\"o}ier} {et~al.}(2005){Sch{\"o}ier}, {van der Tak}, {van
  Dishoeck}, \& {Black}}]{Schoier2005}
{Sch{\"o}ier}, F.~L., {van der Tak}, F.~F.~S., {van Dishoeck}, E.~F., \&
  {Black}, J.~H. 2005, \aap, 432, 369

\bibitem[{{Servoin} \& {Piriou}(1973)}]{Servoin1973}
{Servoin}, J.~L. \& {Piriou}, B. 1973, Physica Status Solidi (b), 55, 677

\bibitem[{{Sitko} {et~al.}(2008){Sitko}, {Carpenter}, {Kimes}, {Wilde},
  {Lynch}, {Russell}, {Rudy}, {Mazuk}, {Venturini}, {Puetter}, {Grady},
  {Polomski}, {Wisnewski}, {Brafford}, {Hammel}, \& {Perry}}]{Sitko2008}
{Sitko}, M.~L., {Carpenter}, W.~J., {Kimes}, R.~L., {et~al.} 2008, \apj, 678,
  1070

\bibitem[{{Tannirkulam} {et~al.}(2008{\natexlab{a}}){Tannirkulam}, {Monnier},
  {Harries}, {Millan-Gabet}, {Zhu}, {Pedretti}, {Ireland}, {Tuthill}, {ten
  Brummelaar}, {McAlister}, {Farrington}, {Goldfinger}, {Sturmann}, {Sturmann},
  \& {Turner}}]{Tannirkulam2008a}
{Tannirkulam}, A., {Monnier}, J.~D., {Harries}, T.~J., {et~al.}
  2008{\natexlab{a}}, \apj, 689, 513

\bibitem[{{Tannirkulam} {et~al.}(2008{\natexlab{b}}){Tannirkulam}, {Monnier},
  {Millan-Gabet}, {Harries}, {Pedretti}, {ten Brummelaar}, {McAlister},
  {Turner}, {Sturmann}, \& {Sturmann}}]{Tannirkulam2008}
{Tannirkulam}, A., {Monnier}, J.~D., {Millan-Gabet}, R., {et~al.}
  2008{\natexlab{b}}, \apjl, 677, L51

\bibitem[{{Tennyson} {et~al.}(2001){Tennyson}, {Zobov}, {Williamson},
  {Polyansky}, \& {Bernath}}]{Tennyson2001}
{Tennyson}, J., {Zobov}, N.~F., {Williamson}, R., {Polyansky}, O.~L., \&
  {Bernath}, P.~F. 2001, J. Phys. Chem. Ref. Data, 30, 735

\bibitem[{{Thi} {et~al.}(2001){Thi}, {van Dishoeck}, {Blake}, {van Zadelhoff},
  {Horn}, {Becklin}, {Mannings}, {Sargent}, {van den Ancker}, {Natta}, \&
  {Kessler}}]{Thi2001}
{Thi}, W.~F., {van Dishoeck}, E.~F., {Blake}, G.~A., {et~al.} 2001, \apj, 561,
  1074

\bibitem[{{Tielens}(2008)}]{Tielens2008}
{Tielens}, A.~G.~G.~M. 2008, \araa, 46, 289

\bibitem[{{Vacca} {et~al.}(2003){Vacca}, {Cushing}, \& {Rayner}}]{Vacca2003}
{Vacca}, W.~D., {Cushing}, M.~C., \& {Rayner}, J.~T. 2003, \pasp, 115, 389

\bibitem[{{van den Ancker} {et~al.}(2000){van den Ancker}, {Bouwman},
  {Wesselius}, {Waters}, {Dougherty}, \& {van Dishoeck}}]{vandenAncker2000}
{van den Ancker}, M.~E., {Bouwman}, J., {Wesselius}, P.~R., {et~al.} 2000,
  \aap, 357, 325

\bibitem[{{van Leeuwen}(2007)}]{vanLeeuwen2007}
{van Leeuwen}, F., ed. 2007, Astrophysics and Space Science Library, Vol. 350,
  {Hipparcos, the New Reduction of the Raw Data}

\bibitem[{{Warren} \& {Brandt}(2008)}]{Warren2008}
{Warren}, S.~G. \& {Brandt}, R.~E. 2008, Journal of Geophysical Research, 113,
  D14220

\bibitem[{{Wassell} {et~al.}(2006){Wassell}, {Grady}, {Woodgate}, {Kimble}, \&
  {Bruhweiler}}]{Wassell2006}
{Wassell}, E.~J., {Grady}, C.~A., {Woodgate}, B., {Kimble}, R.~A., \&
  {Bruhweiler}, F.~C. 2006, \apj, 650, 985

\bibitem[{{Wernli} {et~al.}(2006){Wernli}, {Valiron}, {Faure}, {Wiesenfeld},
  {Jankowski}, \& {Szalewicz}}]{Wernli2006}
{Wernli}, M., {Valiron}, P., {Faure}, A., {et~al.} 2006, \aap, 446, 367

\bibitem[{{Wilson} \& {Bell}(2002)}]{Wilson2002}
{Wilson}, N.~J. \& {Bell}, K.~L. 2002, \mnras, 337, 1027

\bibitem[{{Wisniewski} {et~al.}(2008){Wisniewski}, {Clampin}, {Grady},
  {Ardila}, {Ford}, {Golimowski}, {Illingworth}, \& {Krist}}]{Wisniewski2008}
{Wisniewski}, J.~P., {Clampin}, M., {Grady}, C.~A., {et~al.} 2008, \apj, 682,
  548

\bibitem[{{Woitke} {et~al.}(2009){Woitke}, {Kamp}, \& {Thi}}]{Woitke2009a}
{Woitke}, P., {Kamp}, I., \& {Thi}, W. 2009, \aap, 501, 383

\bibitem[{{Woitke} {et~al.}(2010){Woitke}, {Pinte}, {Tilling}, {M{\'e}nard},
  {Kamp}, {Thi}, {Duch{\^e}ne}, \& {Augereau}}]{Woitke2010}
{Woitke}, P., {Pinte}, C., {Tilling}, I., {et~al.} 2010, \mnras, 405, L26

\bibitem[{{Woitke} {et~al.}(2011){Woitke}, {Riaz}, {Duch{\^e}ne}, {Pascucci},
  {Lyo}, {Dent}, {Phillips}, {Thi}, {M{\'e}nard}, {Herczeg}, {Bergin}, {Brown},
  {Mora}, {Kamp}, {Aresu}, {Brittain}, {de Gregorio-Monsalvo}, \&
  {Sandell}}]{Woitke2011}
{Woitke}, P., {Riaz}, B., {Duch{\^e}ne}, G., {et~al.} 2011, \aap, 534, A44+

\bibitem[{{Woodall} {et~al.}(2007){Woodall}, {Ag{\'u}ndez}, {Markwick-Kemper},
  \& {Millar}}]{Woodall2007}
{Woodall}, J., {Ag{\'u}ndez}, M., {Markwick-Kemper}, A.~J., \& {Millar}, T.~J.
  2007, \aap, 466, 1197

\bibitem[{{Wu} {et~al.}(1983){Wu}, {Ake}, {Boggess}, {Bohlin}, {Imhoff},
  {Holm}, {Levay}, {Panek}, {Schiffer}, \& {Turnrose}}]{Wu1983}
{Wu}, C., {Ake}, T.~B., {Boggess}, A., {et~al.} 1983, NASA IUE Newsl., No.~22,
  2+324 pp., 22

\bibitem[{{Wu} {et~al.}(1992){Wu}, {Reichert}, {Ake}, {Boggess}, {Holm},
  {Imhoff}, {Kondo}, {Mead}, \& {Shore}}]{Wu1992}
{Wu}, C., {Reichert}, G.~A., {Ake}, T.~B., {et~al.} 1992, NASA Reference
  Publication, 1285

\bibitem[{{Yang} {et~al.}(2010){Yang}, {Stancil}, {Balakrishnan}, \&
  {Forrey}}]{Yang2010}
{Yang}, B., {Stancil}, P.~C., {Balakrishnan}, N., \& {Forrey}, R.~C. 2010,
  \apj, 718, 1062

\bibitem[{{Yi} {et~al.}(2001){Yi}, {Demarque}, {Kim}, {Lee}, {Ree}, {Lejeune},
  \& {Barnes}}]{Yi2001}
{Yi}, S., {Demarque}, P., {Kim}, Y.-C., {et~al.} 2001, \apjs, 136, 417

\bibitem[{{Zuckerman} {et~al.}(1995){Zuckerman}, {Forveille}, \&
  {Kastner}}]{Zuckerman1995}
{Zuckerman}, B., {Forveille}, T., \& {Kastner}, J.~H. 1995, \nat, 373, 494

\end{thebibliography}

%\begin{thebibliography}{}
%\end{thebibliography}

\end{document}